\documentclass[12pt,preprint]{aastex}

\shorttitle{Spiral triggering of SF in SA & SAB galaxies}
\shortauthors{Mart\'{\i}nez-Garc\'ia et al.}

\begin{document}

\title{Spiral density wave triggering of star formation in
\\ SA and SAB galaxies}

\author{Eric E. Mart\'inez-Garc\'ia, Rosa Amelia
Gonz\'alez-L\'opezlira\altaffilmark{1,2,3}}
\affil{Centro de Radioastronom\'ia y Astrof\'isica, UNAM, Campus Morelia,
     Michoac\'an, M\'exico, C.P. 58089}
\email{e.martinez@crya.unam.mx, r.gonzalez@crya.unam.mx}

\and

\author{Gustavo Bruzual-A.}
\affil{Centro de Investigaciones de Astronom\'ia, Apartado Postal 264,
     M\'erida 5101-A, Venezuela}

\altaffiltext{1}{Visiting astronomer at Kitt Peak National Observatory, National Optical
Astronomy Observatory, which is operated by the Association of Universities for
Research in Astronomy (AURA), under cooperative agreement with the National
Science Foundation.}
\altaffiltext{2}{Visiting astronomer at Cerro Tololo Inter-American
Observatory, National
Optical Astronomy Observatory, which is operated by the AURA, under contract
with the National Science Foundation.}
\altaffiltext{3}{Visiting astronomer at Lick Observatory, which is
operated by the
University of California.}

\begin{abstract}
Azimuthal color (age) gradients across spiral arms are one of the main
predictions of density wave theory; gradients are the result of
star formation triggering by the spiral waves. In a sample of 13 spiral galaxies
of types A and AB, we find that 10 of them present regions that match the theoretical
predictions. By comparing the observed gradients with stellar population synthesis
models, the pattern speed and the location of major resonances have been determined.
The resonance positions inferred from this analysis indicate that 9 of the
objects have spiral arms that extend to the outer Lindblad resonance (OLR);
for one of the galaxies, the spiral arms reach the corotation radius.
The effects of dust, and of stellar densities, velocities, and metallicities
on the color
gradients are also discussed.
\end{abstract}


\section{Introduction.}

Density wave phenomena have been proposed to explain the spiral structure seen in
disk galaxies \citep{lind63,lin64,too77,ber89a,ber89b}. Observationally,
these phenomena are best studied in the near infrared (near-IR),
especially the $K$-band, that mostly traces the old stars in the disk \citep[e.g.,][]{rix93}.
The old stellar disk, however, has a disordered optical counterpart of young stars,
gas, and dust \citep{zwi55,blo91}, so the question arises whether the spiral
structure, seen in near-IR bands, and the star formation,
seen in the optical bands, are coupled or not.
If disk dynamics and star formation are indeed related, the star formation
rate per unit gas mass should be affected by the presence of the density
wave. Unfortunately, H$_2$ cannot be directly observed, and the relation
between detected CO and total H$_2$ mass is a whole controversial topic
in itself \citep[e.g.,][ and references therein]{allen96}.

The alternative approach of comparing star formation rates, past
(as traced by optical and near-IR surface photometry) and present
(as probed by $H\alpha$ emission), in galaxies
with different Hubble types has been carried out; as a result, both a positive
correlation between disk dynamics and star formation \citep[e.g.,][]{sei02},
and the absence of such
a correlation \citep[e.g.,][]{ryd94,elm86} have been claimed.

\begin{figure}
\epsscale{1.00}
\plotone{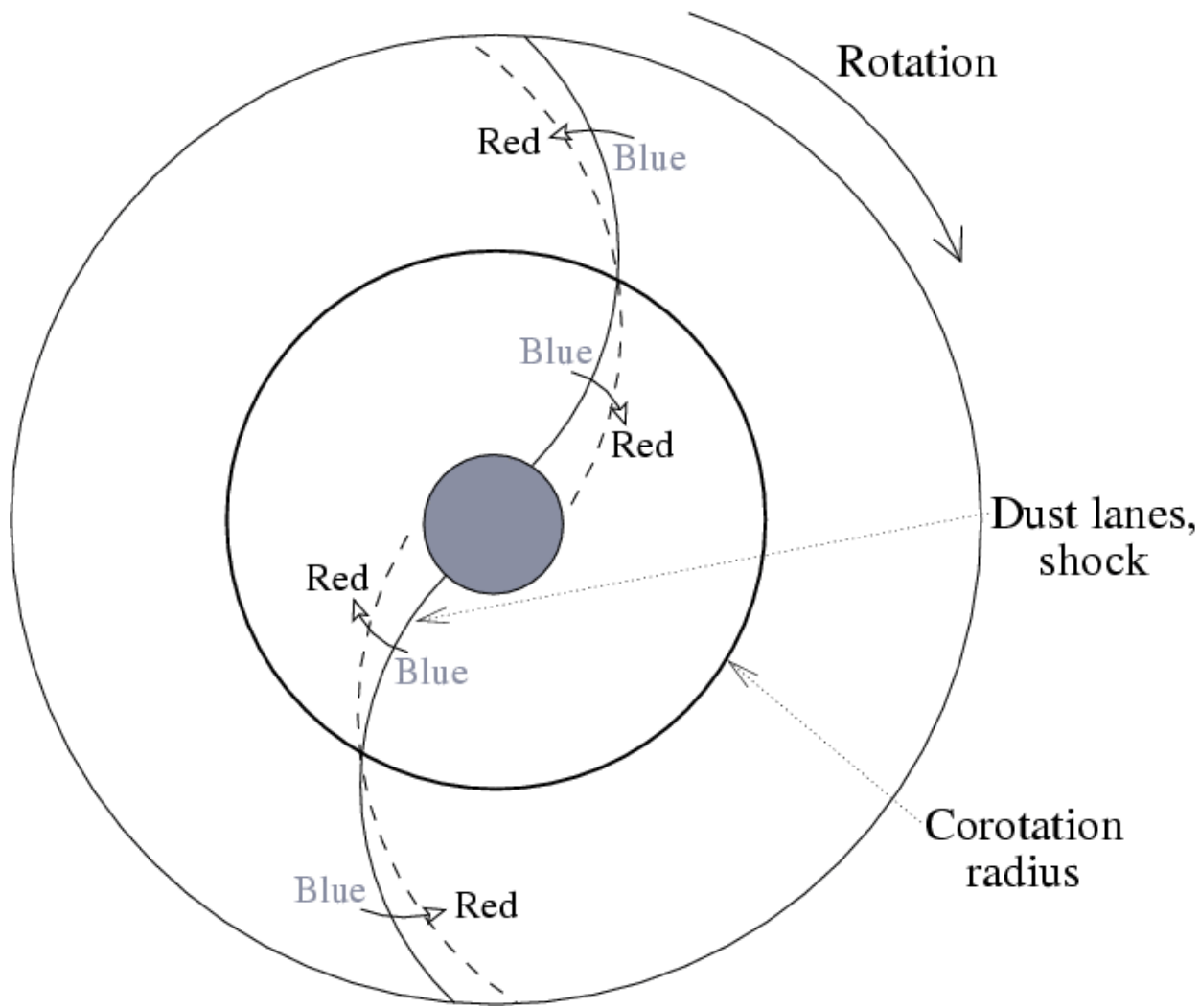}
\caption{Stellar age gradients across the spiral arms are indicated by arrows
that go from blue to red. The azimuthal age gradients are produced by
stars born in the spiral shock, where the shocked interstellar medium
forms a dust lane, that later drift away as they age.
The direction of the gradients changes at the corotation radius,
$R_{\rm CR}$. Inside this radius, the disk material overtakes the spiral wave, and beyond
it the spiral wave catches up with the material 
[see also figure 1 in \citet{pue97}]. \label{colorgrads}}
\end{figure}

In this work, we will focus on the relation between star formation
and density wave phenomena, as proposed in the large scale shock
scenario \citep{rob69,shu72}. Evidence has been gathered, from
observations of dust, gas compression \citep[e.g.,][]{mat72,vis80}, and molecular clouds
\citep{vog88,sch04} near the concave regions of spiral arms, that suggests that
star formation is triggered there. Through a simple interpretation of these
ideas and observations, the existence of azimuthal color-age gradients has been predicted.
As seen in Fig.~\ref{colorgrads},
if we assume that the spiral pattern rotates with constant angular
speed, and that the gas and stars have differential rotation,
a corotation region exists where these two angular speeds are equal.
At smaller radii, bursts of star formation
take place where the differentially rotating gas overtakes the
spiral wave, and show up as brillant HII regions
\citep[e.g.,][]{mor52,elm83a}.
As young stars age, they drift away from their birth site, thus creating a
color gradient. Star formation can occur also beyond the corotation radius,
when the spiral pattern overtakes the gas.
The assumption that the pattern rotates with constant angular speed
can be corroborated with numerical simulations \citep{tho90,don94,zha98},
but the only observational evidence of this premise is the apparent persistence
of the spiral structure for up to a Hubble time in nearby
galaxies \citep{elm83b}, in avoidance of the winding dilemma.

Another way to test the assumption of constant angular speed of the spiral pattern is
by the dynamical consequences it has on the disk environment and material.
In this regard,
the prediction of azimuthal color gradients can be seen as another
test of the spiral density wave theory.
So far, the search for these color gradients has in general yielded
inconclusive results
\citep{sch76,tal79,cep90,hod90,delRio98},
and a few exceptional positive cases like
\citet{efr85} for M31, \citet{sit89} for the
Milky Way, and \citet{gon96}  
for the spiral galaxy M~99.

\begin{figure}
\epsscale{1.00}
\plotone{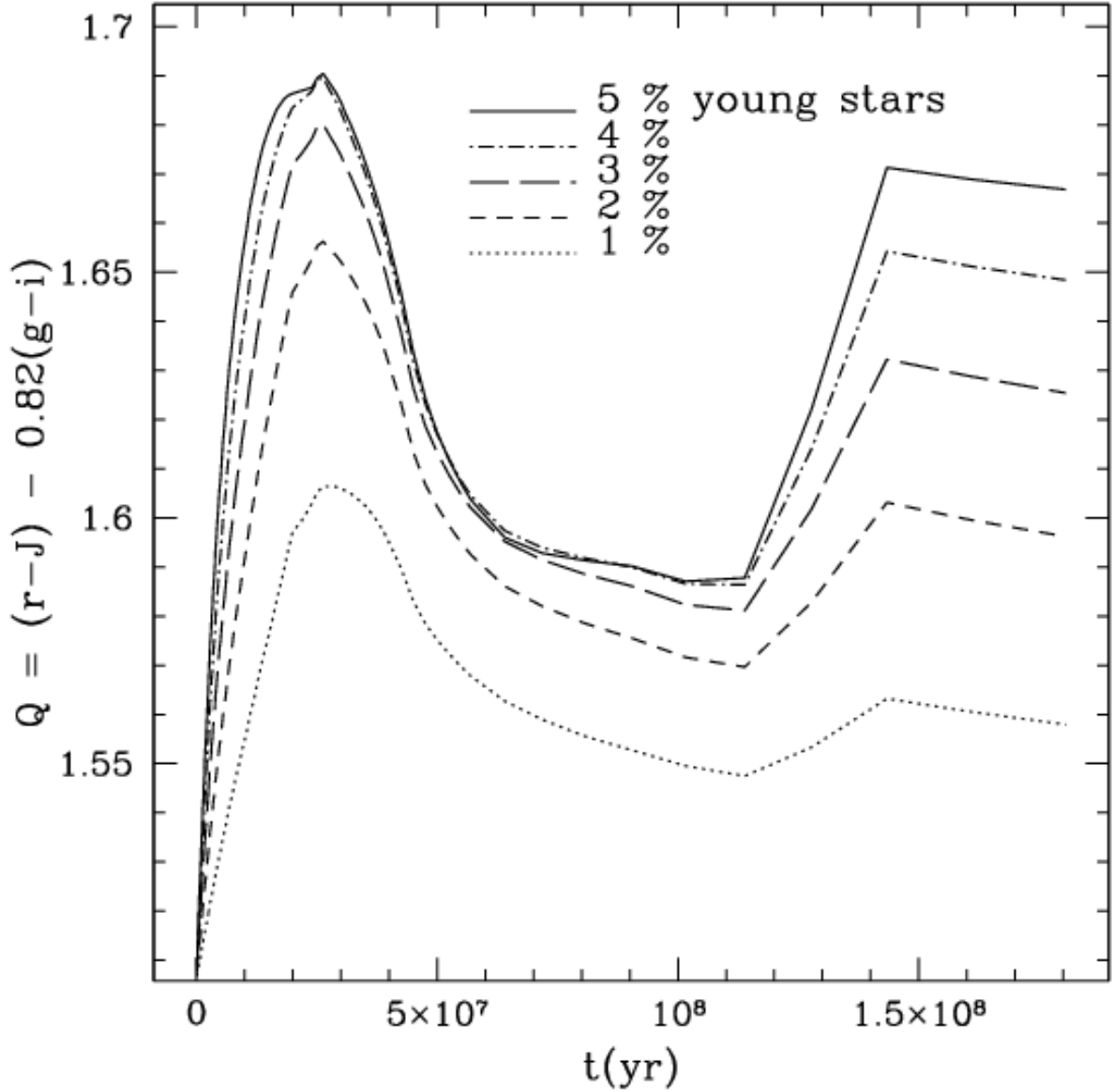}
\caption{Theoretical $Q(rJgi)$ vs.\ time,
CB07 models.
The duration of the burst is
$2 \times 10^7$ years, with a Salpeter IMF; the fraction of young stars
ranges from 1\% to 5\% by mass. Lower and upper mass limits are 0.1 and 10 $M_{\sun}$,
respectively.   \label{Qpercent}}
\end{figure}

\begin{figure}
\epsscale{1.00}
\plotone{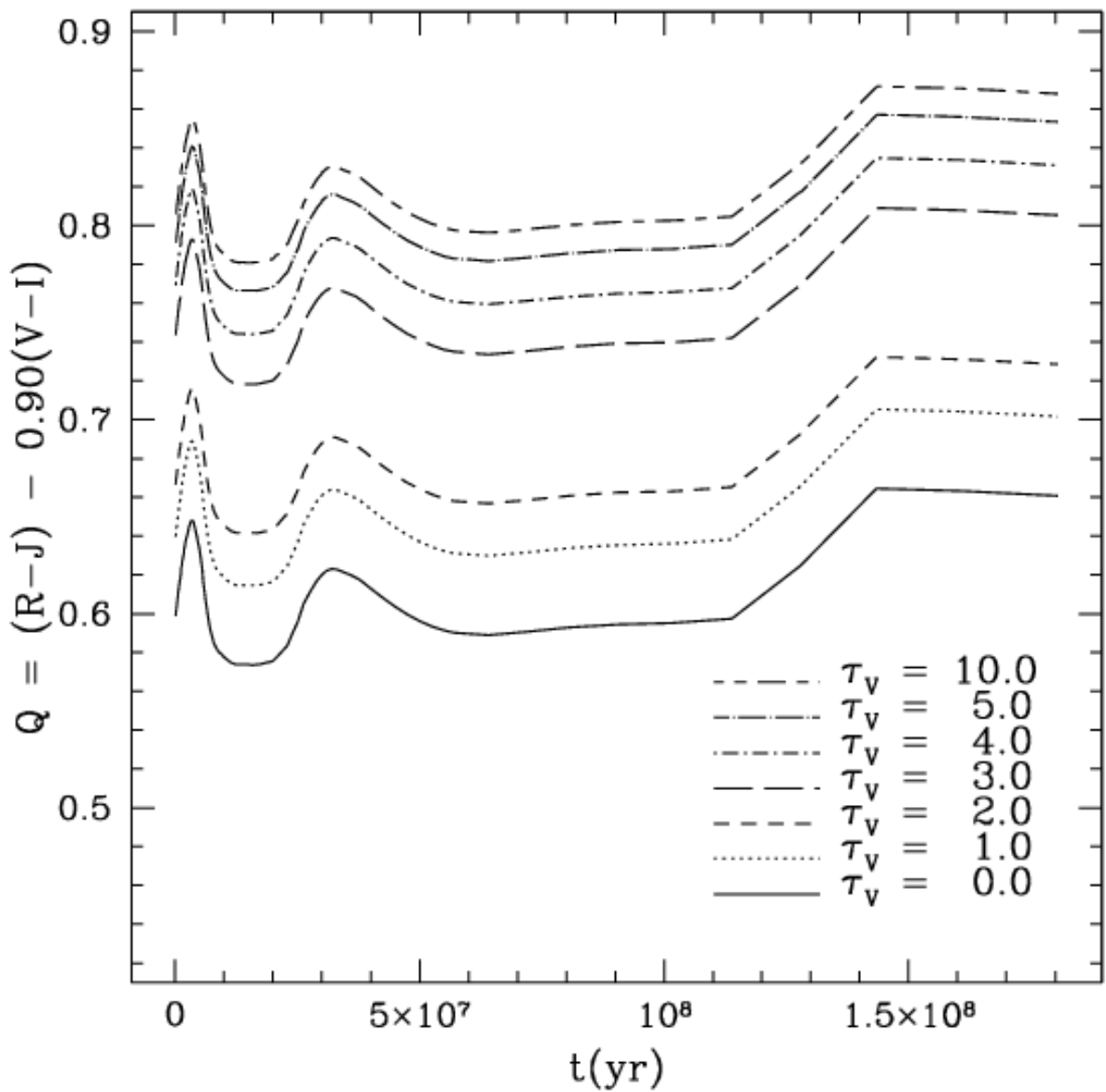}
\caption{
Model Johnson filter $Q(RJVI)$
vs.\ time $(t)$ for CB07 models,
reddened as per the ``starburst galaxy" model of \citet{wit92}.
The duration of the burst is
$2 \times 10^7$ years, with a Salpeter IMF
and 2\% by mass of young stars. Lower and upper mass limits are 0.1 and 100 $M_{\sun}$,
respectively. \label{Qdust}}
\end{figure}

\citet[][ GG96 hereafter]{gon96} used for the first time a supergiant
sensitive and reddening-free photometric index\footnote{
See Binney \& Merrifield (1998), for other reddening-free indices.} to trace star formation,
defined as:

\begin{equation}
  Q(rJgi) = (r-J) - \frac{E(r-J)}{E(g-i)}(g-i).
\end{equation}

\noindent Using the extinction curves
of \citet{sch83}, and \citet{rie85} for a foreground screen, the color excess
term is $E(r-J)/E(g-i) = 0.82$.
According to population synthesis models,
following a star formation burst
the $Q(rJgi)$ photometric index increases its value for
$\sim 2.6 \times 10^7$ years, and then starts to decline, as shown in Fig.~\ref{Qpercent}
for most recent Charlot \& Bruzual (2007, in preparation, CB07 hereafter) models with
different stellar mixtures in which young stars constitute
from 1\% to 5\% by mass. Dust lanes arising from spiral
shocks are expected to precede the azimuthal color gradients, hence the importance of
reddening-free diagnostics.
With the stellar population synthesis (SPS) models of \citet{bru93},
and the radiative transfer models of \citet{bru88} and \citet{wit92},
GG96 investigated the behavior of $Q(RJVI)$ for mixtures of dust and
stars with different relative spatial distributions. The main
effect of a mixture of dust and stars on the $Q$ index,
illustrated in Fig.~\ref{Qdust} for the CB07 population synthesis models,
is to increase
its value relative to the $\tau = 0$ and foreground dust screen cases,
and $Q(RJVI)$ is no longer reddening-free when $\tau_V > 2$.\footnote{The
same conclusion can be applied to $Q(rJgi)$, since the optical
bands $R$, $V$, and $I$ have approximately the same effective
wavelengths, respectively, as the $r$, $g$, and $i$ filters; the main
difference between both sets is that the latter passbands are narrower.}
However, the requirement that $\tau_V < 2$ is likely fulfilled by the disks of
nearly face-on galaxies \citep{pel95,kuc98,xil99}; moreover,
as stars age, dust dissipates, and reddening diminishes
\citep[e.g.,][]{char00}.

Based on their study of M~99, GG96 propose that
real data are best matched by SPS models
with 0.5\% to 2\%, by mass, of young stars. These values are in accordance
with the amount of $B$ light in the arms contributed by young stars, as estimated by \citet{sch76}.
GG96 also hypothesize there is an inverse correlation between high-mass star-forming
regions and detectable azimuthal color gradients. The color gradient in M~99 lies
where no HII regions are identified in $H\alpha$ images of the galaxy. This
counterintuitive finding could help explain the dearth of positive
detections to date, if contamination from bright emission lines produced in
HII regions around the most massive stars masks the color gradients associated
with star formation in spiral shocks \citep{shu97}.
On the other hand,  
evidence has been found recently of star forming regions with very few massive
stars that generate only scant $H\alpha$ emission \citep{inde08}.
In this investigation we apply the GG96 method, using the $Q(rJgi)$ photometric
index, to a new sample of galaxies in order to search for and analize color-age
gradients near spiral arms.

\subsection{Color gradients: the link between star formation and spiral dynamics.} \label{linkdyn}

According to various studies, spiral density waves must propagate
between orbital resonances \citep{lin70,mar76,lin79,too81,con86}.
Of these resonances, the most important ones are the inner Lindblad resonance (ILR),
the 4:1 resonance, corotation (CR), and the outer Lindblad resonance (OLR).\footnote{
At the ILR (OLR), the epicyclic frequency $\kappa = \pm 2(\Omega - \Omega_p)$;
$\kappa = 4(\Omega - \Omega_p)$ at the 4:1 resonance.
}

\noindent
Azimuthal color-age gradients
retain some information about the stellar drift relative to the
spiral shock, allowing us to obtain $\Omega_p$, the angular velocity
of the spiral pattern.
In order to find $\Omega_p$ from the gradient information, we can use:

\begin{eqnarray}
  \label{eqOMEGA_I}
  \Omega_{p} = \frac{1}{t} \left(\int_{0}^{t} \frac{\vec v(t') \cdot \hat{\varphi}(t')}{R(t')} \mathrm{d}t'
             - \left(\theta_{\rm shock} + \Delta\theta \right)\right), \\
  \Delta\theta = \cot(-i) \ln \left(\frac{R(t)}{R(0)}\right),
\end{eqnarray}

\noindent
where $t$ is the age of the young stellar population at an angle
$\theta_{\rm shock}$ away from the shock position ($t'$ is a variable
of integration),
$\vec v(t')$ is the velocity vector of the young stellar population in an
inertial reference system; $\hat{\varphi}(t')$ is the unit vector in plane
polar coordinates $\rho$, $\varphi$, in a non inertial reference system;
$R(t')$ is the orbital radius of the studied region,
measured from the center of the disk to the center of mass
of the young stellar population;\footnote{$R(t)$ corresponds to $\theta_{\rm shock}$.}
and $i$ is the arm pitch angle,\footnote{The angle between a tangent
to the spiral arm at a certain point and a circle, whose center coincides
with the galaxy's, crossing the same point.}
if the azimuthal angle increases in the
direction of rotation and the spiral arms trail.
The angular quantity $\Delta\theta$ accounts for the logarithmic
spiral shape of the shock.
Assuming the departures from circular motion are small, we have:

\begin{equation}
  \Omega_{p} = \frac{1}{R_{mean}} \left(\frac{\int_{0}^{t} v_{rot}(t') \mathrm{d}t' }{t} - \frac{d}{t} \right),
\end{equation}

\noindent
where $R_{mean}$ is the mean orbital radius of the studied region, $v_{rot}$ is the
circular orbital velocity in an inertial reference system, and $t$ is the
age of the young stellar population at the azimuthal distance $d$ from the shock.
Expanding the term $\int_{0}^{t} v_{rot}(t') \mathrm{d}t'$, in a Taylor series, we get :

\begin{equation}
  \Omega_{p} = \frac{1}{R_{mean}} \left(\frac{(v_{rot})|_{0} t +
  \frac{1}{2}  (\frac{\mathrm{d} v_{rot}}{\mathrm{d}t'})|_{0} t^{2}
  + \zeta}{t} - \frac{d}{t} \right),
\label{eqOmega_Ib}
\end{equation}

\noindent
where $\zeta$ represents higher order terms in the expansion.
According to theory \citep[e.g.,][]{rob69,sly03}, the higher order terms may account for 20-30 km s$^{-1}$,
a quantity that is of the same order as the mean rotation velocity error,
after considering the uncertainty due to galaxy inclination (see Table~\ref{tbl-param}).
Hence, we can neglect the higher order terms in eq.~\ref{eqOmega_Ib}, such that $v_{rot} \sim$
constant, and obtain

\begin{equation} \label{eqOMEGA_II}
  \Omega_{p} \cong \frac{1}{R_{mean}} \left(v_{rot} - \frac{d}{t} \right).
\end{equation}

\noindent $\Omega_p$ is found by stretching the model (which gives $Q$ as a function of
$t$) to fit the data (where $Q$ is a function of $d$).

\section{Observations and data reduction.}

Our total sample of objects consists of 31 almost face-on spiral galaxies of
various Hubble types, with angular diameters between $4\arcmin$ and $6\arcmin$.
This sample was chosen from the Uppsala general catalogue of
galaxies \citep[UGC;][]{nil73}, the
ESO-Uppsala survey of the ESO (B) atlas \citep{lau82}, and the Second reference catalogue
of bright galaxies
\citep{dva76}.
From this sample, we further select for their analysis in the present
paper 13 A and AB \citep{dva59} galaxies,
based on the visual inspection of their 2-D images in the  $Q(rJgi)$ photometric index.
In this diagnostic, the disks of some of the galaxies in our original
sample of 31 clearly appear divided in
two halves, each one with a different average value of $Q$.
The analysis of these ``$Q$ effect" galaxies and of the remaining barred galaxies
will be undertaken in subsequent publications.

The data were acquired during 1992-1995 with four different telescopes:
the Lick Observatory 1-m, the Kitt Peak National Observatory (KPNO) 1.3-m, and the Cerro Tololo
Interamerican Observatory (CTIO) 0.9-m and 1.5-m telescopes.
Deep photometric images were taken in the optical filters
$g$, $r$, and $i$, and in the near-IR
$J$, $K_s$ \citep{per98} or $K^\prime$ \citep{wai92}.
Effective wavelengths and widths of all the filters are
listed in Table~\ref{tbl-filters}; the observation log for the 13 galaxies
is shown in Table~\ref{tbl-obslog}.

\begin{deluxetable}{ccc}
\tabletypesize{\scriptsize}
\tablecaption{Filter characteristics \label{tbl-filters}}
\tablewidth{0pt}
\tablehead{
\colhead{Filter} & \colhead{$\lambda_{eff}$} & \colhead{FWHM}
}
\startdata
\emph{$g$}       & 5000\AA & 830\AA \\
\emph{$r$}       & 6800\AA & 1330\AA \\
\emph{$i$}       & 7800\AA & 1420\AA \\
\emph{$J$}       & 1.25\micron & 0.29\micron \\
\emph{$K_s$} & 2.16\micron & 0.33\micron \\
\emph{$K^\prime$}     & 2.11\micron & 0.35\micron \\
\enddata
\end{deluxetable}

The CCD at the Lick 1-m telescope was a Ford $2048^{2}$, with
a pixel scale of $0\farcs185$ pixel$^{-1}$. For the infrared
observations at Lick, the same telescope was fitted with the LIRC-2 camera; it had
a $256^{2}$ NICMOS II detector, with a $1\farcs145$ pixel$^{-1}$ plate scale.
The CTIO 0.9-m optical telescope used a Tek $1024^{2}$
and a Tek $2048^{2}$ CCDs, both with a $0\farcs4$  pixel$^{-1}$
plate scale. The CTIO infrared observations were performed at the 1.5-m
telescope, with the CIRIM instrument, which used a $256^{2}$ NICMOS3 array;
the CIRIM focus was adjusted to give a $1\farcs16$ pixel$^{-1}$ plate scale.
The KPNO infrared observations were made with the IRIM camera, that
employed a $256^{2}$ NICMOS3 array, with a $2\arcsec$ pixel$^{-1}$ plate scale.

The data were reduced with the image processing package IRAF\footnote{
IRAF is distributed by the National Optical Astronomy Observatories,
which are operated by the Association of Universities for Research
in Astronomy, Inc., under cooperative agreement with the National
Science Foundation.} \citep{tody86,tody93}, using standard techniques.
For the optical data, overscan and trimming corrections were first applied.
Bias subtraction and flat field division corrections were used.
Master flats were produced from stacks of twilight flats that
were averaged pixel by
pixel, after scaling each flat by its median value and sigma-clipping
deviant pixels;
the master flats were then divided by their mean pixel value.
For sky subtraction, the sky level was determined by
masking bright objects in the image and subsequently fitting a  
constant value to the remaining pixels, after rejection iterations.
To produce mosaics, individual frames were superpixelated by a factor of 2 in
each dimension, and registered to the nearest half (original) pixel.
Before adding into the final mosaic, cosmic rays were removed by comparing
each individual image with a median one.

A correction for non-linearity was applied to the near-IR data. The correction
was obtained by adjusting a polynomial function, pixel by
pixel, to dome flats of increasing exposure times,  
taking into account variations in the count-rate between exposures.
During observations, frames were mostly taken in the sequence
SKY-SKY-OBJECT-OBJECT-SKY-SKY\dots\
Objects in the sky were masked, and the sky frames were  
median scaled, averaged, and subtracted from the object.
This procedure also takes care of dark current removal.
Flat field corrections were applied with master flats derived from dome flats.
Cosmic ray removal and mosaic registering were done with the same procedure
used for the optical data.

The optical calibration was performed using synthetic photometry in
the Thuan-Gunn system \citep{thu76,wad79}. This system is based on
the standard star BD+17$\degr$4708, to which the magnitude $g = 9.5$,
and the color indices $g - r = r - i = 0$ are assigned. Synthetic magnitudes
for other spectrophotometric standards\footnote{
Feige 15, 25, 34, 56, 92, 98; Kopff 27;
LTT 377, 7987, 9239; EG 21; BD+40$\degr$4032;
and Hiltner 600.} were obtained by means of the following relation:

\begin{equation}    
{\rm mag}_{syn} = -2.5 \log_{10} \left( \frac{\int_{0}^{\infty} f_{\lambda} R(\lambda) d\lambda}
                                             {\int_{0}^{\infty} R(\lambda) d\lambda}\right) + ZP,
\end{equation}

\noindent where $f_{\lambda}$ is the absolute spectral
energy distribution of the star
in ergs s$^{-1}$ cm$^{-2}$ ${\rm \AA}^{-1}$; $ZP$ is the zero
point,
and $R(\lambda)$ is the system response curve,
including the quantum efficiency of the detector, the transmission
function of the filter, and the effect of atmospheric absorption.
The atmospheric transmission curve for the northern hemisphere observations was
taken from \citet{hay70},
and for CTIO from \citet{ham92}. The absolute fluxes were obtained, for
BD+17$\degr$4708
from \citet{oke83}, and for the remaining standard stars
from \citet{sto77}, \citet{mas88},
\citet{mas90}, \citet{ham92}, and \citet{ham94}.
Individual galaxy frames taken under photometric conditions were used for
calibration, and mosaics were scaled to such photometric data.

For the infrared data, we only had one photometric season
at KPNO and none at CTIO. In order to obtain a uniform photometry,
we calibrated our $J$ and $K_s$ data with images from the  
Two Micron All Sky Survey \citep[2MASS,][]{skr97,skr06}.
For the $K^\prime$ data, we use photometric standards from \citet{haw01},
and adopt
$K^\prime  = K + 0.2(H - K)$ \citep{wai92}.
We do not include color correction terms in our infrared calibration,
but take this systematic error into account in the zero point uncertainty.

Finally, the optical images were degraded to the lower resolution of the
infrared images,
and aligned with them in order to proceed.

\begin{deluxetable}{ccccc}
\tabletypesize{\scriptsize}
\tablecaption{Observation Log\label{tbl-obslog}}
\tablewidth{0pt}
\tablehead{
\colhead{Object} & \colhead{Filter} & \colhead{Exposure(s)} & \colhead{Telescope}
& \colhead{Date (month/year)}
}
\startdata

NGC 4939 \dots\dots& $g$     & 2700. & CTIO 0.9 m       &  3/94 \\
                 & $g$     & 300.  & Lick 1 m         &  4/94 \\
                 & $r$     & 2400. & CTIO 0.9 m       &  3/94 \\
                 & $r$     & 300.  & Lick 1 m         &  4/94 \\            
                 & $i$     & 3600. & CTIO 0.9 m       &  3/94 \\
                 & $i$     & 300.  & Lick 1 m         &  4/94 \\            
                 & $J$     & 840.  & Kitt Peak 1.3 m  &  3/94 \\
                 & $K_{s}$ & 515.  &    ''            &   ''  \\            
NGC 3938 \dots\dots& $g$     &  600. & Lick 1 m         &  4/94  \\
                 & $r$     &  300. &    ''            &   ''  \\
                 & $i$     & 1189. &    ''            &  2/94, 4/94 \\
                 & $J$     & 1252. & Lick 1 m         &  2/95 \\
                 & $J$     &  600. & Kitt Peak 1.3 m  &  3/94 \\
           & $K_{s}$ & 560.  &    ''            &   ''  \\
NGC 4254 \dots\dots& $g$     & 4443. & Lick 1 m         & 4/94  \\
                 & $r$     & 2100. &    ''            &  '' \\
                 & $i$     & 3308. &    ''            &  '' \\
                 & $J$     & 501.  & Lick 1 m         & 2/95 \\
           & $J$     & 960.  & Kitt Peak 1.3 m  & 3/94, 11/94 \\
                 & $K_{s}$ & 590.  &    ''            &  '' \\
NGC 7126 \dots\dots& $g$     & 3900. & CTIO 0.9 m       &  9/94  \\
                 & $r$     & 3900. &    ''            &   ''  \\
                 & $i$     & 3900. &    ''            &   ''  \\
                 & $J$     &  300. & CTIO 1.5 m       &  9/95 \\
                 & $K_{s}$ &  390. &    ''            &  9/95 \\                  
NGC 1417 \dots\dots& $g$     & 3600. & Lick 1 m         &  9/93, 10/93  \\
                 & $r$     & 5048. &    ''            &  9/93, 10/93, 11/93 \\
                 & $i$     & 3600. &    ''            &  9/93, 10/93  \\
                 & $J$     &  720. & Kitt Peak 1.3 m  &  9/93 \\
                 & $K_{s}$ &  256. &    ''            &   '' \\
NGC 7753 \dots\dots& $g$     & 4500. & Lick 1 m         &  11/93, 10/94, 11/94 \\
                 & $r$     & 5812. &    ''            &   ''   \\
                 & $i$     & 5307. &    ''            &   ''  \\
                 & $J$     &  626. & Lick 1 m         &  10/94 \\
                 & $K'$    & 113.4 &    ''            &   ''  \\
NGC 6951 \dots\dots& $g$     & 6120. & Lick 1 m         &  6/92, 8/92, 9/93  \\
                 & $r$     & 6900. &    ''            &   ''  \\
                 & $i$     & 6296. &    ''            &  7/92, 8/92, 9/93 \\
                 & $J$     & 2136. & Lick 1 m         &  9/93, 11/94 \\
                 & $J$     & 4140. & Kitt Peak 1.3 m  &  7/94 \\
           & $K_{s}$ &  608. &    ''            &  9/93, 11/94 \\  
NGC 5371 \dots\dots& $g$     & 5100. & Lick 1 m         &  6/92, 3/93, 4/94 \\
                 & $r$     & 6600. &    ''            &   ''  \\
                 & $i$     & 2375. &    ''            &  6/92, 4/94 \\
                 & $J$     &  501. & Lick 1 m         &  2/95 \\
                 & $J$     &  870. & Kitt Peak 1.3 m  &  3/94 \\            
                 & $K_{s}$ & 520.  &    ''            &   ''  \\  
NGC 3162 \dots\dots& $g$     & 5400. & Lick 1 m         &  3/93, 11/93, 4/94, 10/94, 11/94  \\
                 & $r$     & 6000. &    ''            &  4/93, 11/93, 4/94, 10/94, 11/94 \\
                 & $i$     & 1800. &    ''            &  11/93, 4/94, 10/94, 11/94 \\
                 & $J$     &  990. & Kitt Peak 1.3 m  &  3/94 \\
                 & $K_{s}$ &  560. &    ''            &   ''  \\
NGC 1421 \dots\dots& $g$     & 3600. & CTIO 0.9 m       &  9/94 \\
                 & $r$     & 3600. &    ''            &   ''  \\
                 & $i$     & 3550. &    ''            &   ''  \\
                 & $J$     & 2241. & Lick 1 m         &  10/94, 12/94 \\
           & $J$     &  330. & CTIO 1.5 m       &  9/94  \\
           & $J$     &  300. & Kitt Peak 1.3 m  &  11/94 \\
                 & $K_{s}$ &  160. & CTIO 1.5 m       &  9/94  \\
           & $K_{s}$ &   60. & Kitt Peak 1.3 m  &  11/94 \\
NGC 7125 \dots\dots& $g$     & 3900. & CTIO 0.9 m       &  9/94  \\
                 & $r$     & 3900. &    ''            &   ''  \\
                 & $i$     & 3900. &    ''            &   ''  \\
                 & $J$     &  600. & CTIO 1.5 m       &  9/95 \\
                 & $K_{s}$ &  285. &    ''            &   ''  \\      
NGC 918 \dots\dots & $g$     & 4500. & Lick 1 m         &  11/93, 10/94, 11/94 \\
                 & $r$     & 4500. &    ''            &   ''  \\
                 & $i$     & 5400. &    ''            &   ''  \\
                 & $J$     & 2567. & Lick 1 m         &  10/94, 12/94 \\
                 & $K'$    & 897.6 &    ''            &   ''   \\
NGC 578 \dots\dots & $g$     & 3900. & CTIO 0.9 m       &  9/94  \\
                 & $r$     & 3600. &    ''            &   ''   \\
                 & $i$     & 4800. &    ''            &   ''   \\
                 & $J$     &  240. & CTIO 1.5 m       &   9/95 \\
                 & $K_{s}$ &  375. &    ''            &   ''   \\
                                                                                                                                             
\enddata

\end{deluxetable}

Prior to analysis proper, the images were deprojected using position angles ($P.A.$)
and isophotal diameter ratios taken from the Third Reference Catalogue of
Bright Galaxies \citep[RC3,][]{dva91}, unless indicated otherwise in Table~\ref{tbl-param}.
The inclination angle ($\alpha$) was obtained using the simple
approximation  $\cos \alpha = (b/a)$.

The spiral arms where then unwrapped by plotting them in a $\theta$ vs.\ ln $R$
map \citep{iye82,elm92}. Under this geometric transformation,
logarithmic spirals appear as straight lines with slope = cot(-$i$),
where $i$ is the arm pitch angle.
Following the procedure used by GG96,
the phase of $\theta$ was then changed
for each (fixed) value of ln $R$, until the arms appeared
horizontal. This way, selected regions with candidate color gradients
can be easily collapsed in ln $R$ to yield 1-D plots of intensity
vs.\ distance; features in such plots have a higher signal-to-noise ratio
than in 2-D images, and can
be directly compared to the SPS models.  


\begin{deluxetable}{ccccccc}
\tabletypesize{\scriptsize}
\tablecaption{Galaxy parameters\label{tbl-param}}
\tablewidth{0pt}
\tablehead{
\colhead{Name} & \colhead{Type} & \colhead{P.A.\ (degrees)}
& \colhead{$a/b$} & \colhead {$v_{max}$ (km s$^{-1}$) }
& \colhead{Radial velocity (km s$^{-1}$) }
& \colhead{Distance (Mpc) }
}
\startdata
NGC 4939 & SA(s)bc   &  10  & $1.95\pm0.13$ & $207.5\pm7.6$   & $3111\pm5$ & $46.5\pm4.0$  \\
NGC 3938 & SA(s)c    &  52\tablenotemark{a}  & $1.10\pm0.05$ & $39.2\pm3.6$ & $809\pm4$ & $15.8\pm1.4$ \\
NGC 4254 & SA(s)c    &  68\tablenotemark{b}   & $1.15\pm0.05$ & $106.7\pm6.4$ &
                                   $2407\pm3$ &  $16.5\pm1.1$\tablenotemark{c}  \\
NGC 7126 & SA(rs)c   &  80  & $2.19\pm0.15$ & $152.4\pm2.8$   & $3054\pm8$ & $44.6\pm3.8$  \\                        
NGC 1417 & SAB(rs)b  &  10  & $1.62\pm0.11$ & $205.1\pm4.3$   & $4120\pm17$ & $57.1\pm4.8$  \\
NGC 7753 & SAB(rs)bc &  50  & $1.58\pm0.15$ & $164.8\pm9.5$   & $5163\pm4$ & $72.1\pm6.1$  \\
NGC 6951 & SAB(rs)bc & 170  & $1.20\pm0.08$ & $144.2\pm6.0$   & $1426\pm6$ & $24.9\pm2.1$  \\
NGC 5371 & SAB(rs)bc &      8  & $1.26\pm0.09$ & $177.0\pm11.8$   & $2553\pm7$ & $43.5\pm3.7$  \\
NGC 3162 & SAB(rs)bc &  31\tablenotemark{a}  & $1.20\pm0.08$ & $76.6\pm7.8$   & $1298\pm7$ & $23.7\pm2.1$  \\
NGC 1421 & SAB(rs)bc &   0  & $4.07\pm0.28$ & $165.9\pm8.8$   & $2090\pm5$ & $29.3\pm2.5$  \\
NGC 7125 & SAB(rs)c  & 110  & $1.45\pm0.10$ & $104.2\pm18.2$  & $3054\pm8$ & $44.6\pm3.8$  \\
NGC 918  & SAB(rs)c  & 158  & $1.70\pm0.16$ & $112.5\pm6.2$   & $1509\pm4$ & $21.7\pm1.8$  \\
NGC 578  & SAB(rs)c  & 110  & $1.58\pm0.07$ & $117.8\pm4.1$     & $1630\pm4$ & $22.7\pm1.9$  \\

\enddata

\tablecomments{
Col.\ (2) and (3). Types and position angles from RC3.
Col.\ (4). Isophotal diameter ratio derived from the $R_{25}$ parameter
in RC3. Col.\ (5). Maximum rotation velocity obtained from the HI
data of \citet{pat03}, not corrected for inclination.
Col.\ (6). Heliocentric radial velocity from RC3.
Col.\ (7). Hubble distance obtained from the heliocentric radial velocity and
the infall model of \citet{mou00}.
}

\tablenotetext{a}{\citet{pat00}}

\tablenotetext{b}{\citet{pho93}}

\tablenotetext{c}{Distance to Virgo from \citet{mei07}, adopted since
NGC~4254 (M~99) is a member of the Virgo cluster (the procedure used for
the other galaxies yields a Hubble distance  of $40.2\pm3.5$ Mpc).}

\end{deluxetable}


\section{The $Q(rJgi)$ photometric index and the stellar 
population synthesis models.} \label{qnmod}

With the aim of tracing star formation across the spiral arms of disk
galaxies we use the $Q(rJgi)$ reddening-free photometric index, as defined
by GG96. The $Q(rJgi)$ index was chosen between different possible
filter combinations for two reasons: (1) the quotient of relevant
color excesses is close to unity, and hence all bands involved have similar
weights,
and (2) because the SPS models \citep[e.g.,][]{bru03}
predict a detectable gradient. To explain how $Q$ works, GG96 express
it in logarithmic form:

\begin{equation}
  Q(rJgi) = \log_{10} \frac{I^{2.05}_{g} I^{2.50}_{J}}{I^{2.50}_{r} I^{2.05}_{i}},
\label{qeq}
\end{equation}

\noindent where $I_x$ is the light intensity in each filter.
In active star forming regions, blue and red supergiants dominate the scene,
and the $Q(rJgi)$ index has higher values because the $g$ and $J$
bands in the numerator of eq. \ref{qeq} are tracing, respectively, the light from
these stars. Conversely, when star formation activity is poor, the $Q(rJgi)$
index has a relative lower value.

For the present work, we use a preliminary version of the
CB07 models for comparison with the observations.\footnote{
The CB07 code uses the Padova 1994 single stellar population evolutionary
tracks as assembled and described by \citep{bru03}, but includes the  
new prescription for stellar evolution in the thermally-pulsating
asymptotic giant branch (TP-AGB) by \citet{mari07}. Some differences between
the two sets of models are illustrated in \citet{emin08}.  
}
A star formation burst was added to
a background population of old stars with an age of $5 \times 10^9$ years;
both components have a Salpeter IMF. Using an older background population makes no
difference in $Q(rJgi)$.
The burst has a duration of $2 \times 10^7$ years,
in accordance with the age spread of OB associations inferred
from observations \citep[e.g.,][]{elm77,doo85,mas89,bur05}.\footnote{The
star formation timescale in these OB associations is a matter of debate.
The magnetic field-regulated model of star formation \citep{shu87}
predicts timescales in the range 5-10 Myr, while in the supersonic
turbulence-regulated star formation model \citep[e.g.,][]{bal99}
the process up to the  of pre-main-sequence stars
takes around 3 Myr. \citet{mou06} discuss observational
evidence from external galaxies, indicating that the lifetime of
molecular clouds and the timescale of star formation are
$\sim10^{7}$ years. However, \citet{bal07} propose that the large age spread
observed in OB associations is due to successive generations of stars,
triggered by the effects of stellar energy input through photoionization,
stellar winds, and supernovae. This process eventually slows down as gas
gets dispersed and the gravitational potential changes.
Our adopted time for star formation ($2 \times 10^7$ years) is in
accordance with the observational results, although the details
of the star formation processes involved may differ.}

Each of the four bands needed to
produce the model $Q(rJgi)$ index was calculated as shown below for $g$:

\begin{equation} \label{eqGSPM}
   g_{\rm csp}(t) = -2.5\log_{10} (\beta_{I} 10^{-0.4 g_{\rm yng}(t)} + \beta_{II} 10^{-0.4 g_{\rm old}}),
\end{equation}

\noindent where $g_{\rm csp} (t)$ is the $g$-band surface brightness, in mag
arcsec$^{-2}$, of the composite model
stellar population (csp) made up of young and old stars, at time
$t$ after the formation of the young stars. $\beta_{I}$ and $\beta_{II}$ are, respectively, the
fractions of young and old stars by mass ($\beta_{II} = 1 - \beta_{I}$);
$g_{\rm yng} (t)$ is the $g$ surface brightness of the
young single stellar population at time $t$;
and $g_{\rm old}$ is the $g$-band surface brightness
of the old single stellar population. 
The same holds for the $r$, $i$, and $J$ bands.

\section{Results.}

For the time being, we adopt models with constant stellar velocity 
(i.e., stellar orbits are assumed to be circular) and with a constant
fraction of young stars of 2\% by mass.
IMF upper mass limits of both 10 and 100 $M_{\sun}$ are considered. No data were matched by
the models with $M_{upper} = 100 M_\odot$, however.
We also use models with solar metallicity in both populations, and shift them to
the surface brightness level observed from the data for each object. A possible 
reason for the need of this shift is metallicity, as will be explained in \S~\ref{rolmeta}.
The rigorous application of more complex models would require an individual treatment of
each galaxy and region.

Once $\Omega_p$ is known, it is possible to locate major resonances in the
galaxy deprojected images, thus providing a link between the color gradients
and the dynamics of the spiral disk. That is, if the gradients are indeed caused by
star formation in large-scale shocks, the positions of the spiral endpoints
must be consistent with the location of major resonances, as deduced
from a comparison between the derived $\Omega_p$ and the
known orbital velocity.

There are rotation curves
available in the literature for a few objects in our sample, but observed with different criteria.
With the purpose of homogenizing our sample, we use the HI data of
\citet{pat03},
who give the maximum rotation velocity uncorrected for inclination. These
values are shown in Table~\ref{tbl-param}.
An important aspect to notice in equation~\ref{eqOMEGA_II} is that $R_{mean}$,
$v_{rot}$, and $d$ depend on the inclination angle $\alpha$ used to deproject
the images. An independent variable form of this equation is shown
in the Appendix (see equation~\ref{eqOMEGA_pix}),
which we use to obtain the uncertainty in $\Omega_p$.
Also, in order to determine $\Omega_{p}$, the distance to the object must be known.
We use the model of \citet{mou00} to convert galaxy heliocentric
velocities, obtained from the RC3 catalogue, to
Hubble flow velocities; from these we find the distance, adopting
H$_0=71\pm6$ km s$^{-1}$ Mpc$^{-1}$ \citep{mou00}.
At small redshift, the uncertainty in the
model is expected to be $\sim 100$ km s$^{-1}$. The uncertainty for any {\it single}
galaxy is about 100 km s$^{-1}$, plus the field dispersion (perhaps as big
as 250 km s$^{-1}$, and much larger in cluster cores), added
in quadrature. The model works well for groups and clusters
locally, and less well (by definition) for individual galaxies,
unless they are in a quiet part of the flow (Huchra, J.~P. 2008, private communication).
The heliocentric velocities for our objects and the distances derived from this model are shown
in Table~\ref{tbl-param}. To obtain the distance uncertainty we propagate
the errors of the heliocentric and infall velocities in the equations
of the \citeauthor{mou00} model.

All the spiral arms visible in the mosaics were inspected
for profiles similar to those expected from azimuthal
color gradients. Only a few regions across the arms of the analyzed objects
present profiles that match the theoretical expectations.
These regions are marked in the optical mosaics of
each individual object.
The $Q(rJgi)$ index data, tracing star formation; the
$(g-J)$ data, outlining the dust lane location, and the $K_s$ (or $K^\prime$)
data, following the density wave, are shown for each single region.
The site with the highest $(g-J)$ value (i.e., the highest extinction)
is taken as the origin of the (azimuthal) distance (in kpc), which increases
in the sense of rotation; the direction of rotation was inferred assuming that arms are trailing.
The comparison between model and data $Q(rJgi)$ index is shown in separate
figures, where the vertical error bars correspond to $\pm1\sigma$
of the computed uncertainty in $Q$, including read-out noise and sky subtraction.
The horizontal bars represent the uncertainty in the ``stretch'' applied to the 
the model in order to fit the data. This is ultimately an uncertainty in the
stellar drift velocity and, hence, in $\Omega_p$, and is obtained as explained in the
Appendix. We include in this error an estimate of the different stretch that 
would be required by models with 
variable densities, velocities and metallicities discussed in \S~\ref{spshsc}
(see figure \ref{Qvar10}).
The resonance locations determined once $\Omega_p$ was obtained from stretching the stellar
population model to the data are marked on the infrared mosaics.
The observed parameters for each region and the derived
dynamic ones are summarized in Table~\ref{tbl-omegas}.
The $Q$ index magnitude offset, applied to the solar metallicity model with 2\% of young
stars in order to fit the data, is also listed.

\begin{figure}
\epsscale{1.00}
\plotone{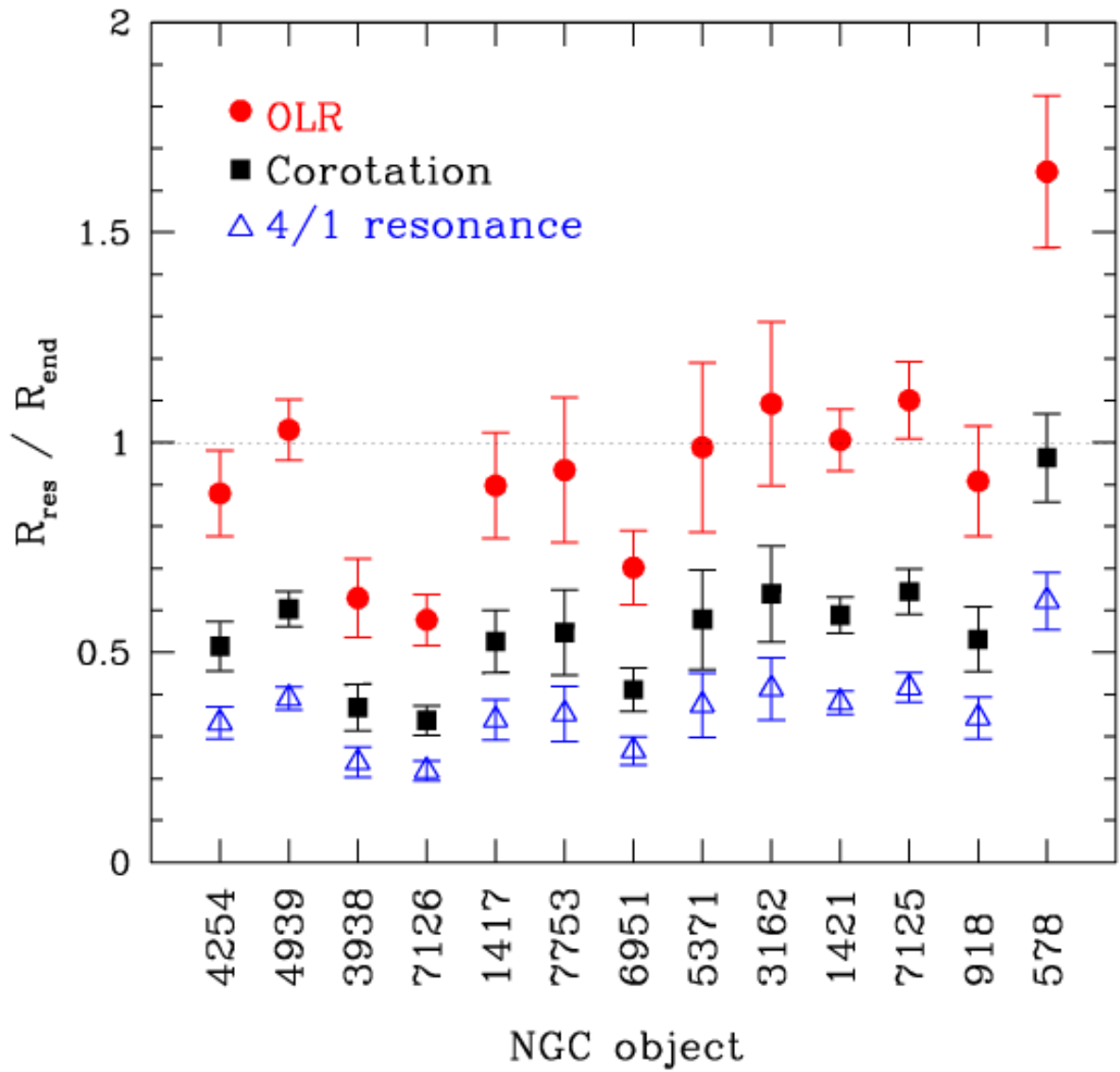}
\caption {
Ratios of resonance positions, $R_{res}$, to spiral endpoints, $R_{end}$.
Except in the case of NGC~4254 (M~99),
objects are organized by their Hubble type; values of the
resonances and spiral endpoints are listed in Table ~\ref{tbl-omegas}.
{\it Filled circles:} OLR; {\it filled squares:} corotation; {\it empty
triangles:} 4:1 resonance. \label{spiralEP}}
\end{figure}

From our $K_s$ and $K^\prime$ data we obtain visually the location
of the spiral endpoints in our objects. In some few cases the signal-to-noise
ratio was too low, and the $J$ or the optical $i$ images were used instead.
The positions of the derived spiral endpoints, as well as the
wavelength used, are listed in Table~\ref{tbl-omegas}.
We compare these spiral endpoints with the locations of major
resonances (4:1 resonance, corotation, and OLR) inferred from the $Q(rJgi)$
index data and the stellar models, as described in \S~\ref{linkdyn}.
For objects with more than one studied region we choose the one with
the lowest uncertainty and the best match with the spiral endpoints.
The selected regions are marked with an asterisk in Table~\ref{tbl-omegas}.
The results of this analysis are summarized in Fig.~\ref{spiralEP};
the vertical axis corresponds to the $R_{res} / R_{end}$ ratio, where
$R_{res}$ is the location of the major resonance derived from the analysis,
and $R_{end}$ is the spiral endpoint obtained from the data.

With the exception of NGC4254 (M~99), objects are organized by Hubble type
(from SAbc to SABc).
Remarks in the caption of Fig.~\ref{REG_4254_A} apply also to
figures~\ref{REG_4254_B} --~\ref{REG_578_B}, unless indicated
in individual captions.

\underline{NGC~4254 :}
[Figures~\ref{REG_4254_A} -~\ref{REG_4254_B}]
Although NGC~4254 is a ``$Q$ effect" galaxy with two
asymmetric halves in this diagnostic, we include it here
as a test of the consistency of our procedure vis-\`a-vis
its first application by GG96.  
The resonance positions computed for region NGC~4254~A (the
same analyzed by GG96) place the spiral endpoint for the
corresponding arm at the OLR.
Even though the $\Omega_p$ for region NGC~4254~B differs from
the value obtained for region A, the spiral endpoint of the
arm to which this region belongs matches the position of
its OLR too!\footnote{Results differ when adopting
$V_{rot} = 140$ km s$^{-1}$ and $\alpha = 42 \degr$, as derived
by \citet{pho93} and used by GG96. With these values, $\Omega_p = 19.0 \pm 2$ km s$^{-1}$ kpc$^{-1}$
and  $R_{\rm CR} = 92\farcs2 \pm 9\farcs0$ for region A.
The resonance positions, however, have similar values
within the errors.}
Not surprisingly for a
``$Q$ effect" galaxy, region B $Q$ values are also 0.03 magnitudes
below those for region A. Given the fact that the two gradients yield
different values for $\Omega_p$ but their corresponding arms
end at their respective OLR,
we hypothesize that the ``$Q$ effect" is related to the dynamics of
the disk.

\underline{NGC~4939 :}
[Figures~\ref{REG_4939_A_I} -~\ref{REG_4939_B_II}]
From the location of the dust lanes in this region, it is
hard to determine the age gradient direction and hence whether it
occurs inside or outside corotation.
If one assumes it is outside corotation (inverse gradient),
the spiral endpoints are located
near the OLR, as shown in Fig.~\ref{REG_4939_A_I}.
If, contrariwise, one assumes the age gradient is
inside corotation,
the spiral endpoints may coincide with
the location of the 4:1 resonance (see Fig.~\ref{REG_4939_B_II}).
Region NGC~4939~A has a shape that differs from
the models, a fact that could be explained if the galaxy
metallicity is different from the one assumed (see \S~\ref{rolmeta}).

\underline{NGC~3938 :}
[Figure~\ref{REG_3938_A}]
The signal-to-noise ratio for the optical images of this
galaxy is lower than for other objects in the sample.
The dynamical parameters inferred from the
data yield resonance positions that do not match the location
of the spiral endpoints.

\underline{NGC~7126 :}
[Figure~\ref{REG_7126_A}]
The inferred locations of major resonances do not match any spiral
endpoint in the $K_s$ mosaic, which has a low signal-to-noise ratio. However,
in the $i$-band mosaic two well defined spiral arms can be seen that
extend beyond the position of the OLR. We conclude that the gradient featured in
region NGC~7126~A may not be due to star formation triggered by the spiral shock.

\underline{NGC~1417 :}
[Figures~\ref{REG_1417_A} -~\ref{REG_1417_C}]
Regions A, B, and C for NGC~1417 give the same corotation
position (8 kpc) within the errors (see Table~\ref{tbl-omegas}).
The spiral arms end at the OLR.

\underline{NGC~7753 :}
[Figure~\ref{REG_7753_A}]
Apparently, the gradient studied in this object coincides
with a dust lane feature. Nevertheless, according to the computed $\Omega_p$,
the spiral endpoints match the OLR within the errors.

\underline{NGC~6951 :}
[Figure~\ref{REG_6951_A}]
Neither of the resonance positions computed from the data
of region NGC~6951~ A matches the spiral endpoints.

\underline{NGC~5371 :}
[Figures~\ref{REG_5371_A} -~\ref{REG_5371_B}]
The dynamic parameters derived for regions A and B
result in different locations of the major resonances.
Those for region A do not coincide with the spiral endpoints of the object.
However, region B yields an OLR position that matches
the spiral endpoints.

\underline{NGC~3162 :}
[Figures~\ref{REG_3162_A} -~\ref{REG_3162_B}]
The computed OLR for region B
coincides with the the spiral endpoints.
The computed errors of the dynamic parameters
for region A are larger than for region B.

\underline{NGC~1421 :}
[Figures~\ref{REG_1421_A} -~\ref{REG_1421_C}]
Region A has an offset
in $Q$ of +0.1 mag when compared to regions B and C.
On the other hand, regions A and C give resonance positions
that agree with the spiral endpoints; the best match is for region C,
which is also an inverse azimuthal color gradient (i.e., situated after
corotation). The feature studied in region B must not be caused
by star formation linked to the dynamics of the disk.

\underline{NGC~7125 :}
[Figures~\ref{REG_7125_A} -~\ref{REG_7125_C}]
Region B displays an inverse gradient.
The results obtained from the
three studied regions agree within their errors.
The spiral endpoints coincide with the OLR in this object.

\underline{NGC~918 :}
[Figure~\ref{REG_918_A}]
The stellar population model stretched to the data from region
A indicates that the spiral arms end at the OLR in this galaxy.

\underline{NGC~578 :}
[Figures~\ref{REG_578_A} -~\ref{REG_578_B}]
Regions A and B, located in two different arms of the four that conform this object,
yield the same value for the spiral pattern speed. In the deprojected mosaics,
the northern arms, including the one harboring region A,
seem to extend beyond the corotation radius without reaching the OLR
within the errors. The other arms, where region B is
located, end at corotation.
This is the only object in the sample that presents
spiral arms ending at corotation.

\clearpage

\begin{deluxetable}{cccccccccc}
\tabletypesize{\scriptsize}
\rotate
\tablecaption{Observed and derived dynamic parameters.\label{tbl-omegas}}
\tablewidth{0pt}
\tablehead{
\colhead{Region number} &
\colhead{Galaxy and region} & \colhead{$R_{\rm mean}$ (arcsec)} & \colhead{$R_{\bf mean}$ (kpc)}
& \colhead{$R_{\rm end}$ (arcsec)} & \colhead{$R_{\bf end}$ (kpc)} &\colhead{$Q_{\bf offset}$ (mag)}
& \colhead{$\Omega_p$ (km s$^{-1}$ kpc$^{-1}$) } & \colhead{$R_{\rm CR}$ (arcsec)}
& \colhead{$R_{\rm CR}$ (kpc)}
}

\startdata
1     & NGC4254 A*& $ 66.4\pm0.25 $ & $5.3\pm0.4$ & 157.5$\pm$7.5 ($K_s$) & 12.6$\pm$0.8  & $-0.13\pm0.05$     &  $33.5\pm8.2$    & $81.0\pm8.6$       & $6.5\pm0.7$         \\
2     & NGC4254 B & $ 49.3\pm0.25 $ & $3.9\pm0.3$ &                       &               & $-0.16\pm0.05$     &  $47.7\pm11.4$   & $57.0\pm5.6$       & $4.6\pm0.4$         \\
&  &  &  &  &  &  &  \\
3     & NGC4939 A*& $ 129.4\pm0.25$ & $29.2\pm2.5$& 145$\pm$5   ($K_s$)   & 32.7$\pm$2.8  & $0.0\pm0.06$       &  $12.3\pm0.8$    & $87.5\pm5.3$       & $19.7\pm1.2$        \\
&  &  &  &  &  &  &  \\
4     & NGC3938 A*& $ 30.7\pm0.25 $ & $2.3\pm0.2$ & 100$\pm$5   ($K_s$)   & 7.7$\pm$0.7   & $+0.09\pm0.06$     &  $33.8\pm12.9$   & $36.9\pm5.2$       & $2.8\pm0.4$         \\
&  &  &  &  &  &  &  \\
5     & NGC7126 A*& $ 34.6\pm0.15 $ & $7.5\pm0.6$ & 118.9$\pm$2.9 ($i$)   & 25.7$\pm$2.2  & $0.0\pm0.09$       &  $19.7\pm2.2$    & $40.2\pm4.1$       & $8.7\pm0.9$         \\
&  &  &  &  &  &  &  \\
6     & NGC1417 A & $ 22.6\pm0.25 $ & $6.3\pm0.5$ & 60$\pm$5     ($J$)    & 16.6$\pm$1.4  & $-0.15\pm0.06$     &  $35.1\pm4.7$    & $26.8\pm2.9$       & $7.4\pm0.8$         \\
7     & NGC1417 B*& $ 25.3\pm0.25 $ & $7.0\pm0.6$ &                       &               & $-0.18\pm0.06$     &  $29.9\pm4.1$    & $31.5\pm3.6$       & $8.7\pm1.0$         \\
8     & NGC1417 C & $ 20.4\pm0.25 $ & $5.7\pm0.5$ &                       &               & $-0.15\pm0.06$     &  $33.6\pm5.2$    & $28.1\pm3.6$       & $7.8\pm1.0$         \\
&  &  &  &  &  &  &  \\
9     & NGC7753 A*& $ 33.7\pm0.15 $ & $11.8\pm1.0$& 95.7$\pm$5.8  ($i$)   & 33.5$\pm$2.8  & $+0.07\pm0.09$     &  $11.7\pm2.8$    & $51.7\pm9.0$       & $18.1\pm3.2$        \\
&  &  &  &  &  &  &  \\
10    & NGC6951 A*& $ 38.8\pm0.25 $ & $4.7\pm0.4$ & 105$\pm$5   ($K_s$)   & 12.7$\pm$1.1  & $+0.14\pm0.07$     &  $49.9\pm12.8$   & $43.2\pm5.0$       & $5.2\pm0.6$         \\
&  &  &  &  &  &  &  \\
11    & NGC5371 A & $ 32.2\pm0.25 $ & $6.8\pm0.6$ & 115$\pm$5   ($K_s$)   & 24.3$\pm$2.1  & $-0.01\pm0.05$     &  $31.2\pm7.6$    & $44.3\pm6.1$       & $9.3\pm1.3$         \\
12    & NGC5371 B*& $ 41.8\pm0.25 $ & $8.8\pm0.7$ &                       &               & $+0.03\pm0.05$     &  $20.8\pm6.5$    & $66.6\pm13.3$      & $14.0\pm2.8$        \\
&  &  &  &  &  &  &  \\
13    & NGC3162 A & $ 33.0\pm0.25 $ & $3.8\pm0.3$ & 75$\pm$5 ($J$)        & 8.6$\pm$0.8   & $0.0\pm0.06$       &  $14.7\pm8.3$    & $81.8\pm33.5$      & $9.4\pm3.8$         \\
14    & NGC3162 B*& $ 35.7\pm0.25 $ & $4.1\pm0.4$ &                       &               & $0.0\pm0.06$       &  $25.0\pm8.3$    & $48.0\pm7.9$       & $5.5\pm0.9$         \\
&  &  &  &  &  &  &  \\
15    & NGC1421 A & $ 34.6\pm0.15 $ & $4.9\pm0.4$ & 92.8$\pm$2.9 ($K_s$)  & 13.2$\pm$1.1  & $+0.05\pm0.09$     &  $16.6\pm3.5$    & $72.6\pm13.9$      & $10.3\pm2.0$        \\
16    & NGC1421 B & $ 57.5\pm0.15 $ & $8.2\pm0.7$ &                       &               & $-0.04\pm0.09$     &  $ 0.5\pm2.2$    & $2230.1\pm15820.0$ & $316.8\pm2247.2$    \\
17    & NGC1421 C*& $ 80.0\pm0.15 $ & $11.4\pm1.0$&                       &               & $-0.05\pm0.09$     &  $22.3\pm1.7$    & $54.0\pm3.6$       & $7.7\pm0.5$         \\
&  &  &  &  &  &  &  \\
18    & NGC7125 A & $ 33.6\pm0.15 $ & $7.3\pm0.6$ & 95.7$\pm$2.9 ($J$)    & 20.7$\pm$1.8  & $0.0\pm0.09$       &  $9.6\pm4.3$     & $69.4\pm22.1$      & $15.0\pm4.8$        \\
19    & NGC7125 B*& $ 79.1\pm0.15 $ & $17.1\pm1.5$&                       &               & $0.0\pm0.09$       &  $10.8\pm1.7$    & $61.7\pm4.8$       & $13.3\pm1.0$        \\
20    & NGC7125 C & $ 45.3\pm0.15 $ & $9.8\pm0.8$ &                       &               & $-0.02\pm0.09$     &  $9.8\pm3.2$     & $68.5\pm12.6$      & $14.8\pm2.7$        \\
&  &  &  &  &  &  &  \\
21    & NGC918 A* & $ 32.1\pm0.15 $ & $3.4\pm0.3$ & 75.4$\pm$5.8 ($K^\prime$)&7.9$\pm$0.7 & $+0.07\pm0.05$     &  $33.4\pm5.7$    & $39.6\pm4.9$       & $4.2\pm0.5$         \\
&  &  &  &  &  &  &  \\
22    & NGC578 A* & $ 75.2\pm0.15 $ & $8.3\pm0.7$ & 95.7$\pm$2.9 ($K_s$)  & 10.5$\pm$0.9  & $-0.2 \pm0.09$     &  $15.0\pm1.9$    & $92.2\pm9.7$       & $10.1\pm1.1$        \\
23    & NGC578 B  & $ 67.8\pm0.15 $ & $7.5\pm0.6$ &                       &               & $-0.16\pm0.09$     &  $15.2\pm2.2$    & $90.8\pm10.9$      & $10.0\pm1.2$        \\

\enddata

\tablecomments{
Col.\ (3) and (4). Mean radius of the studied regions, in arcsec and kpc, respectively.
Col.\ (5) and (6). Radius of the spiral endpoint, in arcsec and kpc, respectively, and bandpass used to determine it.
Col.\ (7). $Q$ index magnitude offset, applied to the solar metallicity model with 2\% of young
stars in order to fit the data.
Col.\ (8). Spiral pattern speed.
Col.\ (9) and (10). Corotation radius, in arcsec and kpc, respectively.
}

\end{deluxetable}

\section{Expected $Q(rJgi)$ index profiles in a spiral shock scenario.} \label{spshsc}

In this section, we consider in a qualitative way
a more realistic situation within 
the density wave scenario.

The relative density distributions of young and old stars after the shock, as
well as non-circular stellar velocities produce changes in the azimuthal 
light profiles with respect to the idealized case discussed earlier.
Also, since we are dealing with different generations of stars, metallicity 
variations should be considered. In what follows, we discuss each one of
these issues at moderate length, and make an estimate of their 
impact on the disk dynamic parameters derived through the comparison
of model and observed $Q$ profiles.

\subsection{Post-shock density and velocity distributions.}\label{denveldist}

\begin{figure}
\epsscale{1.00}
\plotone{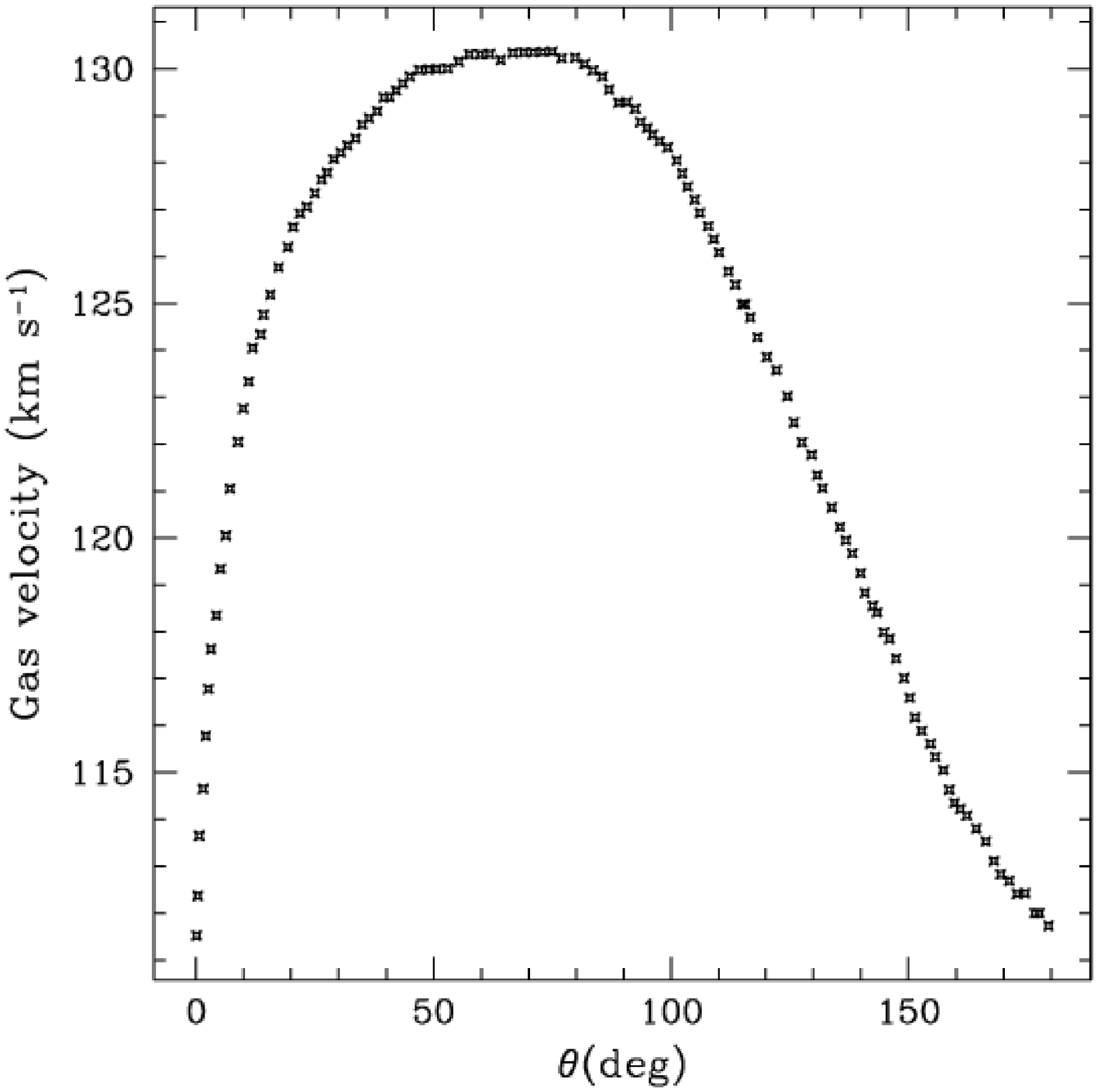}
\caption{Gas circular velocity vs.\ angular distance obtained from the solution by \citet{rob69}.
The distance is measured relative to the spiral shock location, and the velocity, relative
to the non-inertial frame of reference of the spiral pattern.
We assume that
the gas circular velocity is approximately equal to the
velocity parallel to the spiral equipotential.
\label{gasvel}}
\end{figure}

With the purpose of increasing our understanding of gas dynamics in the presence of
spiral density waves, stationary spiral shock patterns have been studied with both semi-analytical approaches
\citep[e.g.,][]{rob69,git04} and numerical simulations \citep[e.g.,][]{sly03,mar04,yan08}.
The post-shock density and velocity profiles obtained from such studies show that they
depend on many physical parameters. It is reasonable to assume, however, that the newborn stars
product of these shocks have densities and velocities that are similar to those of the collapsed
gas clouds where they form, at least in the early stages of their evolution, and even if
only a few percent of the gas will form stars.
Also, it is commonly accepted that dust lanes trace the location of spiral shocks. As newborn
stars move away from their birth site, different distances are reached due to
accelerated movements. In figure~\ref{gasvel} we show the gas velocity parallel to the
spiral equipotential curve from the semi-analytical solution of \citet{rob69};
this solution corresponds to a radius of 10 kpc; a pattern speed,
$\Omega_{p}$, of 12.5 km s$^{-1}$ kpc$^{-1}$; an arm pitch angle
$i = 8\fdg13$; a spiral field strength of 5\% that of the axisymmetric field;
and a mean gaseous dispersion of 10 km s$^{-1}$.
The distance is measured relative to the spiral shock location, and the velocity,
in the non-inertial frame of reference of the spiral pattern.
The variable velocity in an inertial frame of reference goes from
235 to 255 km s$^{-1}$.
For the time being, we assume that the circular gas velocity is approximately
equal to this velocity, and investigate the effects of such variable
stellar velocity on the 1-D profiles of the photometric index $Q(rJgi)$.

\begin{figure}
\epsscale{1.00}
\plotone{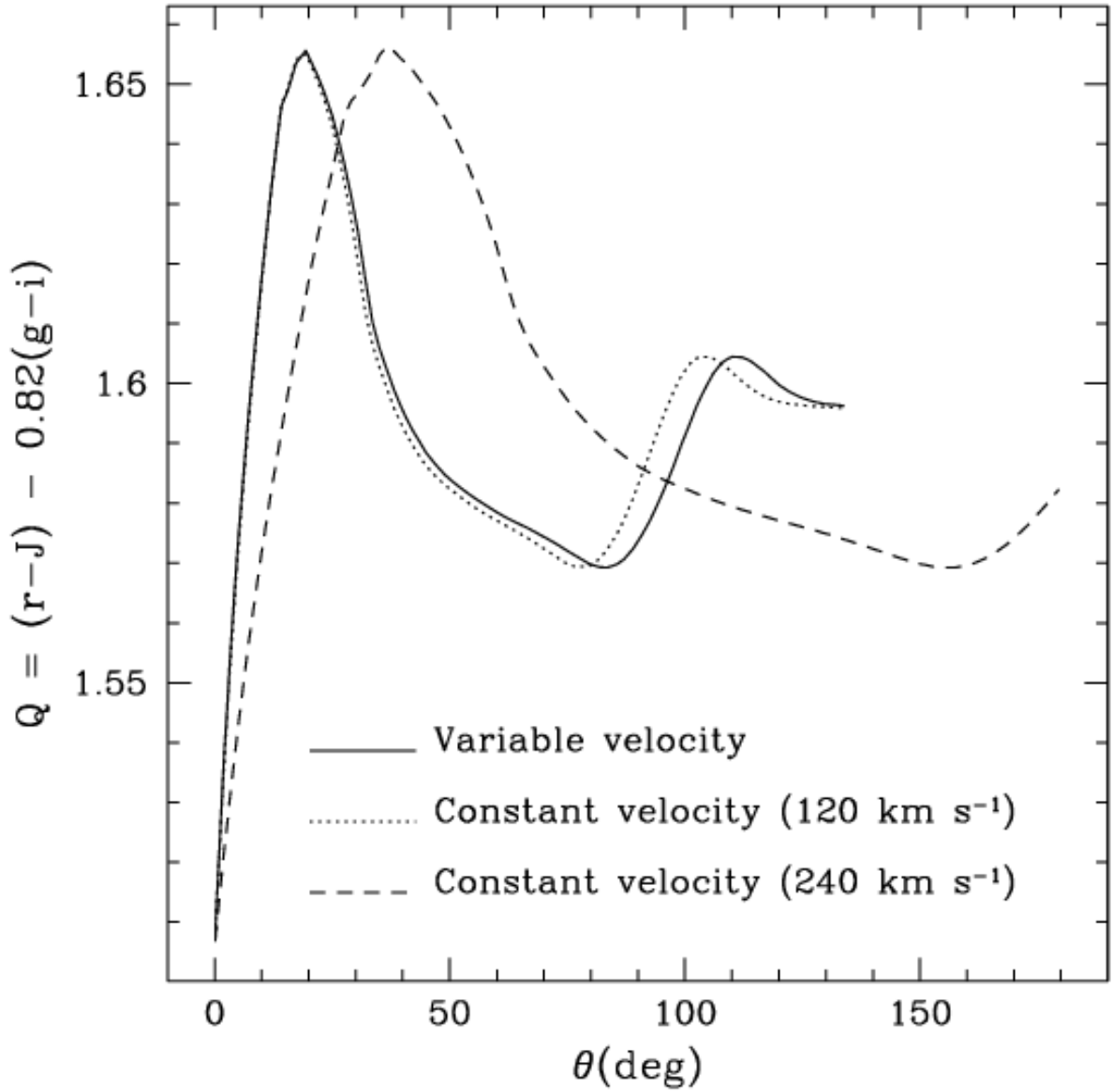}
\caption{Theoretical profiles for the $Q(rJgi)$ photometric index, CB07 models.
{\it Solid line:} stars have a variable orbital velocity,
as shown in Fig.~\ref{gasvel}; {\it dotted line:} stars have a constant circular
velocity of 120 km s$^{-1}$; {\it dashed line:} stars with a constant orbital
velocity of  240 km s$^{-1}$.
\label{Qvel}}
\end{figure}

Figure~\ref{Qvel} shows theoretical profiles for the $Q(rJgi)$ index, obtained
with the variable velocity plotted if Fig.~\ref{gasvel}. The models
have an IMF upper mass limit of 10 $M_{\sun}$, and a constant fraction
of young stars of 2\% throughout.
A comparison with constant velocities of 120 km s$^{-1}$
and 240 km s$^{-1}$ is also shown.

\begin{figure}
\epsscale{1.00}
\plotone{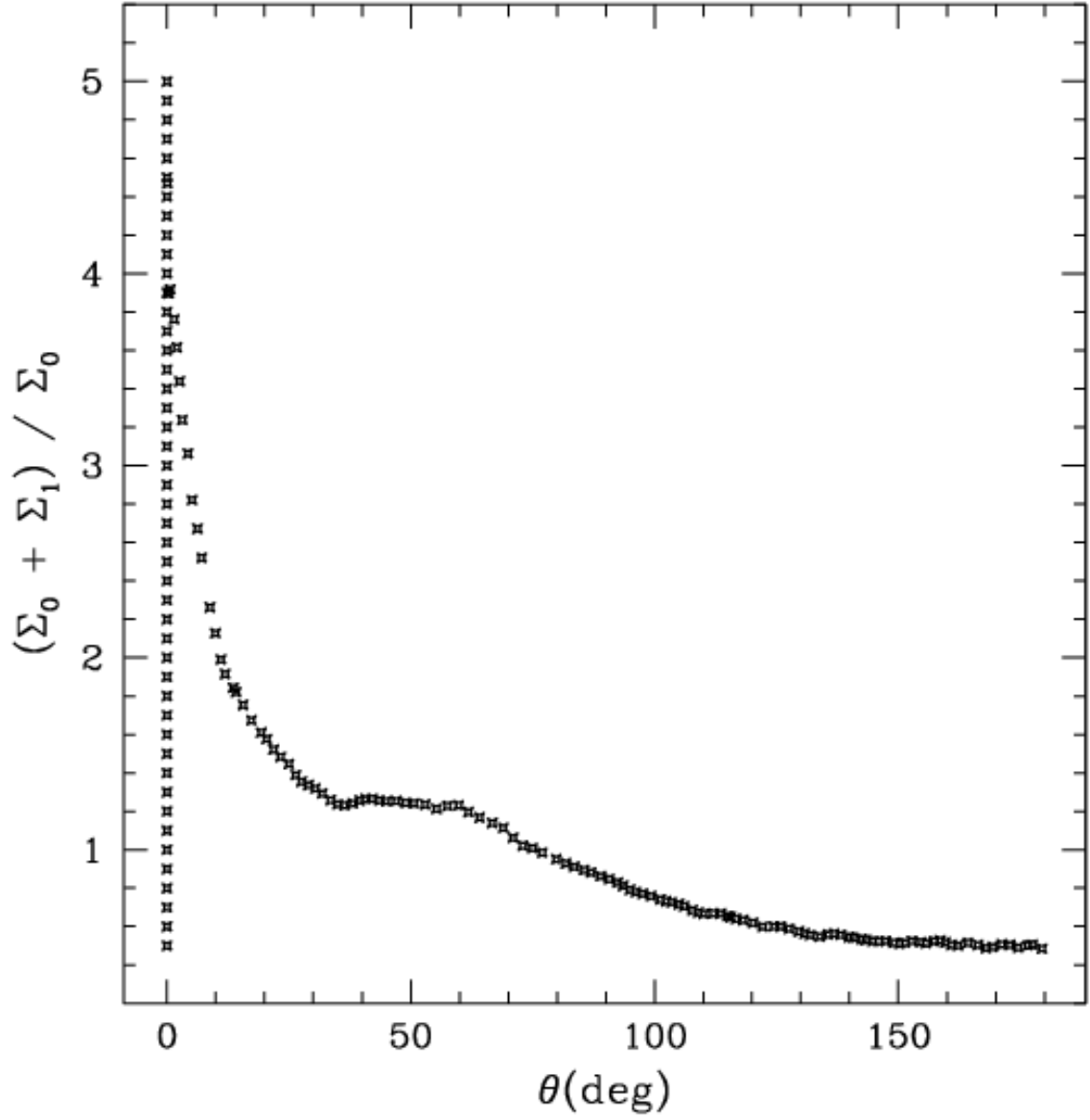}
\caption{Gas relative density from the shock solution of \citet{rob69}. $\Sigma_{0}$
is the unperturbed density in the axisymmetric potential, and $\Sigma_{1}$
is the perturbed density due to the spiral gravitational field.
The resulting total gas density is $\Sigma = \Sigma_{0} + \Sigma_{1}$. \label{gasden}}

\end{figure}

\begin{figure}
\epsscale{1.00}
\plotone{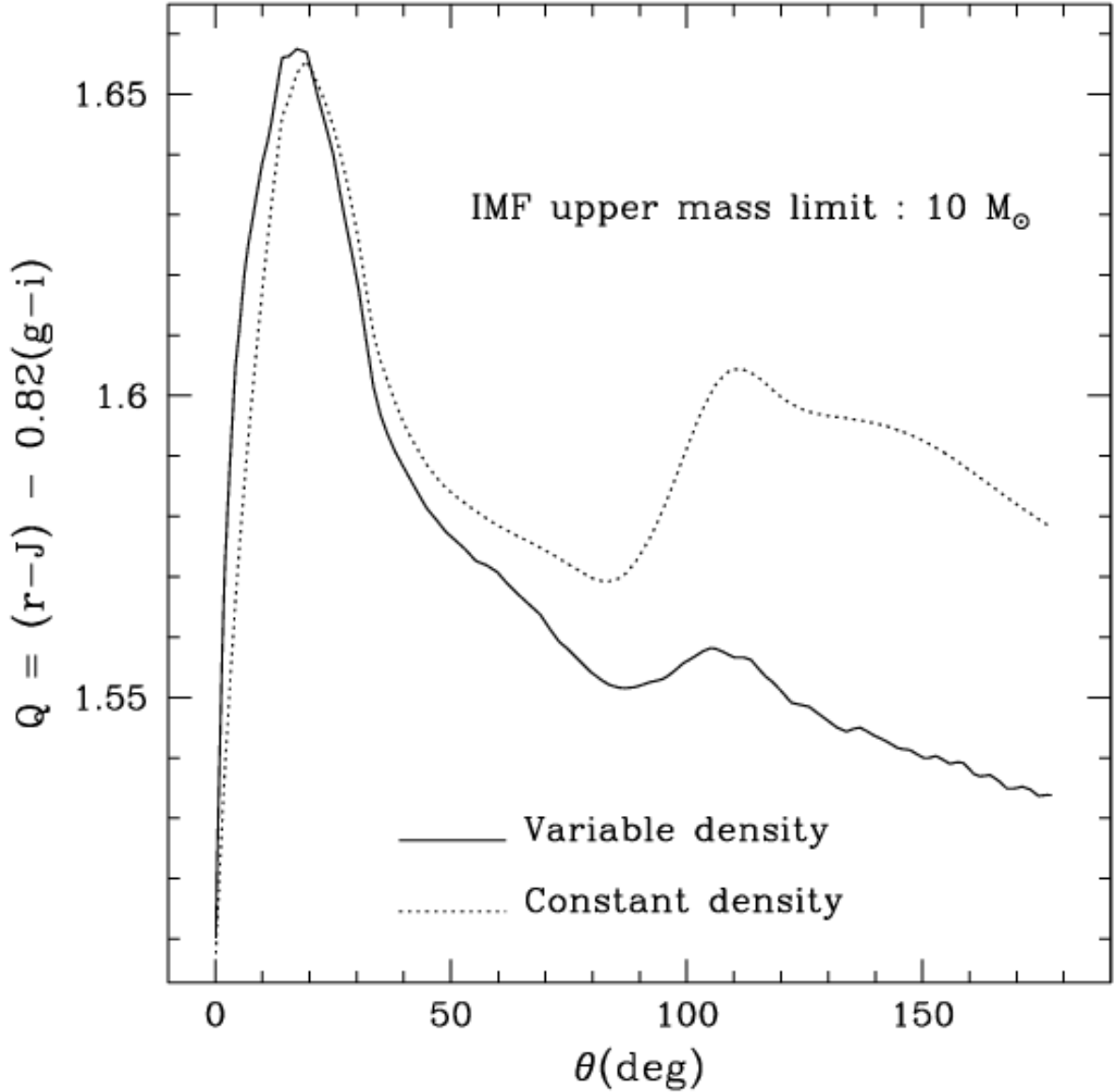}
\caption{Model profiles for the photometic index $Q(rJgi)$, low IMF upper mass limit.
{\it Solid line:} variable
stellar velocity and density as shown in figures~\ref{gasvel} and~\ref{gasden},
respectively. {\it Dotted line:}
variable stellar velocity and constant density (2\% of young stars by mass). The
IMF lower and upper mass limits are 0.1 and 10 $M_{\sun}$, respectively,
for both models. \label{Qden10}}
\end{figure}

\begin{figure}
\epsscale{1.00}
\plotone{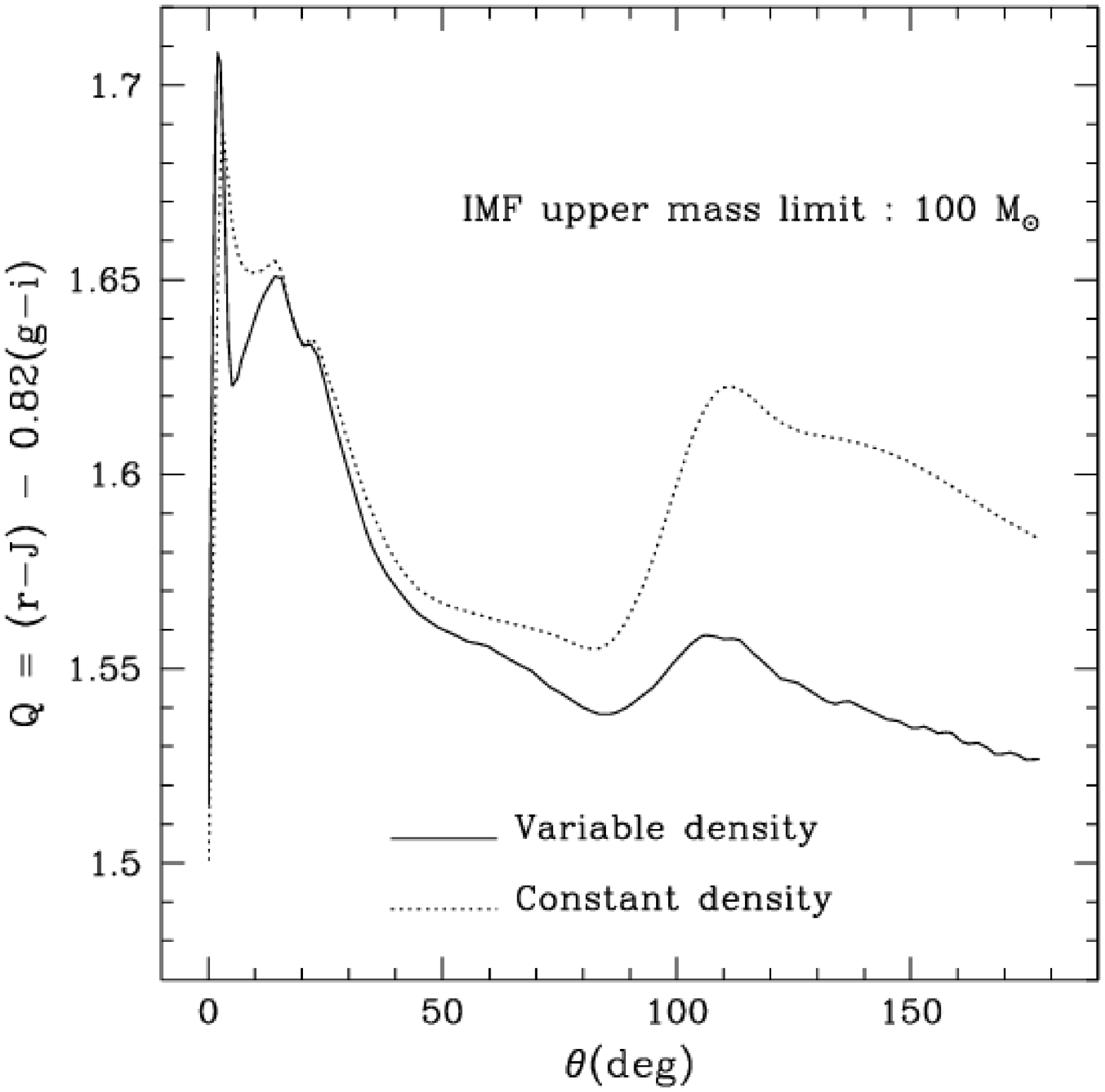}
\caption{Model profiles for the photometic index $Q(rJgi)$, high IMF upper mass limit.
{\it Solid line:}
variable stellar velocity and variable stellar density; {\it dotted line:}
variable stellar velocity and constant density. The IMF
lower and upper mass limits are 0.1 and 100 $M_{\sun}$, respectively,
for both models. \label{Qden100}}
\end{figure}

Next, we investigate the effect of including variable, more realistic, mass fractions of
old and young stars. To this end, we use the relative gas density
shown in Fig.~\ref{gasden}. This density was derived by \citet{rob69}, using the
same parameters listed above for his velocity solution.
We then assume that the fraction of young stars must be 2\% at an age of $3 \times 10^7$
years ($\theta \sim 20\degr$),
and propagate this fraction to other locations, supposing that the
young stars share the gas density distribution (for example, the young stars fraction
would be about 2.5\% at an age of $1.5 \times 10^7$ years or $\theta \sim 10\degr$).  
As already stated, for the old stars mass fraction we use $\beta_{II}$ = 1 - $\beta_{I}$
(see equation \ref{eqGSPM}). 
The theoretical $Q$ profiles involving variable stellar velocities, and both variable and constant
stellar densities are
shown in figures~\ref{Qden10} and~\ref{Qden100}, for IMF upper mass limits of 10 and 100
$M_{\sun}$, respectively.

\begin{figure}
\epsscale{1.00}
\plotone{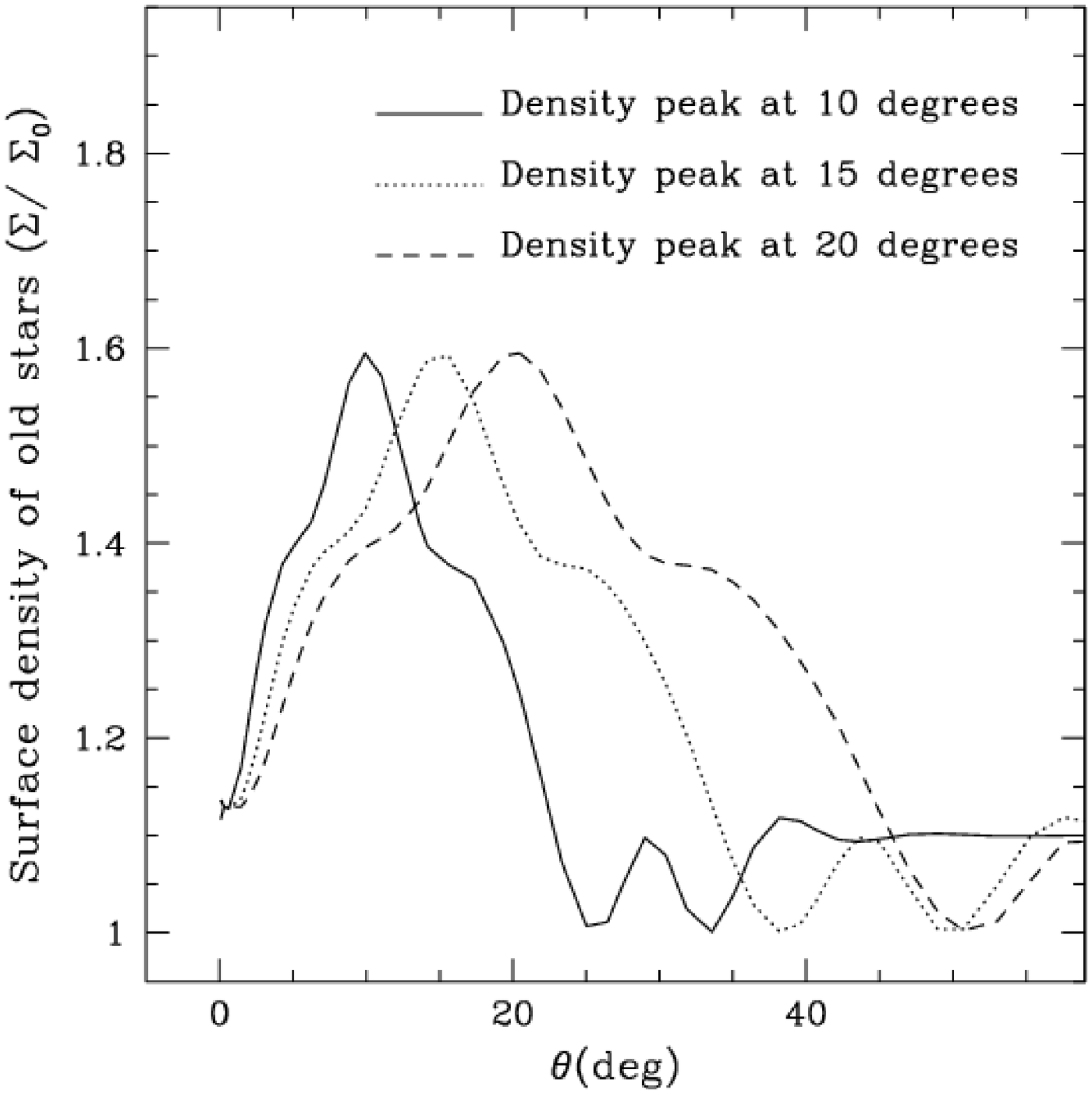}
\caption{Density profiles of old stars for different locations of the density wave
relative to the spiral shock. {\it Solid line:} old population density peak at 10$\degr$;
{\it dotted line:} density peak at 15$\degr$; {\it dashed line:}
density peak at 20$\degr$.  $\Sigma_0$ is the unperturbed density in
an axisymmetric potential; $\Sigma_1$ corresponds to the density wave perturbation;
$\Sigma = \Sigma_0 + \Sigma_1$. Angular locations
are the same as in previous figures. \label{denwaves}}
\end{figure}

\begin{figure}
\epsscale{1.00}
\plotone{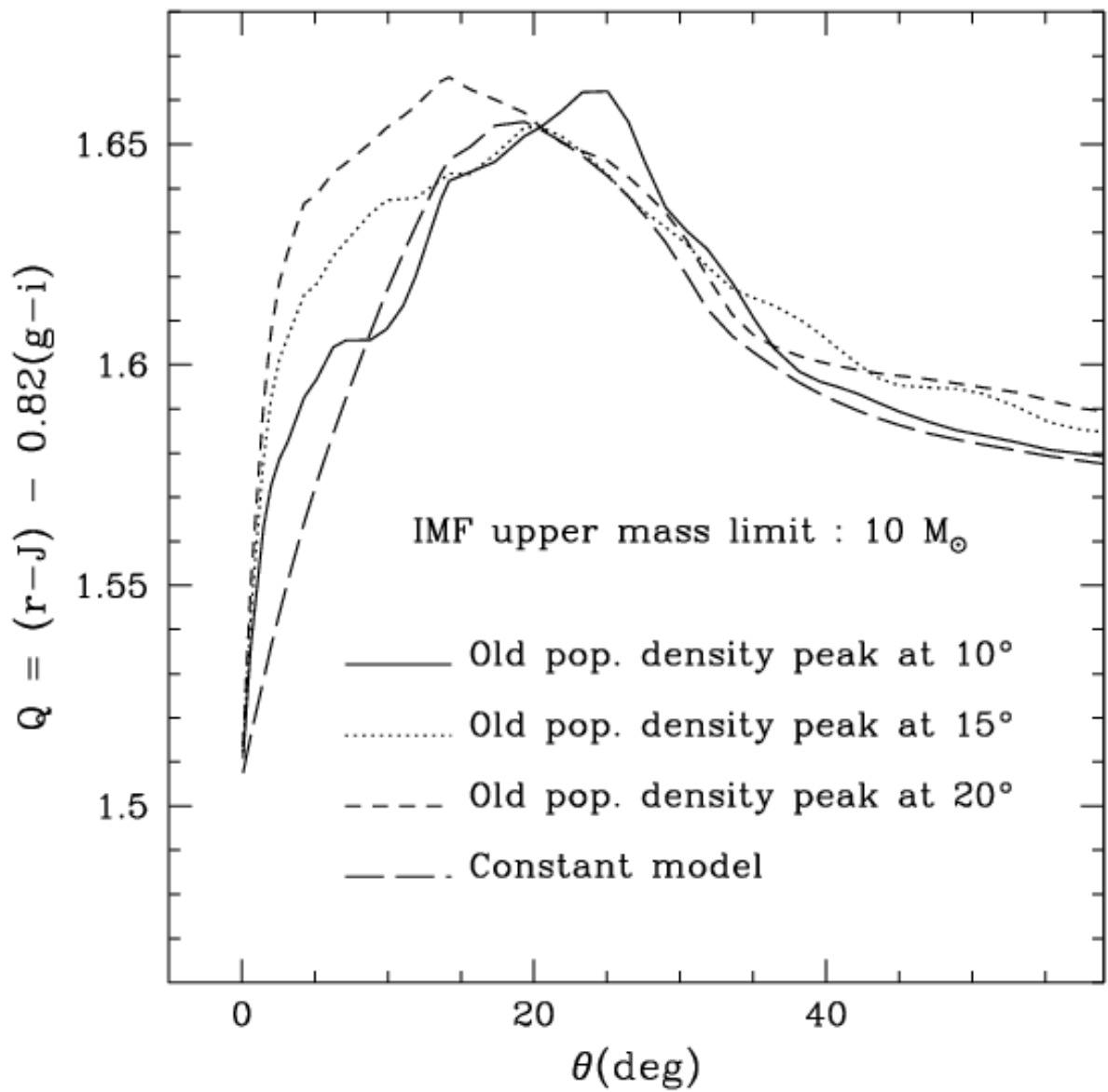}
\caption{Theoretical $Q$ profiles for variable stellar densities and velocities, in the
case of an IMF upper mass limit of 10 $M_{\sun}$.
Each model corresponds to a different location of the stellar density wave
peak, as shown in Fig.~\ref{denwaves}.
Lines have same meaning as in Fig.~\ref{denwaves},
except for the {\it long-dashed line:} constant stellar density and velocity. \label{Qvar10}}
\end{figure}

\begin{figure}
\epsscale{1.00}
\plotone{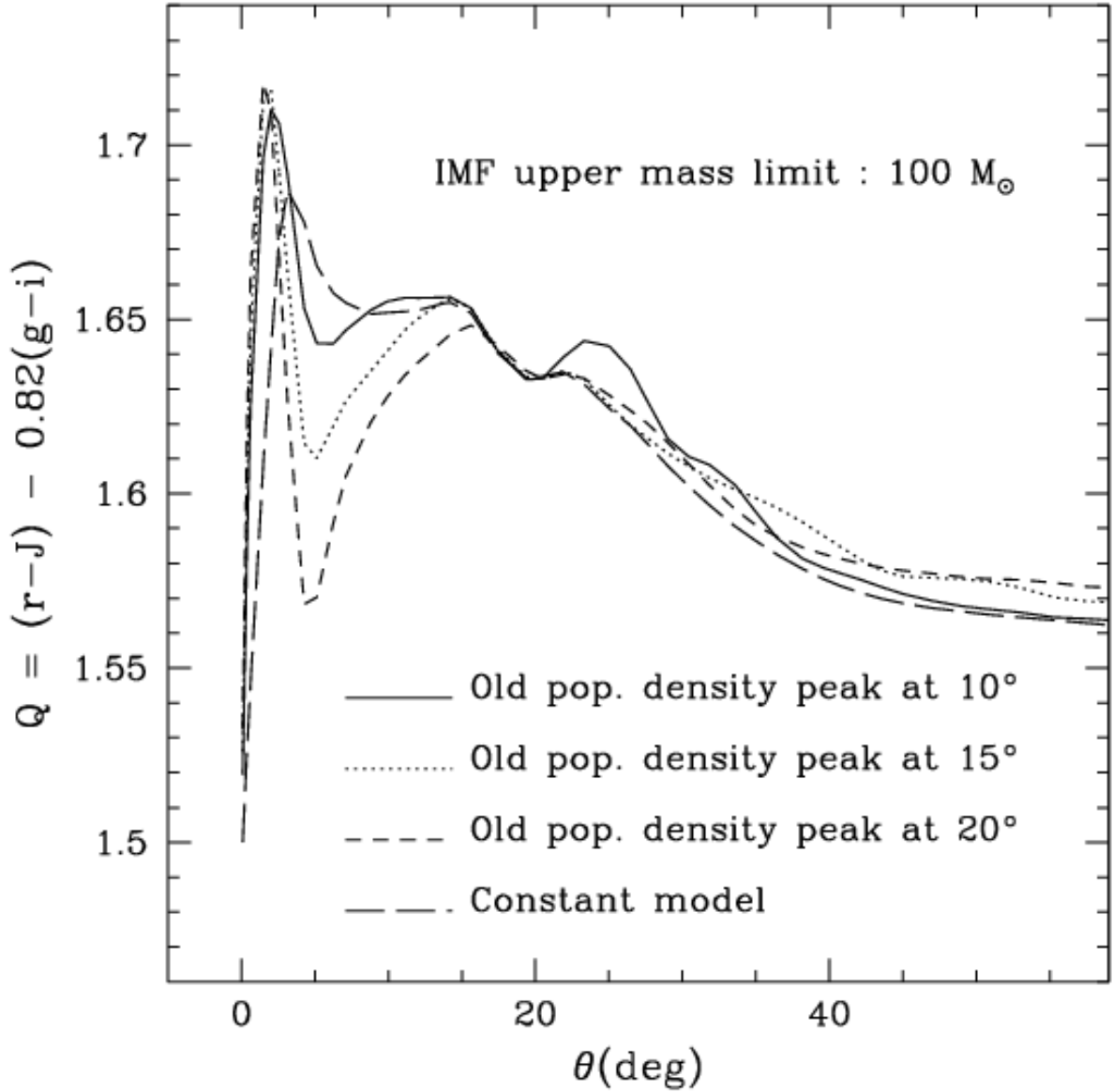}
\caption{Same as Fig.~\ref{Qvar10}, for an IMF upper mass limit of 100
$M_{\sun}$.
The main difference with respect to Fig.~\ref{Qvar10} is the sharp peak at $\theta \sim 0\degr$.
\label{Qvar100}}
\end{figure}

A further possible refinement to the models concerns the relative location of
the shock and the potential minimum.
According to \citet{git04}, the shock location moves to different azimuthal
values for tightly wound spirals. At small radii (inside
corotation), the shock occurs near the potential minimum; at larger radii, though,
the shock weakens and moves upstream towards the potential maximum.
In real galaxies, the maximum surface density of old stars and the potential
minimum will not coincide exactly \citep{zha07}. The gas responds to the
potential minimum, while the maximum observed surface density of old stars
marks the peak of the density wave. The old
stars' surface density is commonly inferred from observations at 2$\mu$m, although
red supergiants may contribute 20\% of the flux in this wavelength \citep{rix93}.
Assuming that the onset of star formation occurs almost immediately after the
shock, the resulting total stellar density (i.e., considering both young and old
stars) will be affected by the relative positions of each component.
In Fig.~\ref{denwaves}, we show a possible shape for a density wave taken
from the $K_s$ data of NGC 7125, assuming all the emission
comes from an old population with a constant mass-to-light ratio.
We try three positions of the density wave peak at, respectively, 10, 15, and 20 degrees away from the shock,
with increasing widths. With these density distributions of old stars,
and the variable velocity and relative density distributions for young stars
discussed previously
(the fraction of young stars is taken to be 2\% at an age of $3 \times 10^7$ years,
and propagated to other positions following the gas density distribution shown
in Fig.~\ref{gasden}), we produce once again theoretical profiles for the $Q(rJgi)$ index.
These are shown in figures~\ref{Qvar10} and~\ref{Qvar100}, for IMF upper mass limits of 10 and 100 $M_{\sun}$,
respectively. The comparison between the more complex models and those
with constant stellar densities and velocities
are also shown in this figures, for a fraction of young stars of 2\% by mass.
From this analysis it is clearly seen that galaxy azimuthal light profiles
may suffer deformations depending on the density and velocity
distributions of the underlying stars, being the deformation due to density
the most important one.

\begin{figure}
\epsscale{1.00}
\plotone{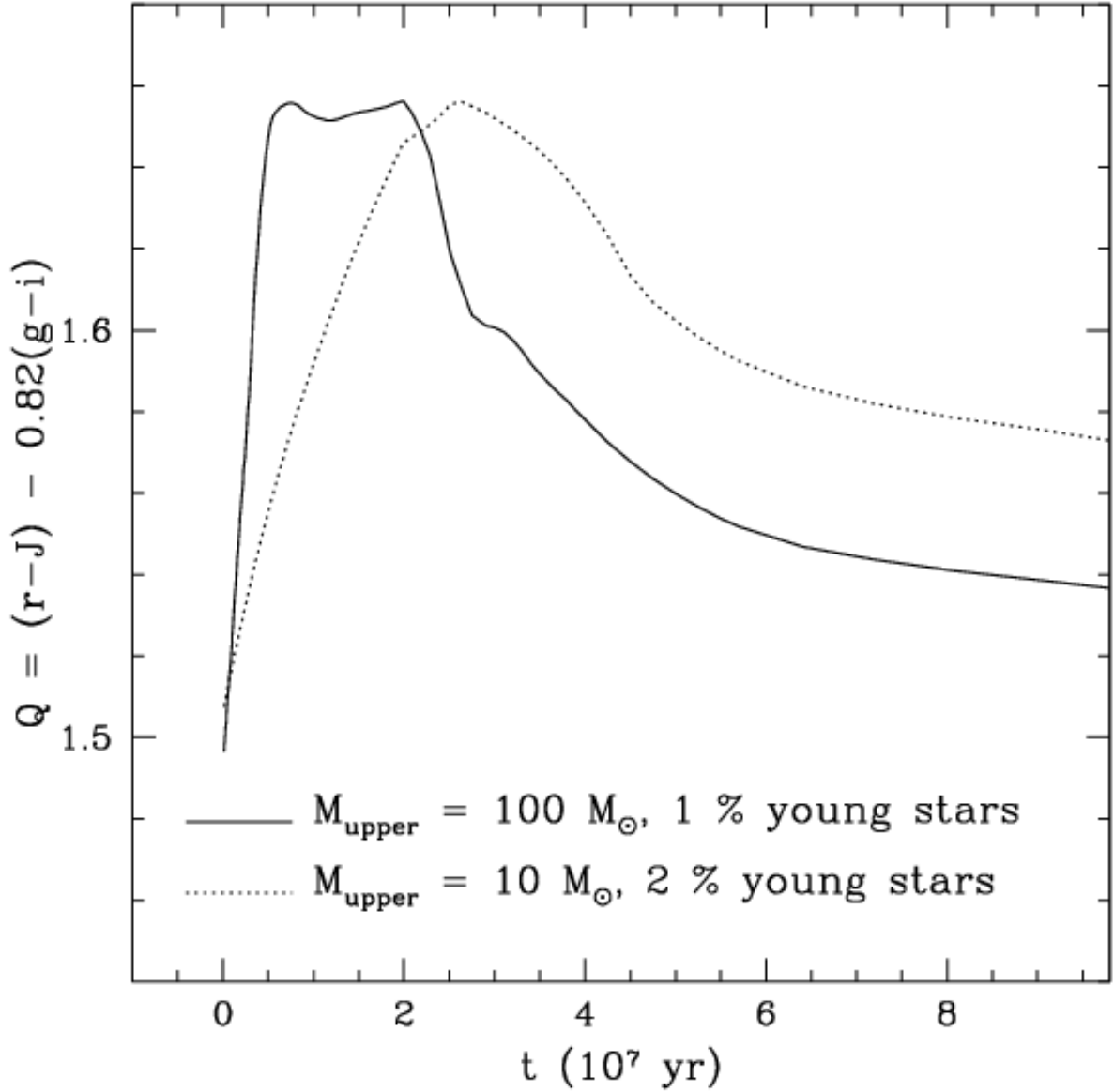}
\caption {
Degeneracy in constant stellar density models between IMF upper mass limit and fraction of
young stars. {\it Solid line:} $M_{upper} = 100 M_{\sun}$ and 1\% (by mass) of young stars;
{\it dotted line:} $M_{upper} = 10 M_{\sun}$ and 2\% of young stars.
Burst duration is $2 \times 10^7$ years.
The difference in the derived $\Omega_p$ from the two
models would be $\sim$ 3 km s$^{-1}$ kpc$^{-1}$. \label{degene}}
\end{figure}

It is worth noticing that in models with constant stellar density and velocity
a degeneracy occurs between IMF upper mass limit and fraction of young stars (GG96),
where models with low IMF upper mass limit and higher young star fraction
resemble those with high upper mass limit and lower young star fraction (1 \%). This degeneracy
is shown in Fig.~\ref{degene}.  
\footnote{
A similar degeneracy exists between IMF upper mass limit and length of star formation
burst (GG96), in the absence of independent constraints on their values.
}
In models with a fraction of young stars that varies with azimuthal position, however,
this degeneracy is broken, as is shown in figures~\ref{Qvar10} and~\ref{Qvar100}. The
models with an upper mass limit of 100 $M_{\sun}$ show a peak close to the
shock that is absent in those with $M_{upper}$ = 10 $M_{\sun}$, although it is
true that the peak might be lost in data with a poor signal-to-noise ratio.

\subsection{Non-circular motions.} \label{non_circ}

The motions of young stars under the assumption of spiral
density wave triggering have been studied by, among others, 
\citet{yua69,wie73,wie78,wie79,fer08}. 
\citet{wie79} has emphasized that, in a galaxy with spiral
density waves, newly-born stars have not had time to
reach dynamical equilibrium with the
galaxy potential. Hence, instead of responding to
a stationary wave like the old population, young stars
migrate out of the arms following complicated, non-circular, orbits.
The aforementioned trajectories are not necessarily closed;
also, they can run almost parallel to the arms for significant stretches,
with the consequence that stars are not seen leaving the arms
until they have somewhat aged and at a radius where
the dust lane location does not mark the site of star formation;
stars might even move to locations upstream of the
shock. The color gradients discussed before would 
overlap with stars drifting back to the arms. 

\begin{figure}
\epsscale{1.00}
\plotone{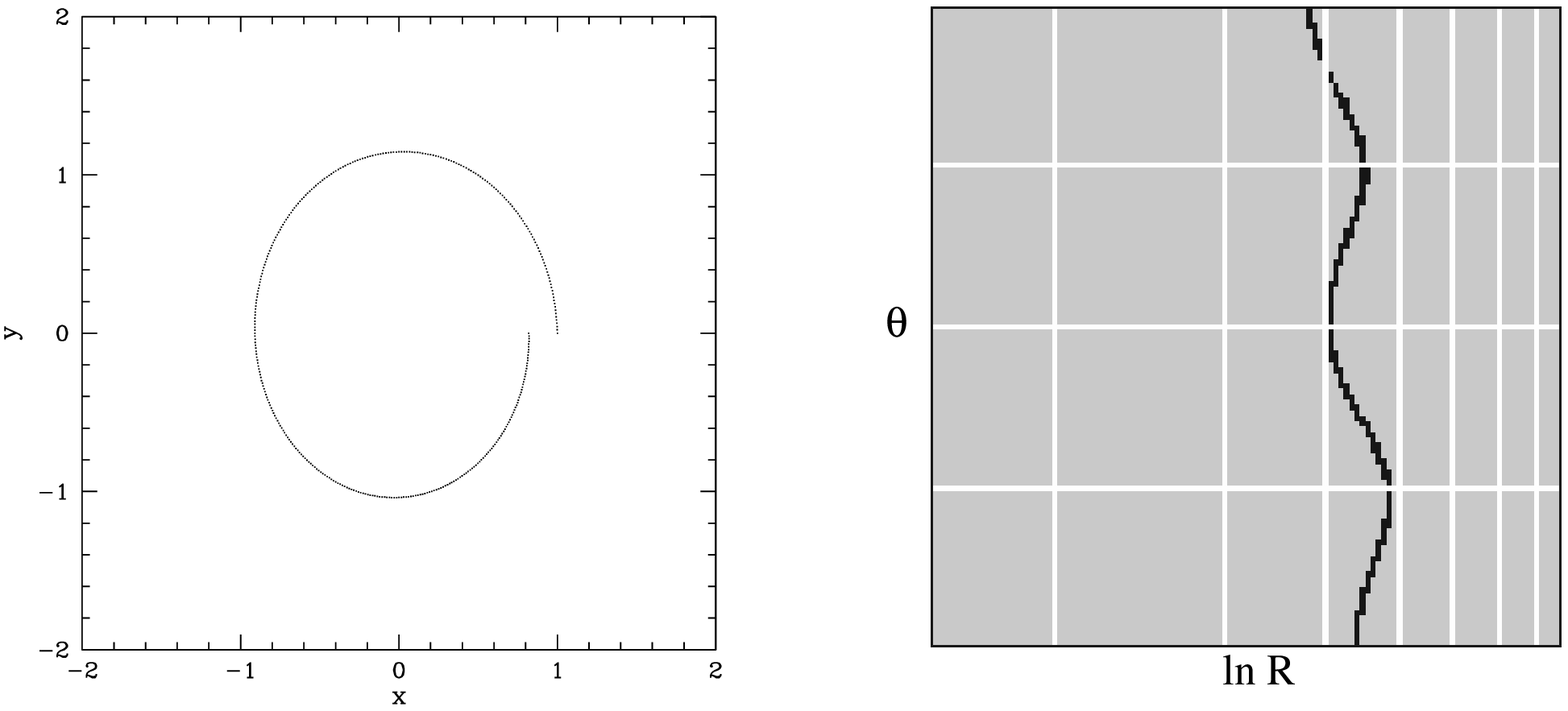}
\caption{{\it Left:} Orbit in cartesian coordinates (arbitrary units).
{\it Right:} The same orbit in the $\theta$ vs. $\ln R$ plane. Vertical lines
represent circles every $\sim 0.3$ arbitray units ($R=\sqrt{x^{2}+y^{2}}$).
Horizontal lines are distributed every $90\degr$.
\label {wavy_fig}}
\end{figure}

Another consequence in real galaxies is that orbits are no
longer vertical lines in the $\theta$ vs.\ $\ln R$ plane; rather,
they are oblique and wavy, as shown in Fig.~\ref{wavy_fig}.
Given that we have collapsed our data in radius to improve
the signal-to-noise ratio, our analysis is insensitive to
this effect.

Finally, if molecular clouds suffer the effects
of the spiral shock, drift velocities will be
smaller and stars will take longer to
migrate away from the arms, as shown in Fig.\ 5 of \citet{wie79};
there, lines of equal stellar age are closer together than in his Fig.\ 4,
where clouds and young stars keep their pre-shock velocities.

Qualitatively, these phenomena are akin to the variable
stellar velocities and densities considered in \S~\ref{denveldist},
and whose impact on the derived density wave parameters
is minimal, as briefly discussed in \S~\ref{simpvscmpx}. Numerical simulations beyond the scope of
this paper would be required to assess their repercussions
in a quantitative way.
      
\subsection{The role of metallicity.} \label{rolmeta}

\begin{figure}
\epsscale{1.00}
\plotone{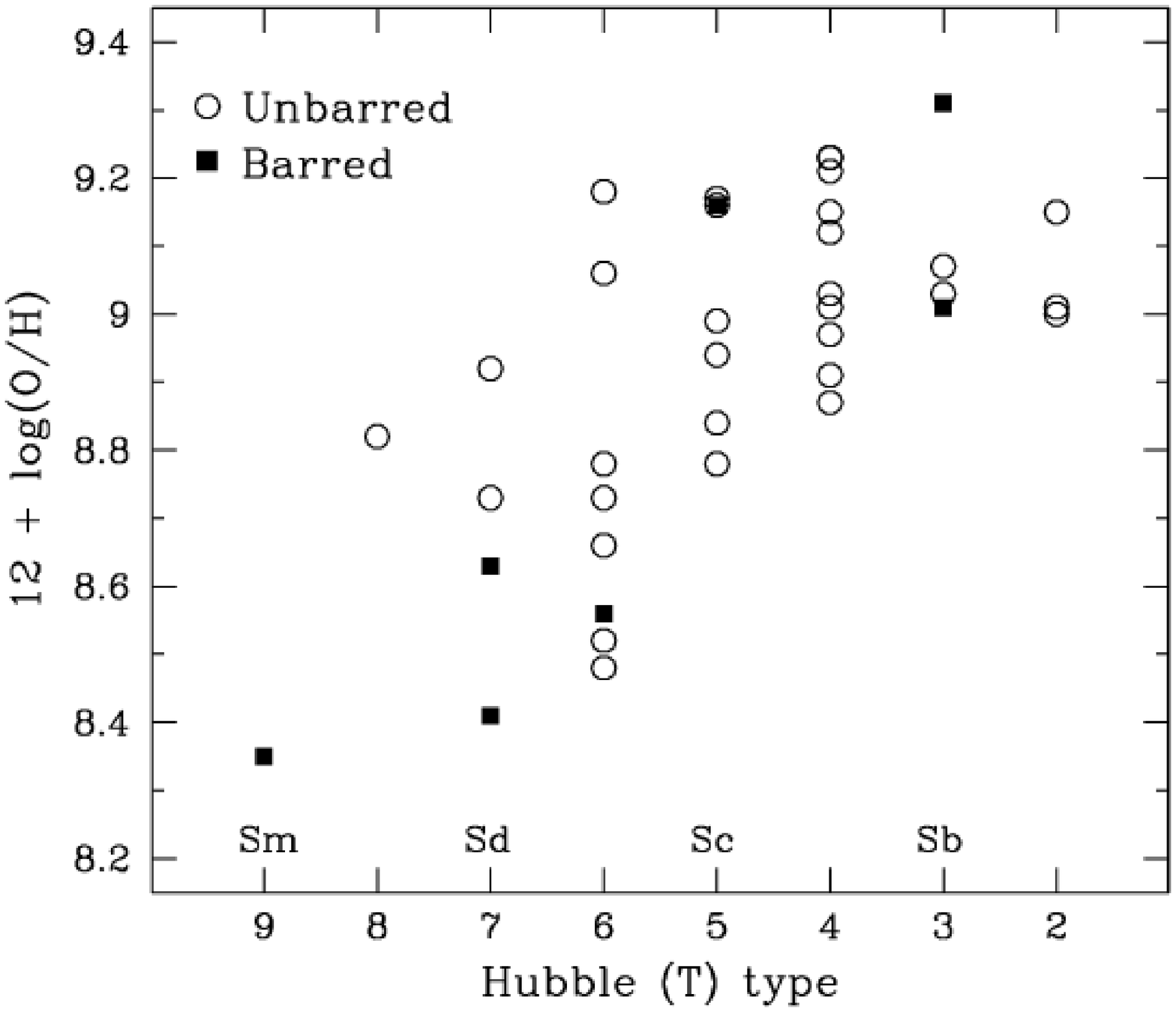}
\caption{Characteristic abundance vs.\ Hubble type (T-type as defined by RC2) from
\citet{zar94}. Smaller T-types correspond to earlier type galaxies.
\label {zari_fig}}
\end{figure}

The chemical composition of spiral disks has been studied by several authors,
mainly through spectroscopic studies of HII regions \citep{vil92,zar94,van98,gar02}.
\citet{zar94} measured the oxygen abundance,\footnote{Defined as 12 + log(O/H), where (O/H) is
the number ratio of oxygen to hydrogen atoms.
The present day local ISM has the value
12 + log(O/H) = $8.79 \pm 0.08$ \citep{car08}.
} using a photoionization
model for calibration, in 39 disk galaxies at the characteristic radius
$r=0.4\rho_{0}$, where $\rho_{0}$ is the $B$ = 25 mag arsec$^{-2}$ isophotal radius.
Their main result for various Hubble types is shown in Fig.~\ref{zari_fig}.

Adopting the usual normalization $X$ + $Y$ + $Z$ = 1, where $X$, $Y$, and $Z$ are, respectively,
the abundances
per unit mass of hydrogen, helium, and the remaining elements,
we have:

\begin{equation}
  \frac{Z(O)}{X} = 16(O/H),
\end{equation}

\noindent
where $Z(O)/Z$ is the fraction of $Z$ due to oxygen. This fraction
varies with metallicity between 41\% and 53\% in the Local Group of galaxies \citep{pei03}.
Taking the values $X$ = 0.75 and $Z(O)/Z$ = 0.45, the expression for $Z$ in terms of the oxygen
abundance is:

\begin{equation}
  \log(Z) \simeq 1.43 + \log(O/H).
\end{equation}

\begin{figure}
\epsscale{1.00}
\plotone{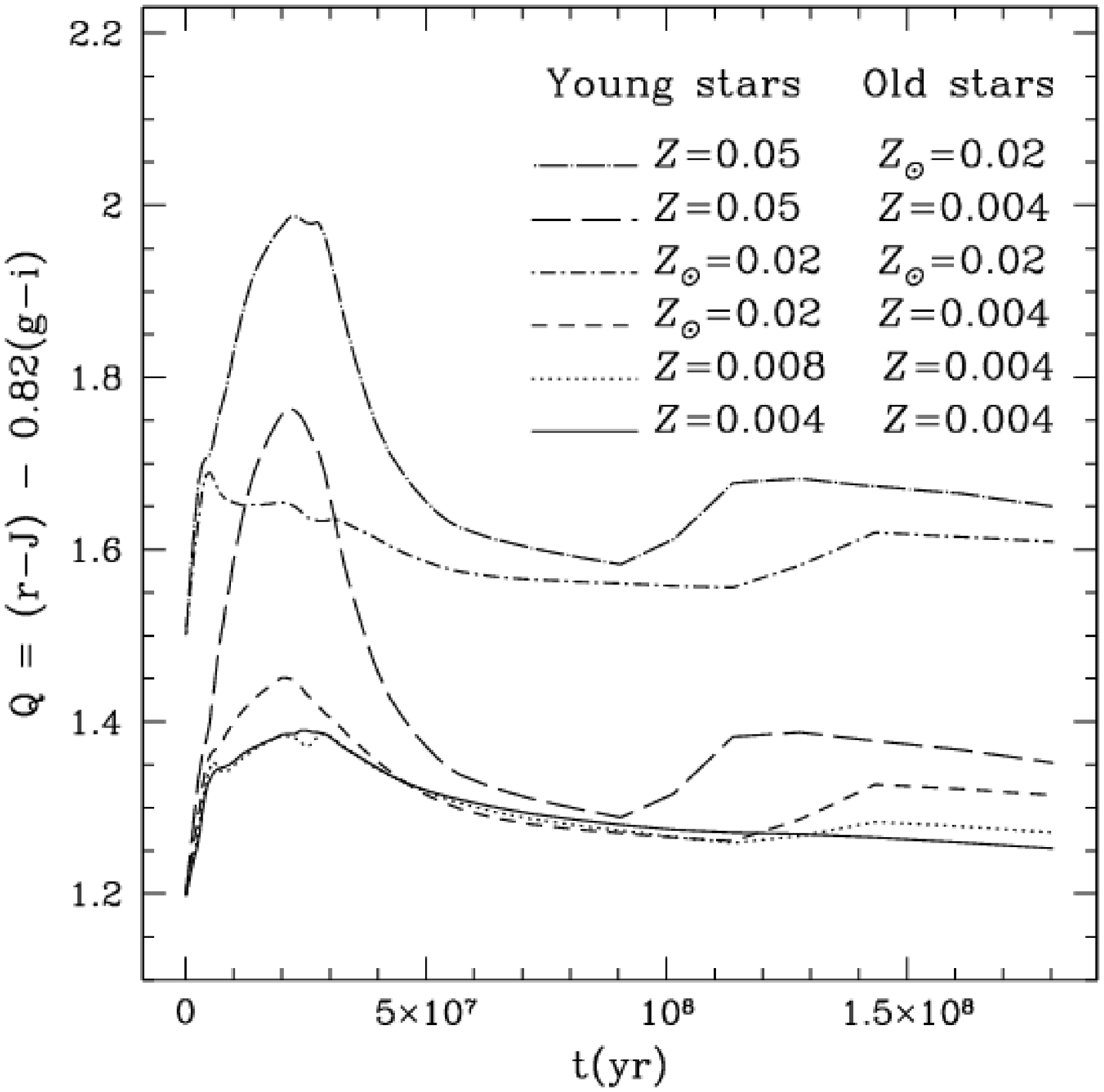}
\caption{$Q$ index behavior with different combinations of metallicities for the young and old stars.  
{\it Solid line:} $Z_{\rm young} = 0.004$, $Z_{\rm old} = 0.004$;
{\it dotted line:} $Z_{\rm young} = 0.008$, $Z_{\rm old} = 0.004$;
{\it dashed line:} $Z_{\rm young} = 0.02$, $Z_{\rm old} = 0.004$;
{\it dashed-dotted line:} $Z_{\rm young} = 0.02$, $Z_{\rm old} = 0.02$;
{\it long-dashed line:} $Z_{\rm young} = 0.05$, $Z_{\rm old} = 0.004$;
{\it long-dashed-dotted line:} $Z_{\rm young} = 0.05$, $Z_{\rm old} = 0.02$.
\label {Qmeta}}
\end{figure}

Assuming that young stars have metallicity values around the ones inferred
from abundance ratios, we obtain that $Z$ is in the range $[0.006,0.05]$
for the disk stars in the Zaritsky et al.\ (1994) sample.
In Fig.~\ref{Qmeta} we show the behavior of the $Q$ index for various combinations of metallicities
for the young and old stars, using the CB07 models.
In these models the adopted solar metallicity is $Z_{\sun} = 0.02$. The duration of the burst
is $2 \times 10^7$ years, both young and old stars have a Salpeter IMF with $M_{lower} = 0.1
M_{\sun}$ and $M_{upper} = 100 M_{\sun}$. The fraction of young stars is 2\% by mass,
and the old stellar background population is $5 \times 10^9$ years old.

When young stars have $Z > Z_{\sun}$,
$Q$ reaches much higher values than in models where young stars have
subsolar metallicities.
Models that differ in the metallicities of the young population show
different $Q$ values mainly between 1$\times 10^7$ and 3$\times 10^7$ yr,
when the young stars are most prominent; if, on the other hand,
the young stars have the same $Z$ but the old populations do not, the
models run basically parallel, with a roughly constant offset in the
$Q$ values at all ages.

\subsection{Simple vs. complex models.} \label{simpvscmpx}

Through the comparison between examples of the more
complex models presented and our data,
we estimate that ignoring the deformations produced
by variable densities, variable velocities, and different
metallicities (see \S~\ref{denveldist}, \S~\ref{non_circ} and \S~\ref{rolmeta})
would translate into a maximum error of approximately 
$\pm$1 km s$^{-1}$ kpc$^{-1}$ in $\Omega_p$ (see also the Appendix).
This quantity is well within the random errors computed for 
$\Omega_p$ shown in Table \ref{tbl-omegas}.

The small size of this error might signal a sort of selection bias:
if non-linear effects were very important, it would be unlikely
to find a gradient. On the other hand, remarkably, we have been able to
detect gradients in an unprecedented 10 out of 13 galaxies.
Not only have we quadrupled the historic number of detections,
but we have found gradients in more than 75\% of the subsample
analyzed in this paper, that does not even comprise
(with the exception of M~99) the ``$Q$ effect" galaxies,
i.e., the objects with the strongest features. This means that
non-linear effects that would deform all gradients beyond 
recognition might not be too significant
in galaxies with spiral density waves. When GG96 found the gradient in
M~99, an obvious question that arose was whether NGC~4254 was an exception
or if an adequate technique to search for the gradients had been finally
devised. The present work answers that question unambiguously.

Furthermore, the comparison of the gradients with the simplest
of models yields orbital resonance positions that match the observed spiral endpoints.
These results agree with the predictions of density wave theory, and
establish a strong link between disk dynamics and
large scale star formation, as we argue below.

\section{Discussion and conclusions.}

If star formation is related to disk dynamics then, according to theory, the
$R_{res} / R_{end}$ ratio must be close to one, for either the 4:1 resonance,
the corotation radius, or the OLR.
\citet{con86}, based on orbital calculations,
determined that the spiral pattern of strong spirals must extend to the 4:1 resonance.
\citet{pats91} defined strong spirals as those with large pitch angles (i.e., Sb and Sc galaxies,
that are not tightly wound); their theory is considered
``non linear". Conversely, the ``linear" theory of spiral density waves
concerns itself with tightly wound spirals and has concluded that
the arms of some normal (nonbarred) spirals reach corotation \citep{lin70}
and are stationary, while those of others grow to the OLR \citep{mar76,lin79,too81}.
In the \citet{con86} treatment, the ``linear" theory
is recovered when spiral arms are not strong (as in Sa galaxies).

In Fig.~\ref{spiralEP}, we can see that most spirals in our sample
extend to the OLR and one of them (NGC~578) reaches corotation.
The mean of the ratio $R_{OLR}/R_{end}$, for all 13 points, is
0.95$\pm$0.03. The reduced $\chi^{2}/n$ of this result is 7.12;\footnote{
For an expected value of $R_{OLR}/R_{end}=1$.}
the probability of this result being
due to chance for 13 degrees of freedom is less than 1 out of 10,000.
If we remove the points corresponding to NGC~3938, NGC~7126, NGC~6951, and NGC~578,
$R_{OLR}/R_{end} = 0.98\pm0.04$, with a reduced $\chi^{2}/n$ = 0.48, for $n=9$.
This last result may indicate that some of our errors are overestimated.
It is interesting to note that none of our objects is an Sa (weak spiral)
galaxy. From this we may conclude that the ``linear" result
for the extent of spiral patterns applies to strong spirals too!

On the other hand, for most objects the spiral begins at the location of the ILR,
as expected.
For the analysis presented here, though,
the location of this resonance should be taken with care, since we
are employing flat rotation curves, even at small radii. When using real
rotation curves, the positions of the ILRs for our sample may vary.
The possibility also exists that there is no ILR in some objects.

In some regions, the downstream values of the $Q$ index are
lower than the models. This is the case for
NGC~4939~A, NGC~3162~B, NGC~1421~A, NGC~1421~C,
NGC~7125~A, NGC~7125~C, NGC~578~A, and NGC~578~B.
For the present analysis we have assumed
that every star-forming region maintains its previous circular motion after the
spiral shock. As already stated, in real galaxies the situation is different: variations
in the velocity vector are present due to the shock itself and the resulting loss
of angular momentum \citep{yua81,fer08}. This effect could explain the
``downstream fall" of the gradients.
Metallicity is another factor that can play a significant part: 
a lower metallicity of the older stars may produce a steep drop of the gradients
(see \S~\ref{rolmeta}).

We have shown that azimuthal color gradients are common in spiral arms of
disk galaxies. Even inverse color gradients have been found in the
arms of NGC~1421, NGC~4939, and NGC~7125. The observed picture is, of course, not as clean
as originally envisioned, for several reasons. Dust and HII regions may mask the
gradients (GG96); indeed,
all the detected gradients have been satisfactorily fitted with
models where $M_{upper}$ = 10 $M_\odot$.
Other star formation mechanisms, like self-propagating star
formation, may take place simultaneously with density wave triggering in spiral arms and disks
in general. Likewise, there may be substructure formation, as a result of non-linear
and chaotic effects associated to the wave phenomenon itself \citep{cha03,dob06,kim06,she06}.
In the future,
three aspects can be improved when carrying out this type of study.
(1) A better determination of the inclination angle of the galaxy, and of
the rotation curve at different radii; this will reduce the
uncertainty in $\Omega_p$.
(2) The inclusion in the models of the variable density distributions of young and
old stars, as well as of the stellar velocity changes near the spiral shocks.
This will account for non-circular motions near the spiral
shocks and for the higher order terms
in equation \ref{eqOMEGA_I}.
Numerical simulations or semi-analytical treatments may be needed, particularly
for the young stars;
also, the contribution of red supergiants
to the observed $K$-band surface brightness should be considered
when modeling the distribution of old stars.
(3) Spectroscopic studies could be very helpful as diagnostics of the
population properties, particularly of the metallicity along the
gradients.

According to the dynamic parameters derived from our analysis,
spiral arms mostly extend to the OLR and
sometimes to the corotation radius. Ten of thirteen objects (77\%) support
this conclusion. Similar results have been obtained by other authors as
well \citep[e.g.,][]{elm92,zha07}. In view of the consistency between the
pattern speeds and resonance positions determined from the azimuthal stellar gradients,
and the predictions of density wave theory, we conclude that disk dynamics do
play an important role in large scale star formation in some spiral galaxies.

\acknowledgments

We are grateful to the anonymous referee for her/his  
helpful comments.
It is a pleasure to thank A.\ Watson, L.\ Loinard,
and S.\ Lizano for their advice and motivation.
We thank M.\ Peimbert
for useful discussions about chemical abundances,
and J.\ Huchra for kindly responding our inquires about Hubble distances.
We also acknowledge CONACyT and DGEP (UNAM), M\'exico, for financial support during
the development of this work.

\appendix

\section{Pattern speed and corotation radius error calculation.}

In principle, the three contributors to the random error in the
pattern speed, $\sigma_{\Omega_{p}}$ (km s$^{-1}$ kpc$^{-1}$), are the
uncertainties in:
the inclination angle, $\sigma_{\alpha}$; the
rotation velocity, $\sigma_{v_{max}}$; and the distance to the galaxy,
$\sigma_{D}$.

In order to compute
$\sigma_{\Omega_{p}}$ from equation~\ref{eqOMEGA_II},
and hence the uncertainty in the resonance positions, we replace $R_{mean}$, $v_{rot}$, and $d$
with  their corresponding expressions in terms of independent variables.
We obtain:

\begin{equation} \label{eqOMEGA_pix}
\Omega_{p} = \frac{k_{1} v_{max} }{r D (\sin\alpha) }
- \frac{k_{2}}{t} \cos^{-1} \left( \frac{\frac{x_{1}x_{2}}{\cos^{2} \alpha} + y_{1}y_{2}}{r^{2}} \right),
\label{eq_omegaind}
\end{equation}

\begin{equation}
r = \frac{\sqrt{\frac{x^{2}_{1}}{\cos^{2} \alpha} + y_{1}^{2}}+
         \sqrt{\frac{x^{2}_{2}}{\cos^{2} \alpha} + y_{2}^{2}}} {2},
\end{equation}

\noindent
where constant $k_{1} = \frac{648000}{p_{s} \pi}$ (pixels), $p_{s}$ is
the adopted plate scale in arsec pixel$^{-1}$, constant $k_{2} =
9.766 \times 10^{8}$ (yr km s$^{-1}$ kpc$^{-1}$),
$r$ is the mean deprojected radius of the region in pixels;
$v_{max}$ is the maximum rotation velocity in km s$^{-1}$;
$(x_{1}, y_{1})$ and $(x_{2}, y_{2})$ are the cartesian coordinates (in pixels)
of the extremes of the curved
line segment corresponding to the studied region in the non-deprojected image
(the origin is located at the center of the object and
the galaxy major axis runs along the y coordinate);
$D$ is the distance to the object in kpc;
$\alpha$ is the inclination angle used to deproject the image;
and $t$ is the stellar model age in years.

The corotation radius, $R_{\rm CR}$ (kpc), is:

\begin{equation}
R_{\rm CR} = \frac{v_{max}}{(\sin\alpha) \Omega_{p}}.
\label{eq_rcr}
\end{equation}

\noindent
The expressions for each contributor to the error calculation are:

\begin{eqnarray}
\Delta \Omega_{p} [\alpha] = \Omega_{p}(\alpha+\sigma_{\alpha},v_{max},D,t) -
\Omega_{p}(\alpha,v_{max},D,t) \\
\Delta \Omega_{p} [v_{max}] = \Omega_{p}(\alpha,v_{max}+\sigma_{v_{max}},D,t) -
\Omega_{p}(\alpha,v_{max},D,t) \\
\Delta \Omega_{p} [D] = \Omega_{p}(\alpha,v_{max},D+\sigma_{D},t) -
\Omega_{p}(\alpha,v_{max},D,t) \\
\sigma_{\Omega_{p}} = \sqrt{(\Delta \Omega_{p} [\alpha])^{2}  +
                          (\Delta \Omega_{p} [v_{max}])^{2} +
                    (\Delta \Omega_{p} [D])^{2} }.      
\end{eqnarray}

\begin{figure}
\centering
\epsscale{1.0}
\plotone{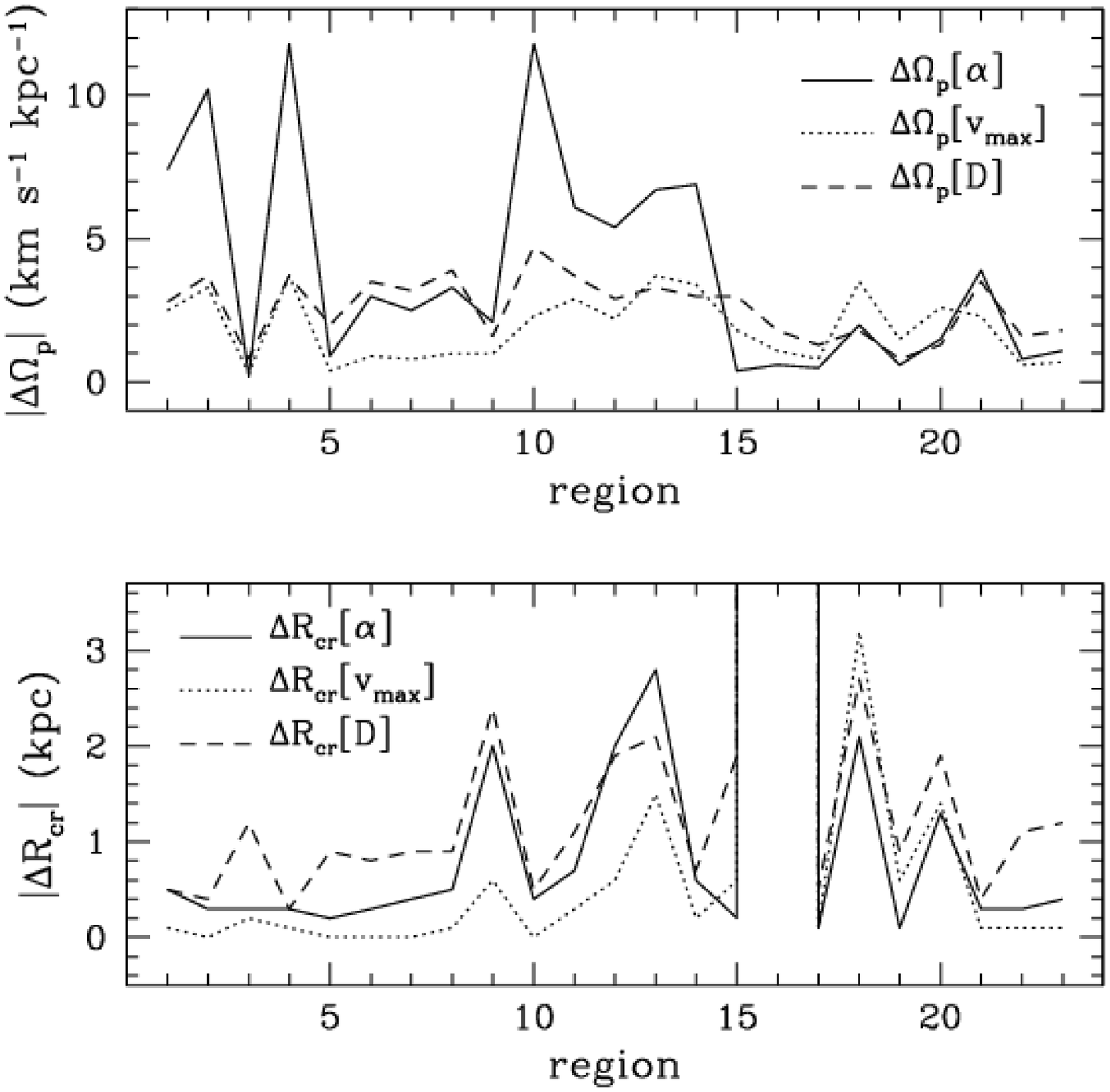}
\caption[f17.eps]{Absolute $\Delta \Omega_{p}$ and $\Delta R_{\rm CR}$ average values for
galaxy regions. Regions are identified by their numbers in Table~\ref{tbl-omegas}. {\it Solid line:}
contribution from inclination angle, $\alpha$; {\it dotted line:} contribution
from rotation velocity, $v_{max}$; {\it dashed line:} contribution from distance
to the galaxy, $D$.
The measurements are shown as continuous lines for easiness of reading; in reality, they are
discrete points.\label{sigmas_OM}}
\end{figure}

\noindent
Values for $\pm \sigma$ were calculated, and the absolute $\Delta \Omega_{p}$ values were averaged.
Similar expressions hold for $\sigma_{R_{\rm CR}}$. Figure~\ref{sigmas_OM} shows the $\Delta \Omega_{p}$
and $\Delta R_{\rm CR}$ absolute average values for each region listed in Table~\ref{tbl-omegas}. 
The systematic error due to the choice of stellar population model was estimated by comparing
the $\Omega_p$ values derived with the simple models of \S~\ref{qnmod}, on the one hand,
with examples of the more complex models discussed in \S~\ref{simpvscmpx}, on the other.

\clearpage

\begin{figure}
\centering
\epsscale{1.0}
\plotone{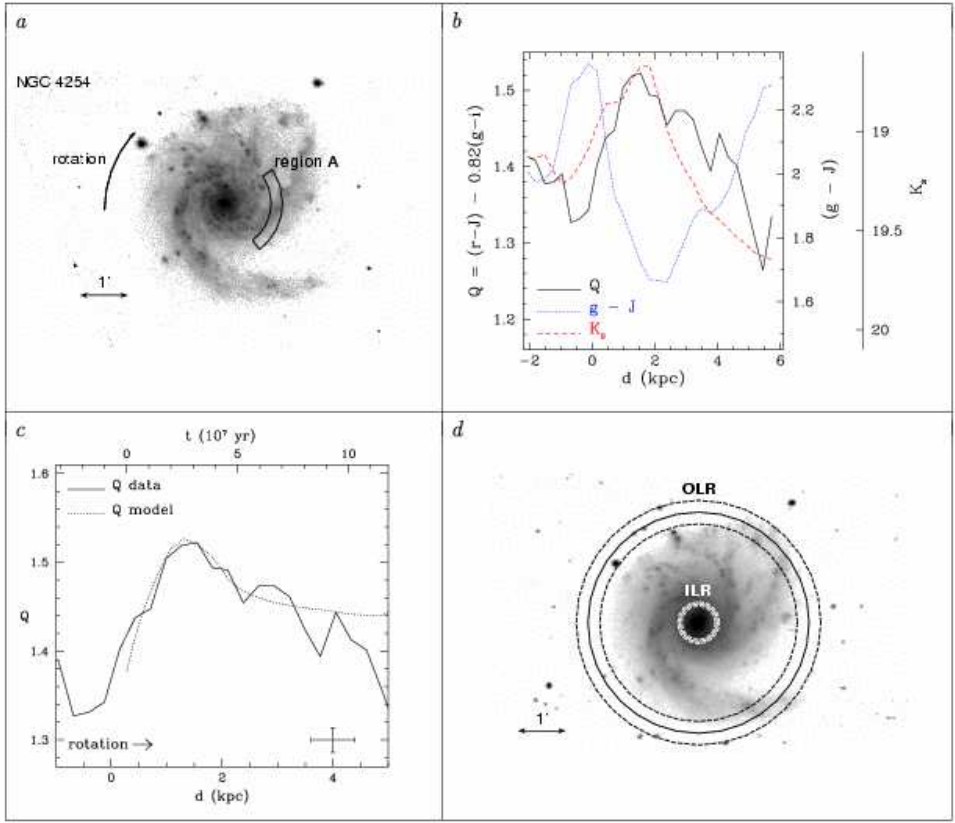}
\caption[f18.eps]{Region NGC 4254 A.
Panel $a$: Deprojected spiral galaxy NGC~4254 in the optical $g$-band.
The display is in logarithmic scale.
The studied regions are marked on the figure.
Panel $b$: 1-D data vs.\ azimuthal distance for region NGC~4254~A.
{\it Solid line:} $Q$ index; {\it dotted line:}
($g - J$) color; {\it dashed line:} $K_s$-band profile.
Panel $c$: Zoom-in of region, and
1-D $Q$ index vs.\ azimuthal distance.
{\it Solid line:} data; {\it dotted line:} model.
Data error bars and direction of rotation are indicated.
Panel $d$: $K_s$-band deprojected mosaic of spiral galaxy NGC~4254.
{\it Solid line circles:}
location of the ILR and the OLR,
as obtained from the comparison between data and stellar population
SPS model shown in panel $c$.
{\it Dashed line circles:} $\pm1\sigma$ errors. \label{REG_4254_A}}
\end{figure}

\begin{figure}
\centering
\epsscale{1.0}
\plotone{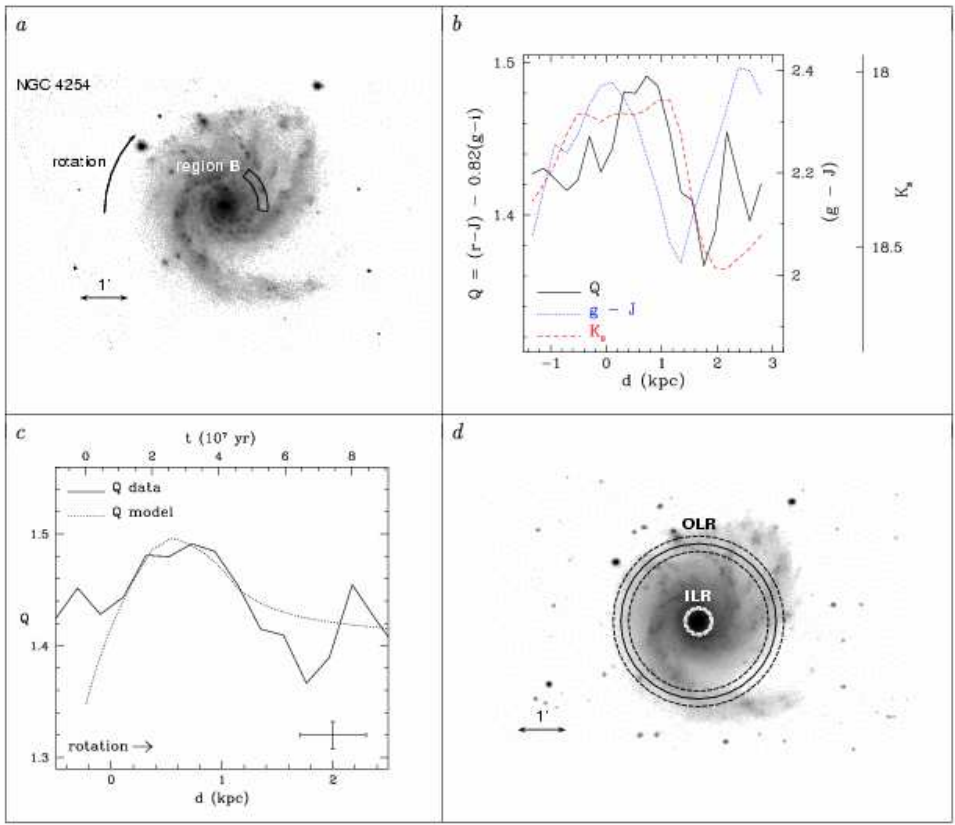}
\caption[f19.eps]{Region NGC 4254 B.
\label{REG_4254_B}}
\end{figure}

\clearpage

\begin{figure}
\centering
\epsscale{1.0}
\plotone{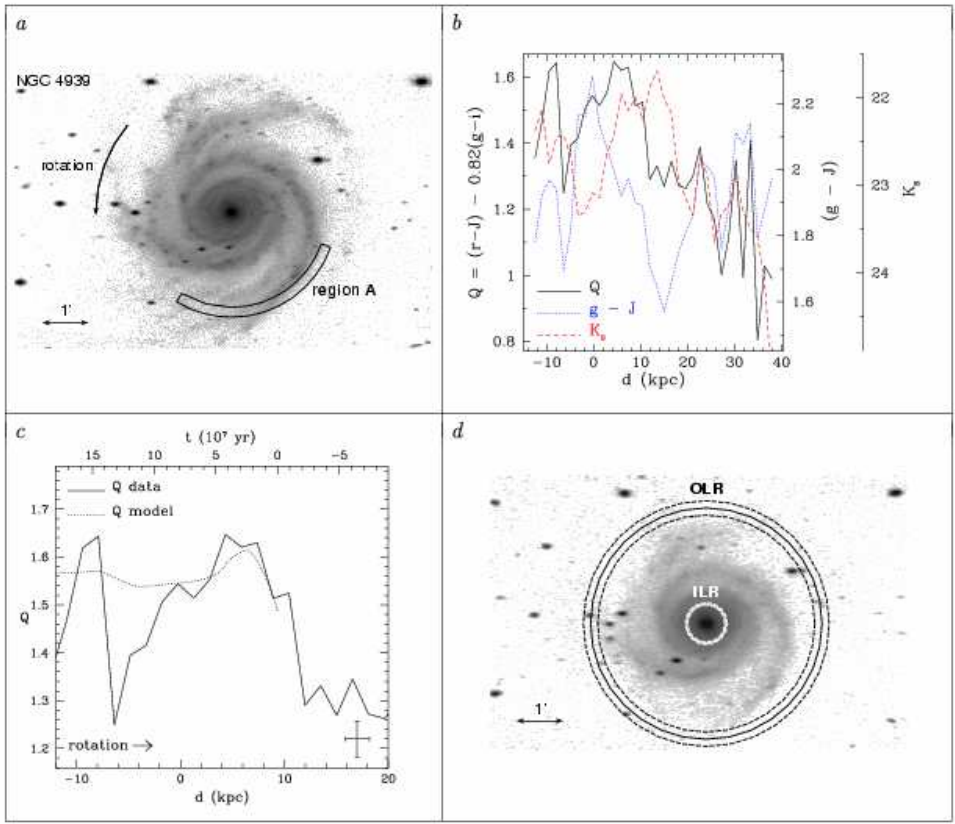}
\caption[f20.eps]{Region NGC 4939 A.
In order to fit the model to the data we assume
the gradient is inverse, occurring beyond corotation.
This can also be seen from the fact that time evolution
(top axis in panel $c$) and rotation have opposite directions. \label{REG_4939_A_I}}
\end{figure}

\begin{figure}
\centering
\epsscale{1.0}
\plotone{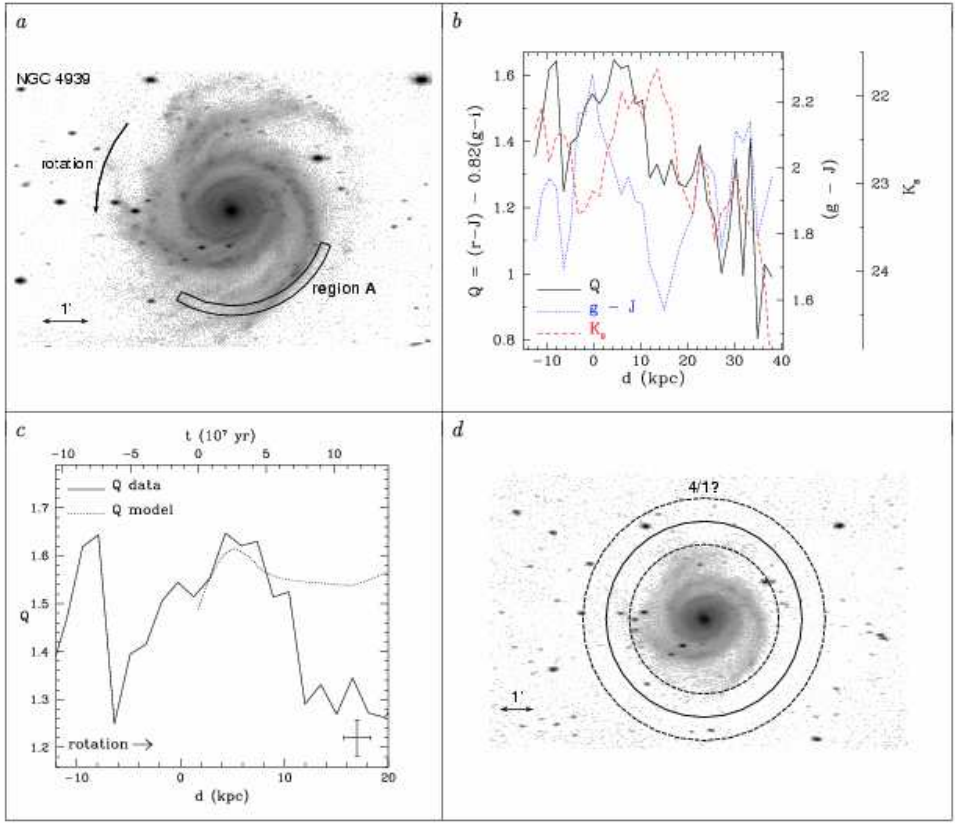}
\caption[f21.eps]{Region NGC 4939 A.
In order to fit the model to the data we assume
now that the gradient occurs inside corotation.
In this case, time evolution (top axis in panel $c$) and
rotation have the same direction.
Panel $d$,
{\it solid line circle:}
location of the 4:1 resonance,
as obtained from the comparison between data and
SPS model shown in panel $c$.
\label{REG_4939_B_II}}
\end{figure}

\clearpage

\begin{figure}
\centering
\epsscale{1.0}
\plotone{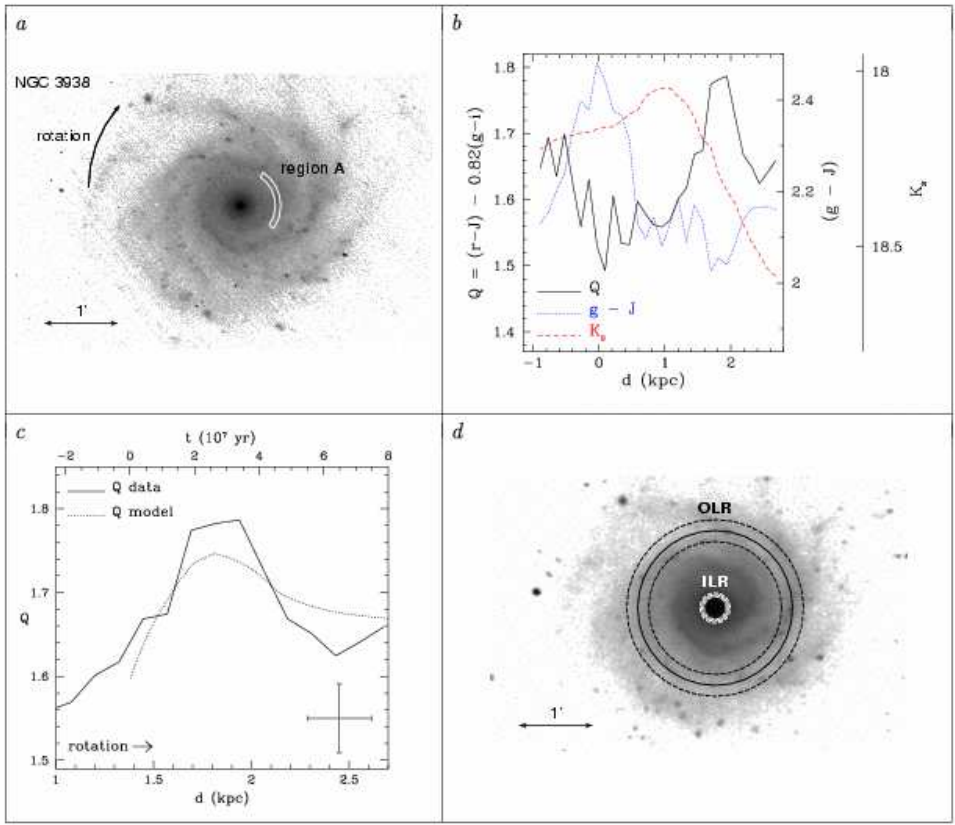}
\caption[f22.eps]{Region NGC 3938 A.
Panel $a$: Deprojected spiral galaxy NGC~3938
in the optical $r$-band. \label{REG_3938_A}}
\end{figure}

\clearpage

\begin{figure}
\centering
\epsscale{1.0}
\plotone{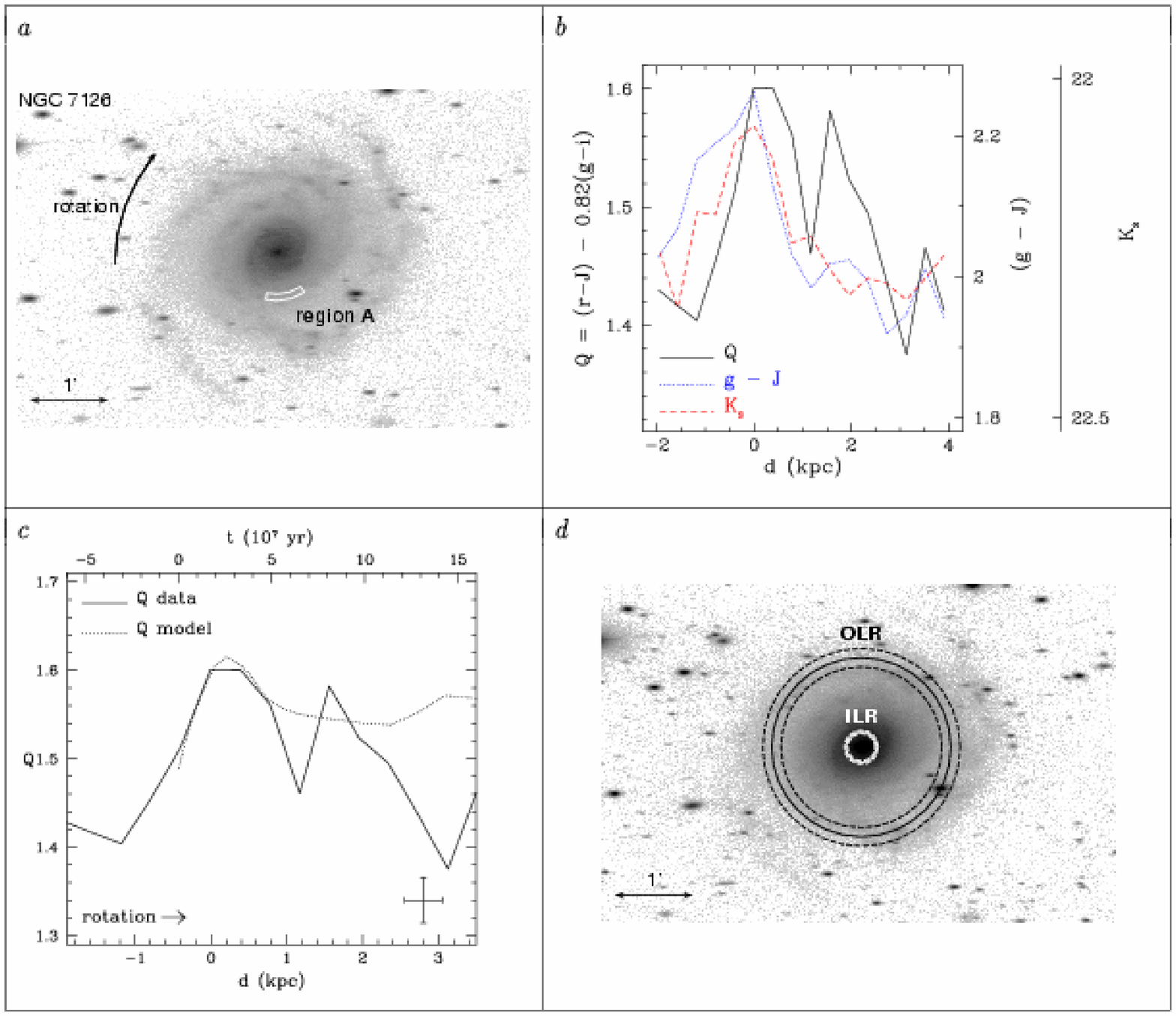}
\caption[f23.eps]{Region NGC 7126 A.
Panel $d$: $i$-band deprojected mosaic
of spiral galaxy NGC~7126. \label{REG_7126_A}}
\end{figure}

\clearpage

\begin{figure}
\centering
\epsscale{1.0}
\plotone{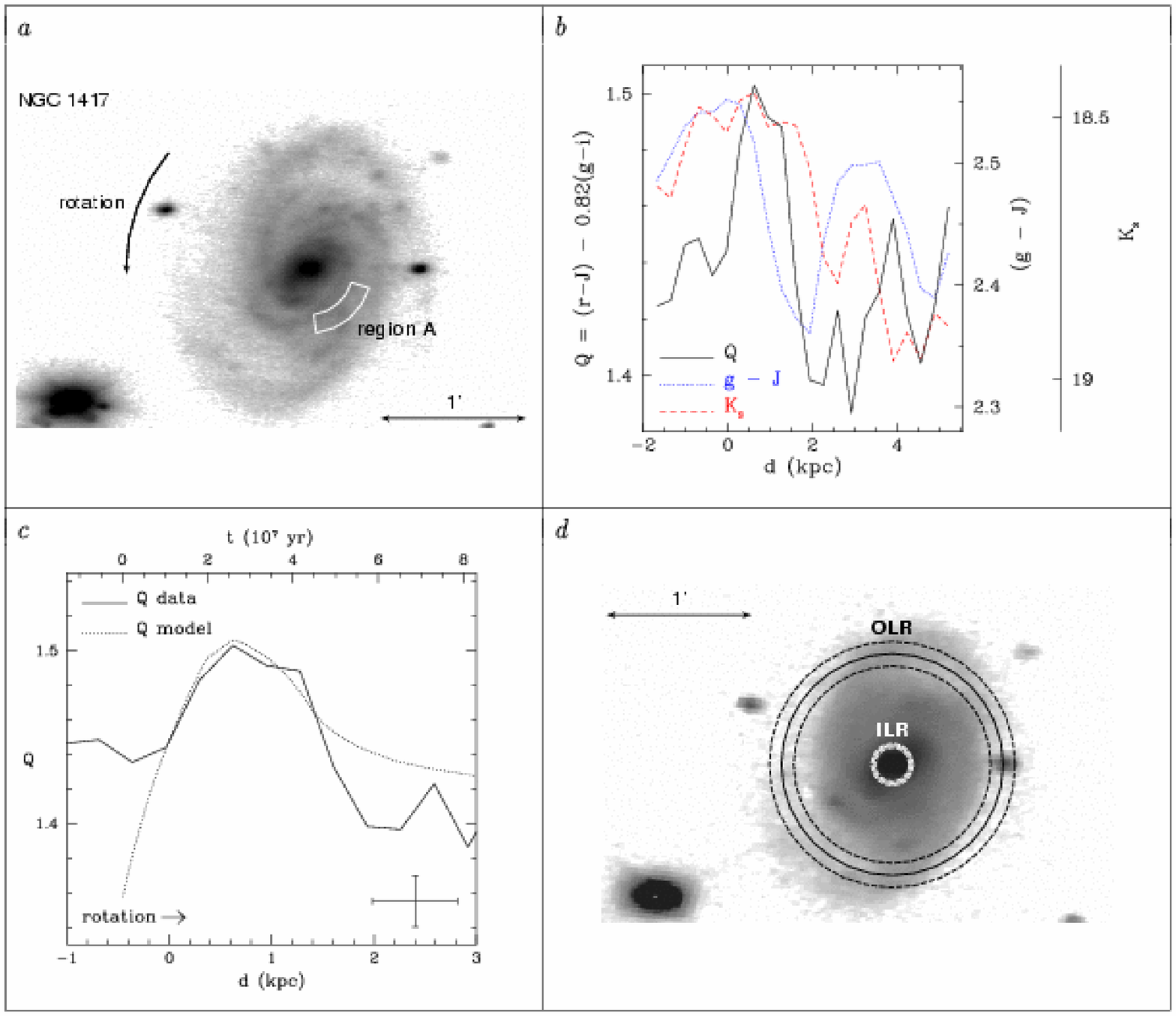}
\caption[f24.eps]{Region NGC 1417 A.
Panel $d$: $J$-band deprojected mosaic of spiral galaxy NGC~1417.
\label{REG_1417_A}}
\end{figure}

\begin{figure}
\centering
\epsscale{1.0}
\plotone{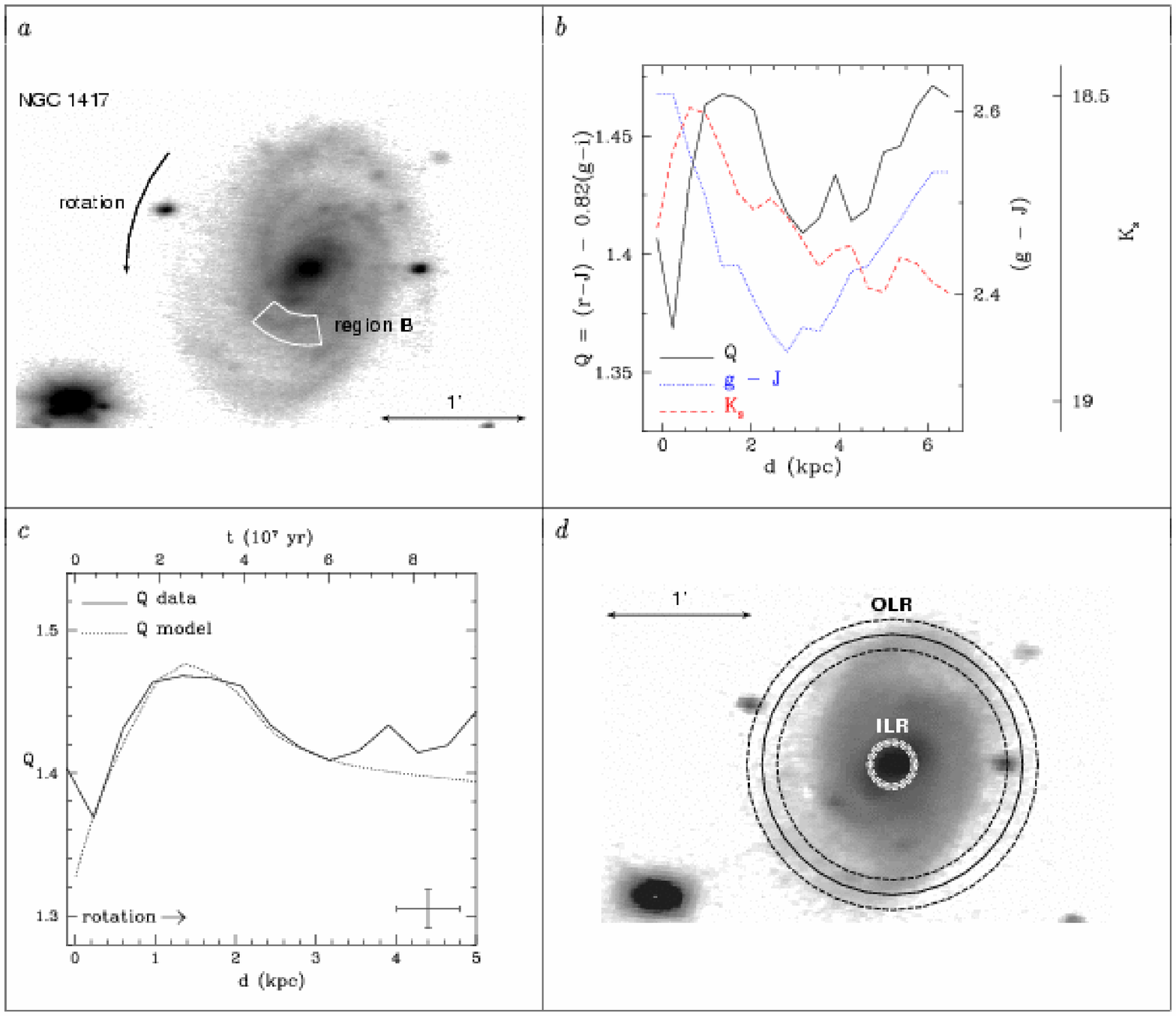}
\caption[f25.eps]{Region NGC 1417 B.
Panel $d$: $J$-band deprojected mosaic of spiral galaxy NGC~1417.
\label{REG_1417_B}}
\end{figure}

\begin{figure}
\centering
\epsscale{1.0}
\plotone{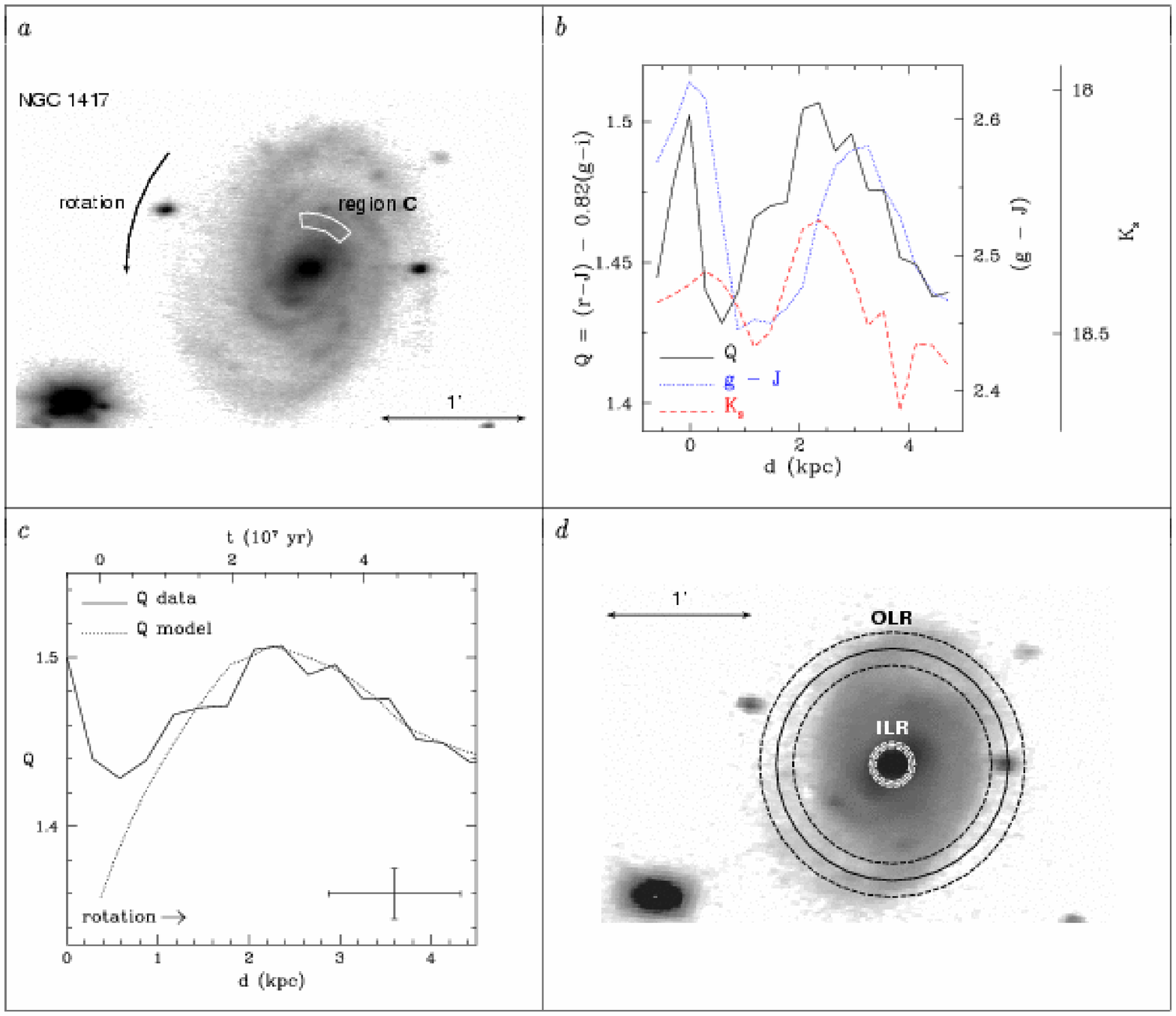}
\caption[f26.eps]{Region NGC 1417 C.
Panel $d$: $J$-band deprojected mosaic of spiral galaxy NGC~1417.
\label{REG_1417_C}}
\end{figure}

\clearpage

\begin{figure}
\centering
\epsscale{1.0}
\plotone{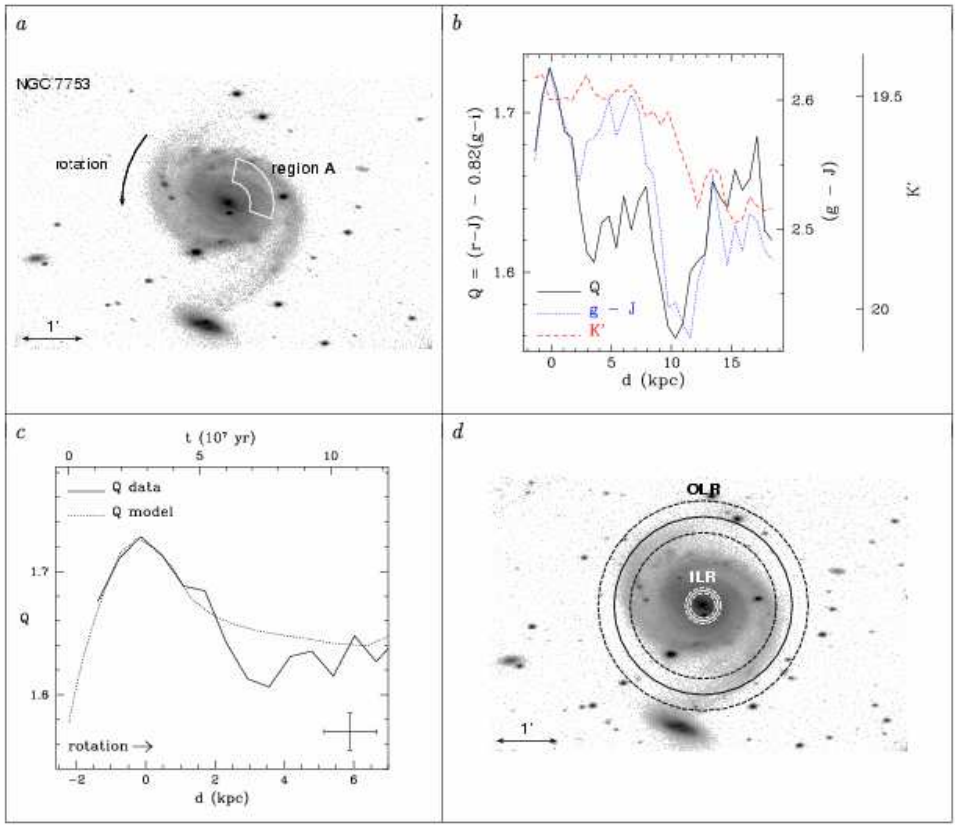}
\caption[f27.eps]{Region NGC 7753 A.
Panel $d$: $i$-band deprojected mosaic of spiral galaxy NGC~7753.
\label{REG_7753_A}}
\end{figure}

\clearpage

\begin{figure}
\centering
\epsscale{1.0}
\plotone{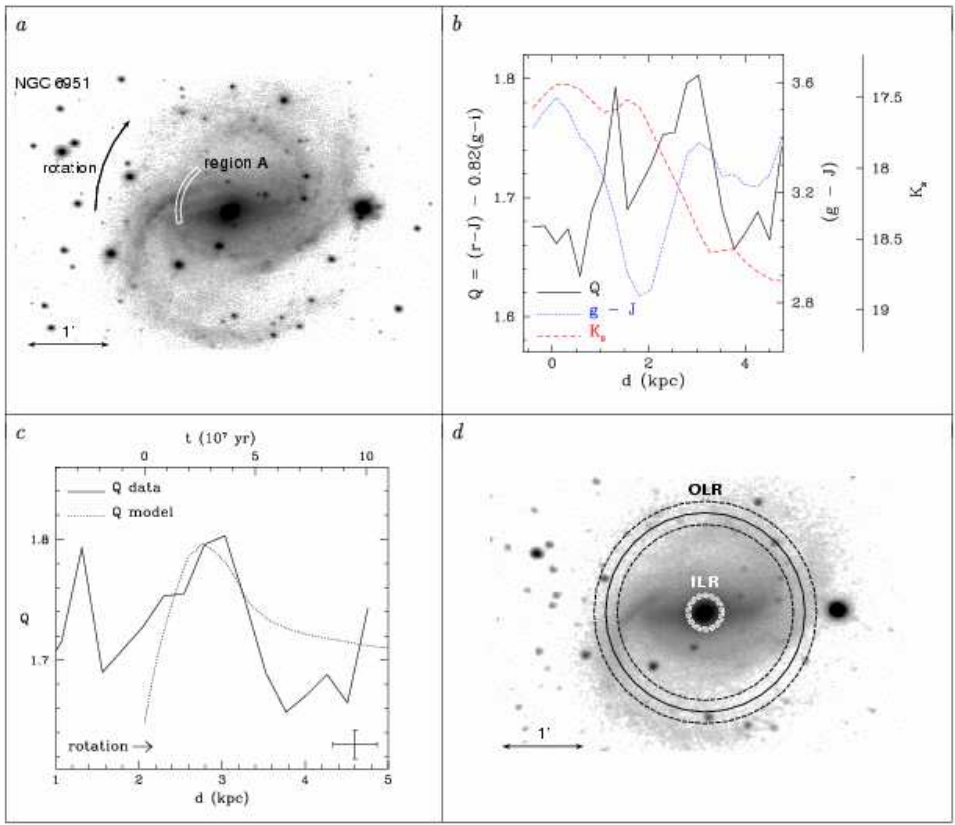}
\caption[f28.eps]{Region NGC 6951 A. \label{REG_6951_A}}
\end{figure}

\clearpage

\begin{figure}
\centering
\epsscale{1.0}
\plotone{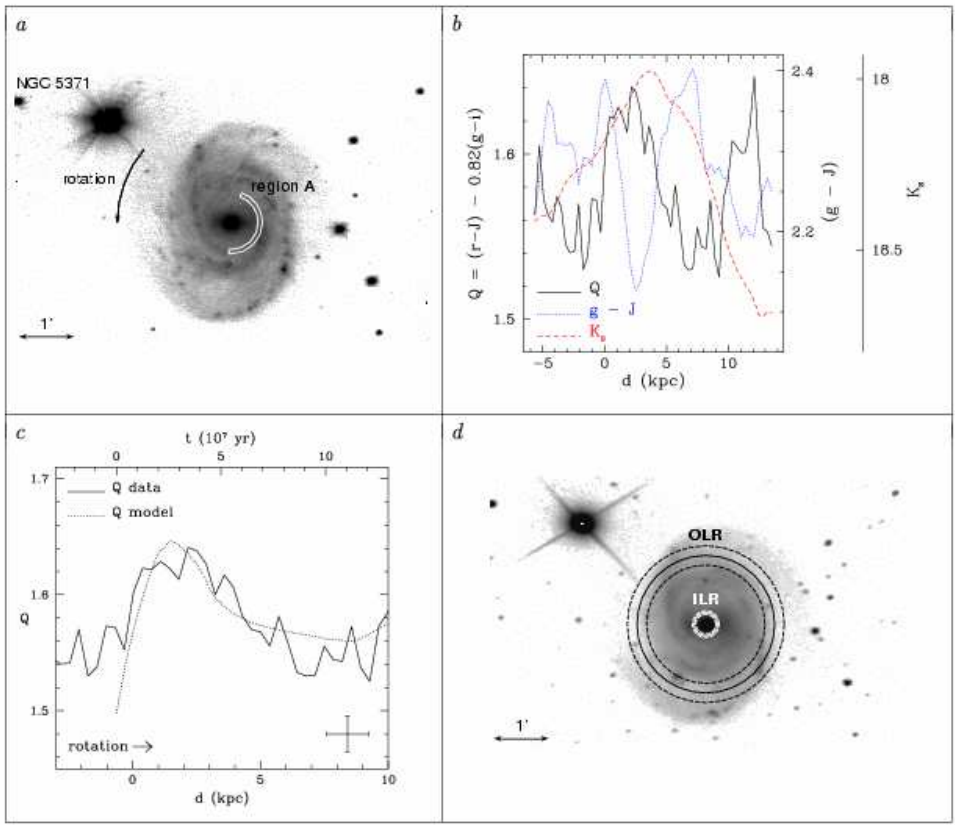}
\caption[f29.eps]{Region NGC 5371 A. \label{REG_5371_A}}
\end{figure}

\begin{figure}
\centering
\epsscale{1.0}
\plotone{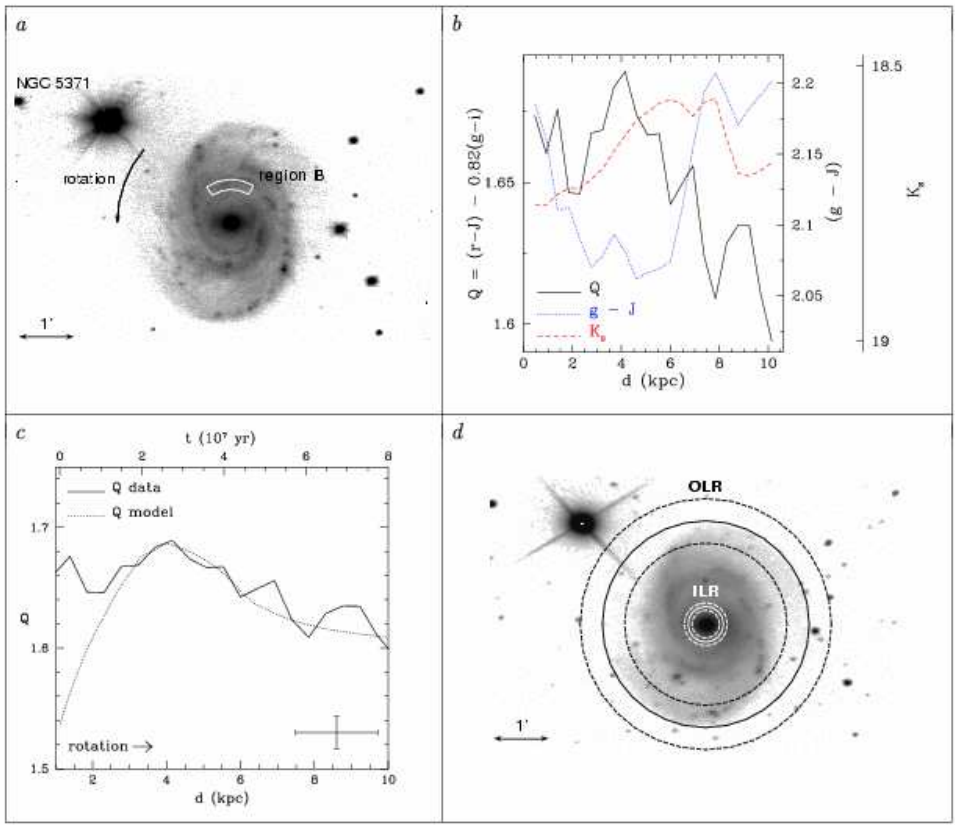}
\caption[f30.eps]{Region NGC 5371 B. \label{REG_5371_B}}
\end{figure}

\clearpage

\begin{figure}
\centering
\epsscale{1.0}
\plotone{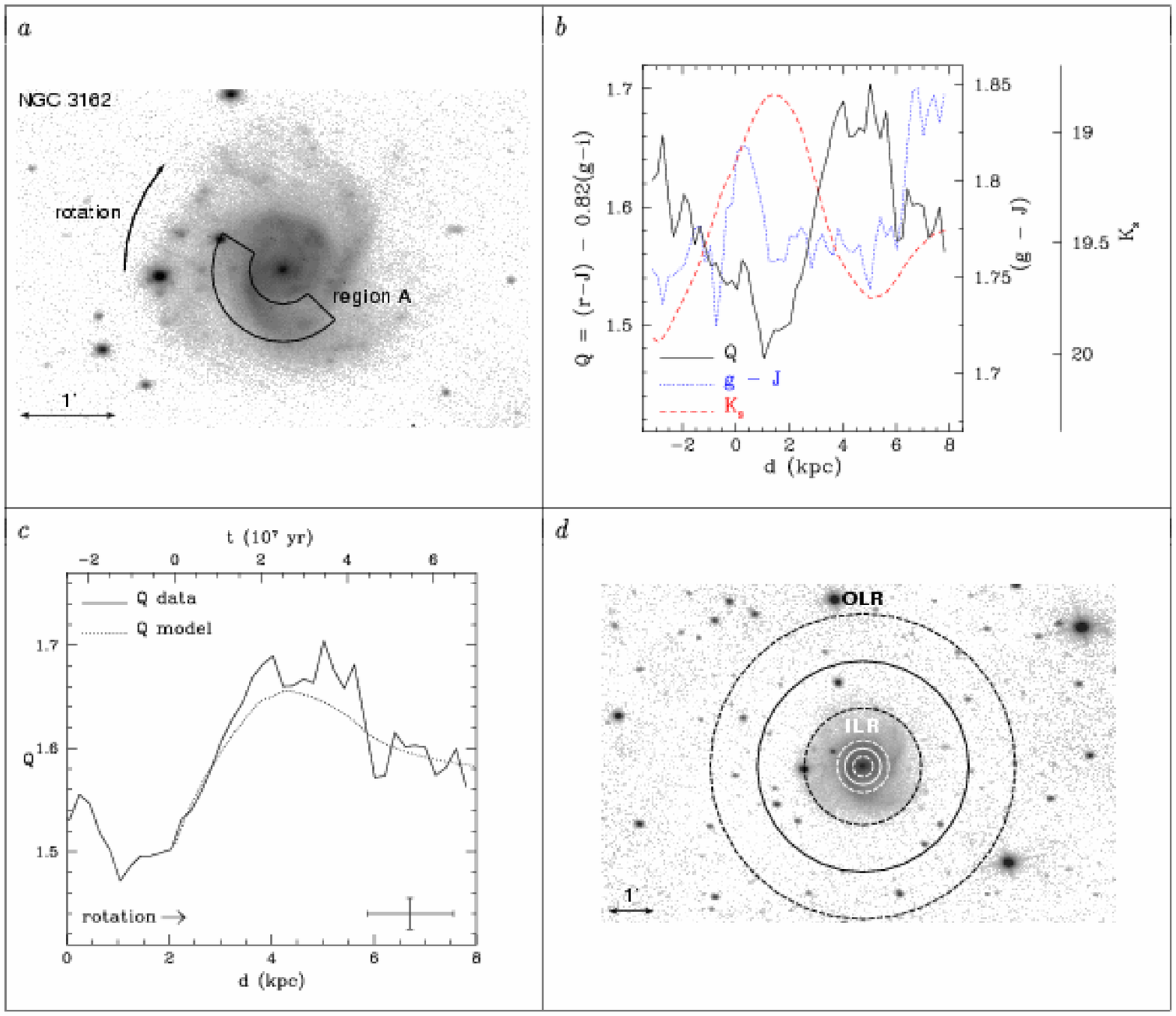}
\caption[f31.eps]{Region NGC 3162 A.
Panel $d$: $J$-band deprojected mosaic of spiral galaxy NGC~3162.
\label{REG_3162_A}}
\end{figure}

\begin{figure}
\centering
\epsscale{1.0}
\plotone{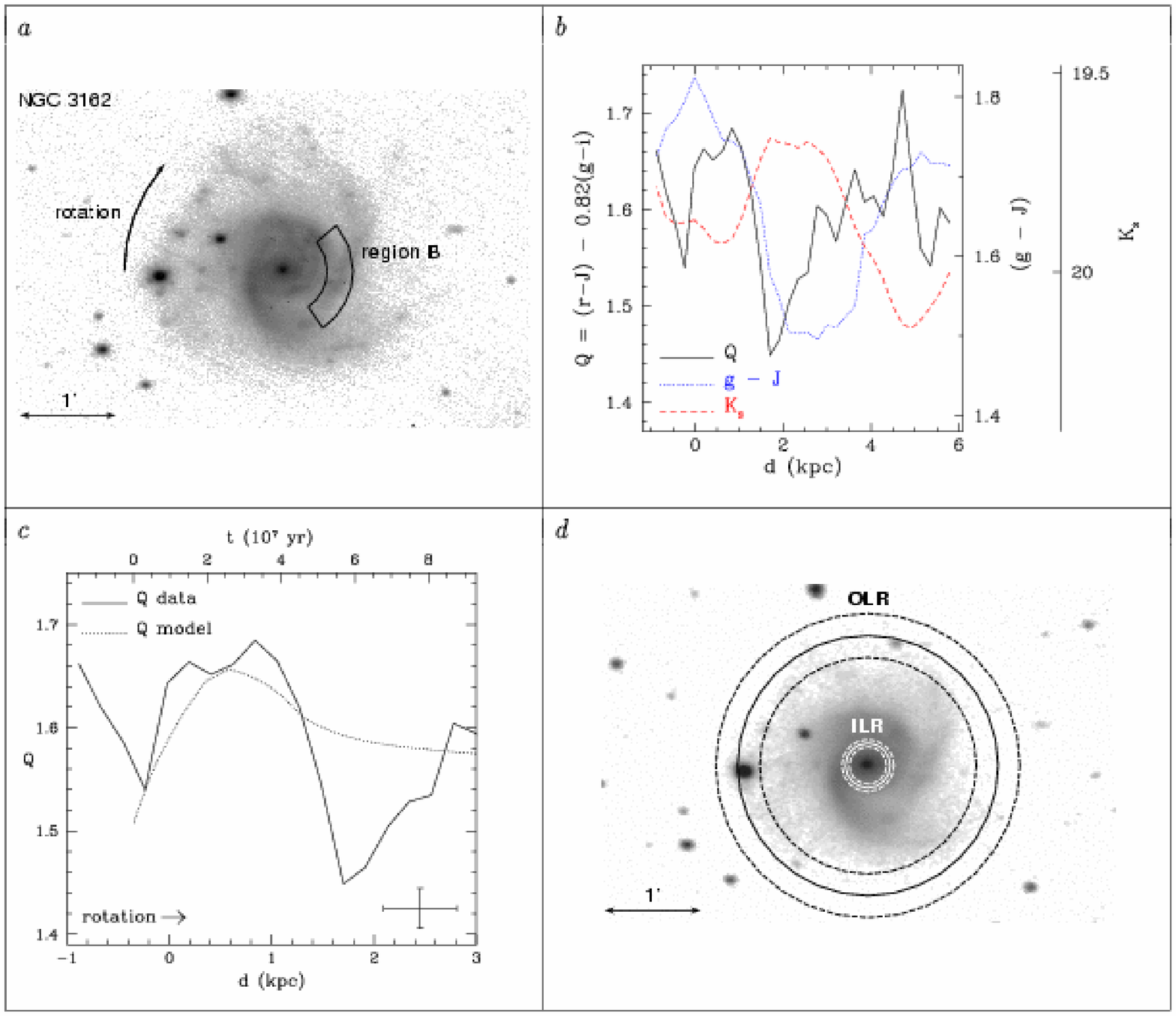}
\caption[f32.eps]{Region NGC 3162 B.
Panel $d$: $J$-band deprojected mosaic of spiral galaxy NGC~3162.
\label{REG_3162_B}}
\end{figure}

\clearpage

\begin{figure}
\centering
\epsscale{1.0}
\plotone{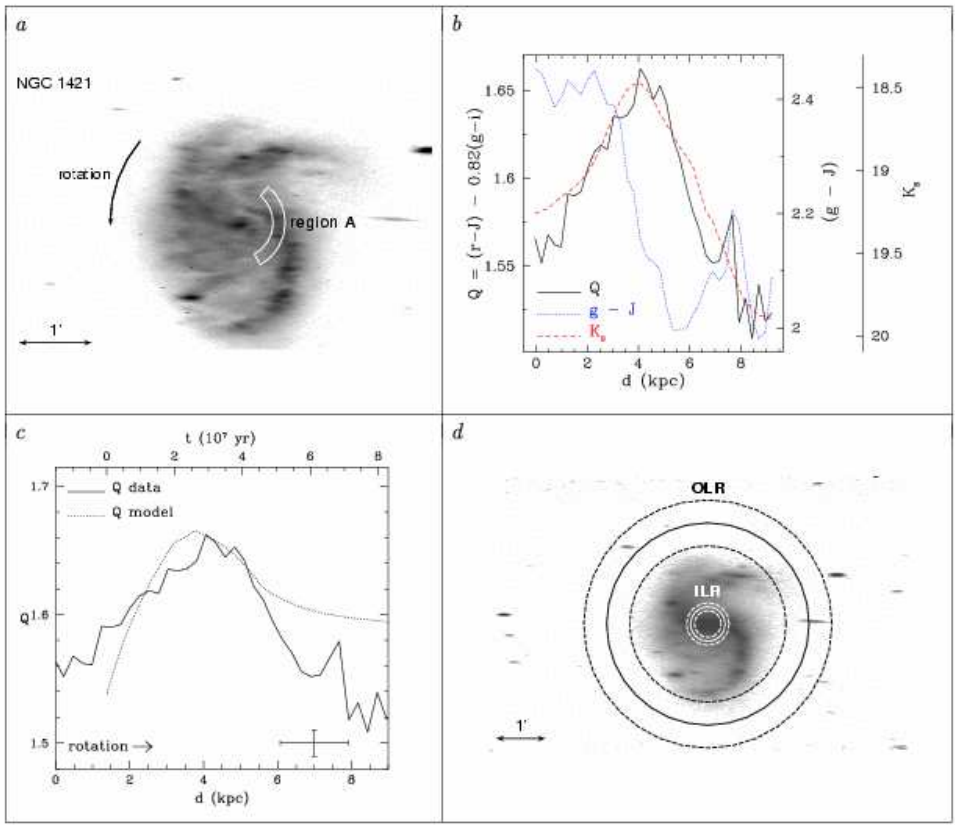}
\caption[f33.eps]{Region NGC 1421 A. \label{REG_1421_A}}
\end{figure}

\begin{figure}
\centering
\epsscale{1.0}
\plotone{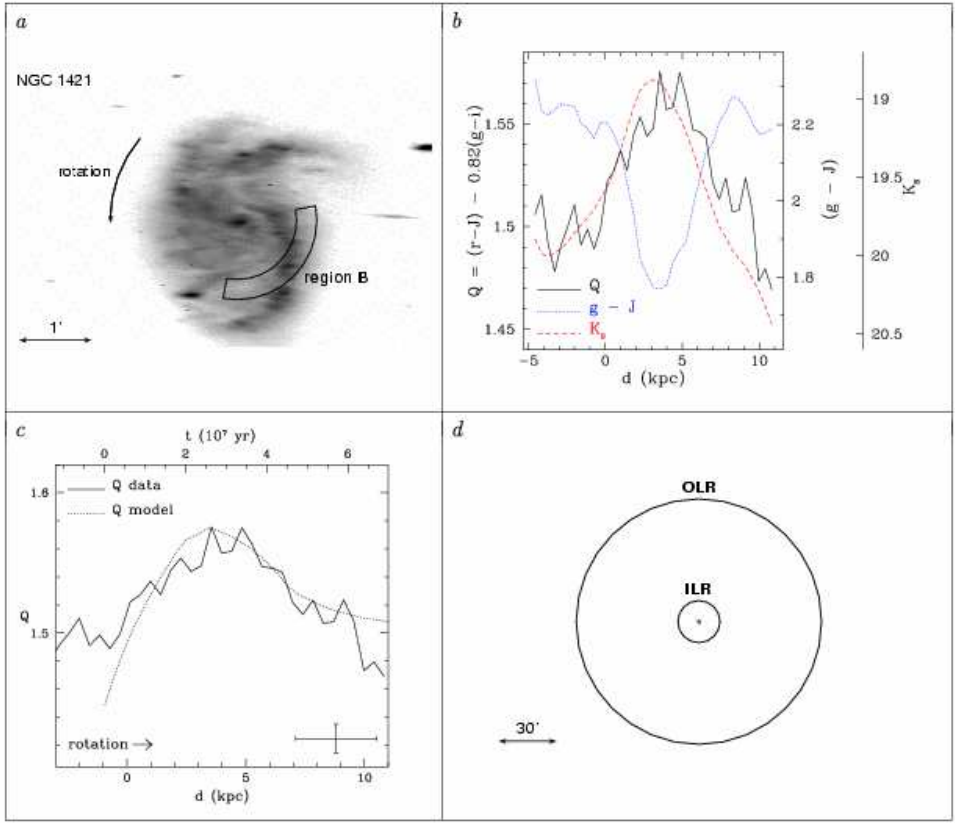}
\caption[f34.eps]{Region NGC 1421 B.
Panel $d$,
{\it solid line circle:}
location of the ILR and the OLR,
as obtained from the comparison between data and
SPS model shown in panel $c$.
The errors for the resonance positions are much greater
than the computed resonance radii themselves (see Fig.~\ref{sigmas_OM}).
This is due to the fact that region B
lies right on the corotation position obtained from regions A \& C.
\label{REG_1421_B}}
\end{figure}

\begin{figure}
\centering
\epsscale{1.0}
\plotone{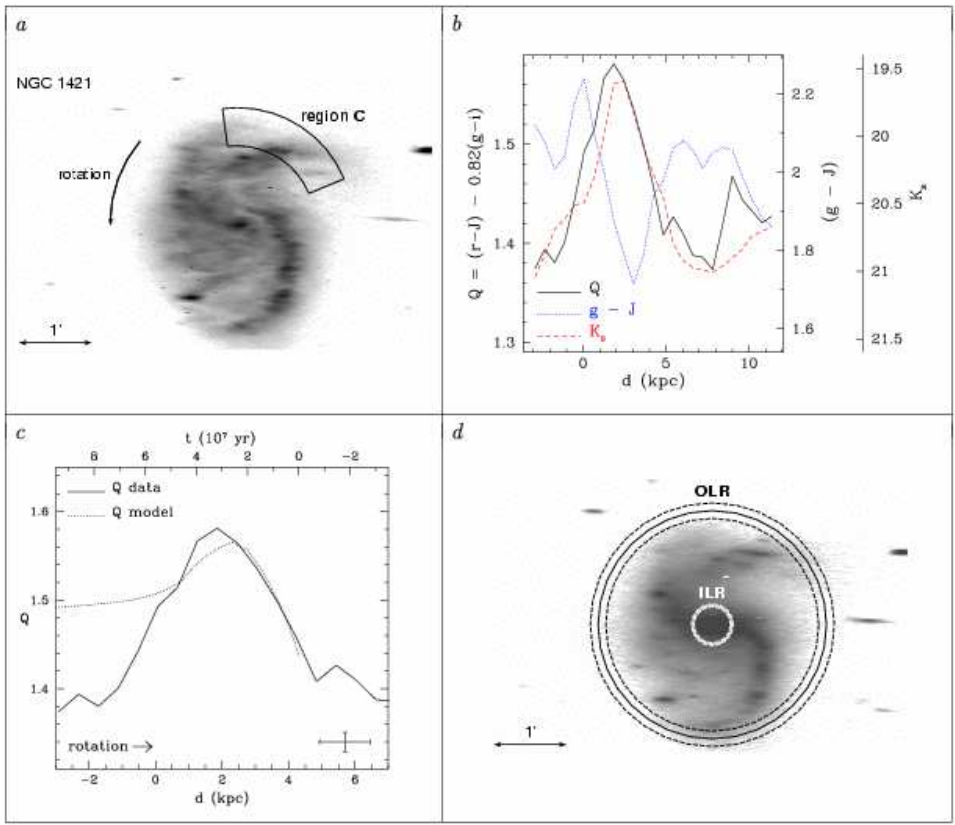}
\caption[f35.eps]{Region NGC 1421 C. \label{REG_1421_C}}
\end{figure}

\clearpage

\begin{figure}
\centering
\epsscale{1.0}
\plotone{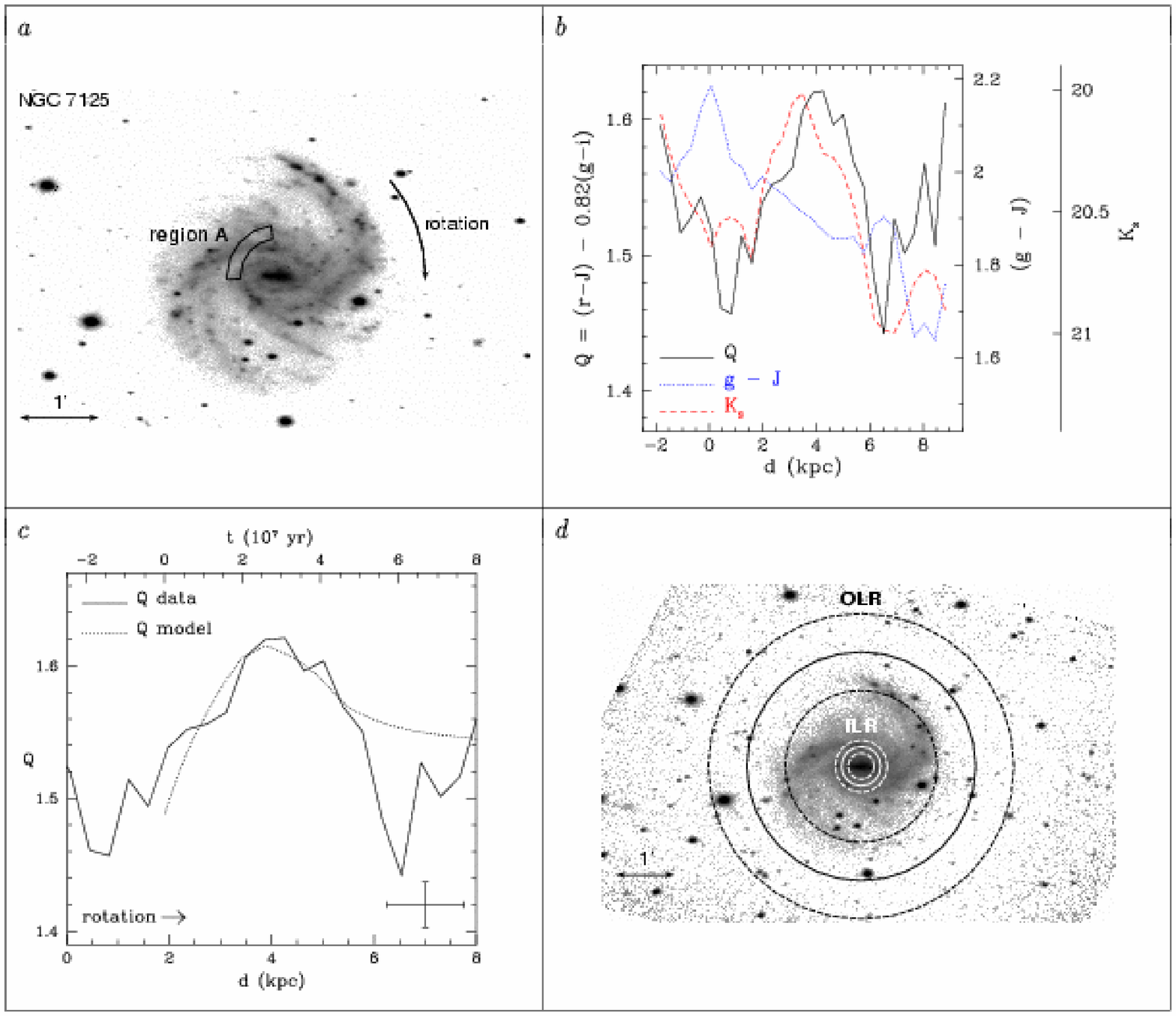}
\caption[f36.eps]{Region NGC 7125 A.
Panel $d$: $J$-band deprojected mosaic of spiral galaxy NGC~7125.
\label{REG_7125_A}}
\end{figure}

\begin{figure}
\centering
\epsscale{1.0}
\plotone{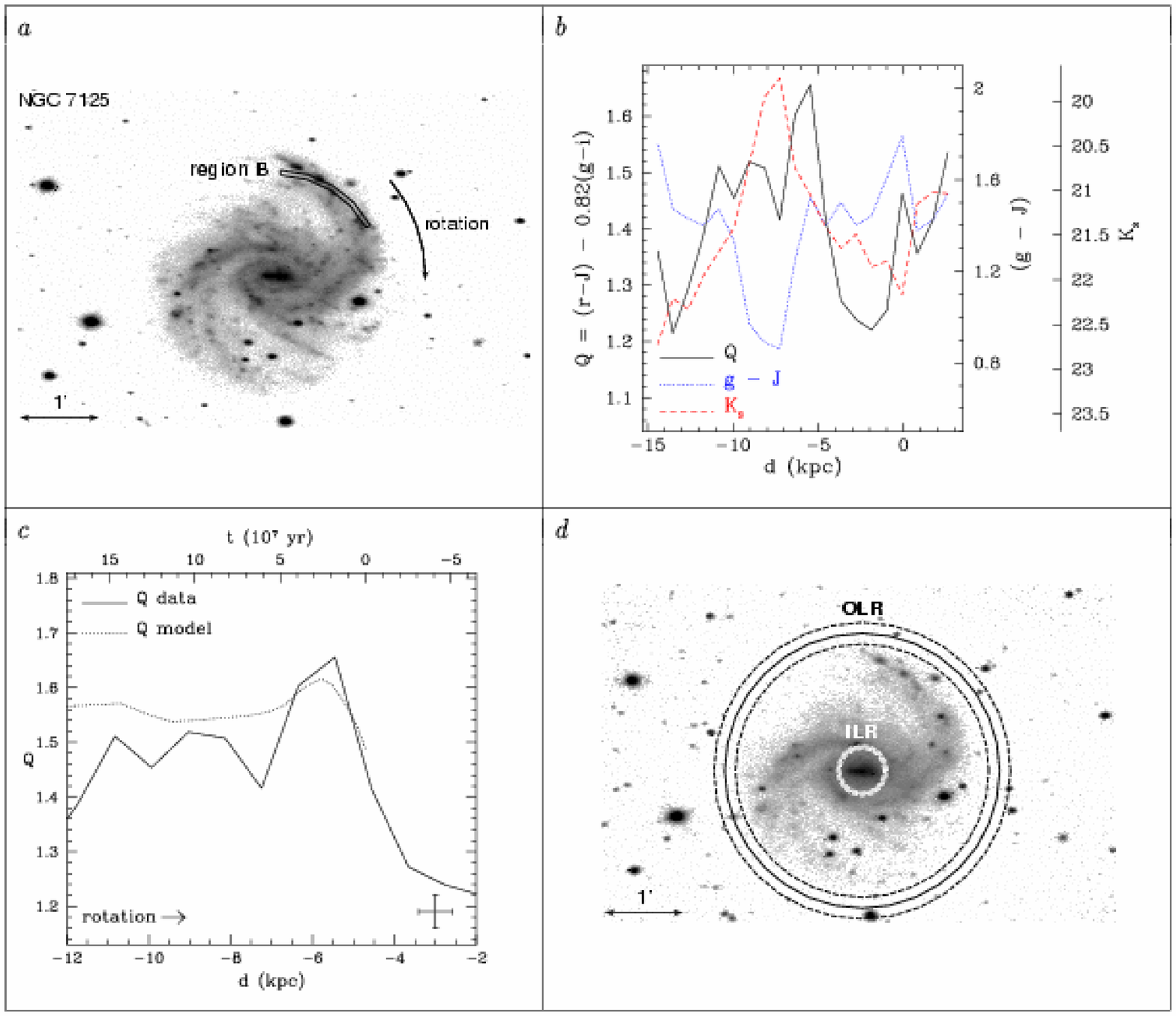}
\caption[f37.eps]{Region NGC 7125 B.
Panel $d$: $J$-band deprojected mosaic of spiral galaxy NGC~7125.
\label{REG_7125_B}}
\end{figure}

\begin{figure}
\centering
\epsscale{1.0}
\plotone{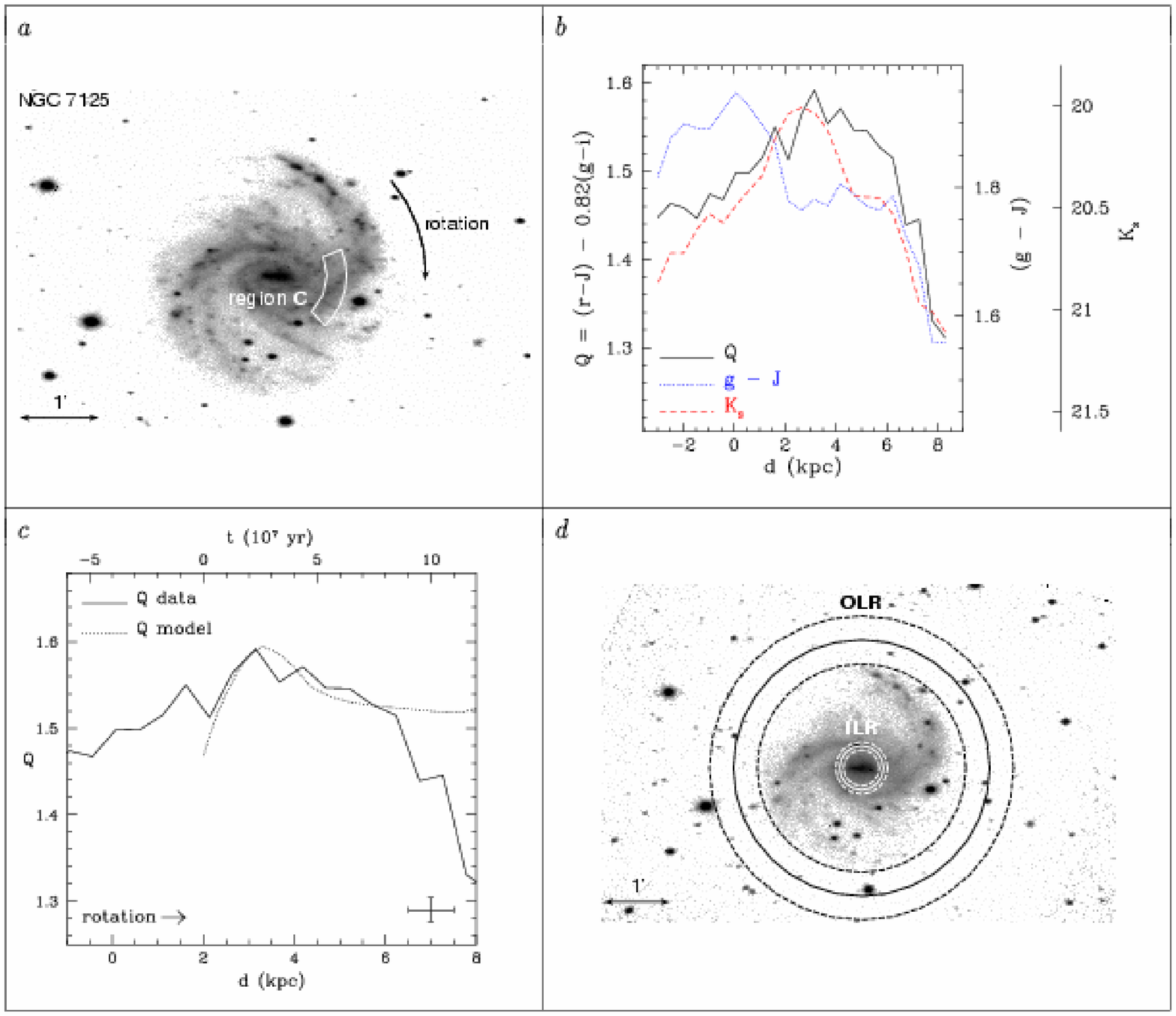}
\caption[f38.eps]{Region NGC 7125 C.
Panel $d$: $J$-band deprojected mosaic of spiral galaxy NGC~7125.
\label{REG_7125_C}}
\end{figure}

\clearpage

\begin{figure}
\centering
\epsscale{1.0}
\plotone{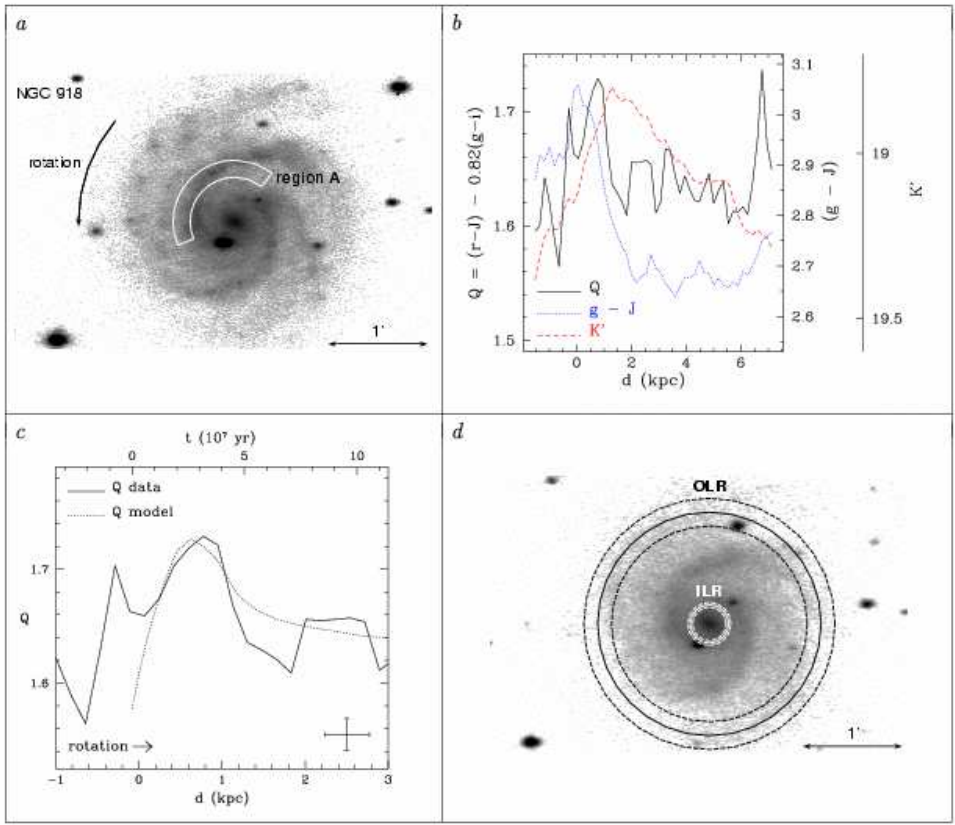}
\caption[f39.eps]{Region NGC 918 A.
Panel $d$: $K'$-band deprojected mosaic of spiral galaxy NGC~918.
\label{REG_918_A}}
\end{figure}

\clearpage

\begin{figure}
\centering
\epsscale{1.0}
\plotone{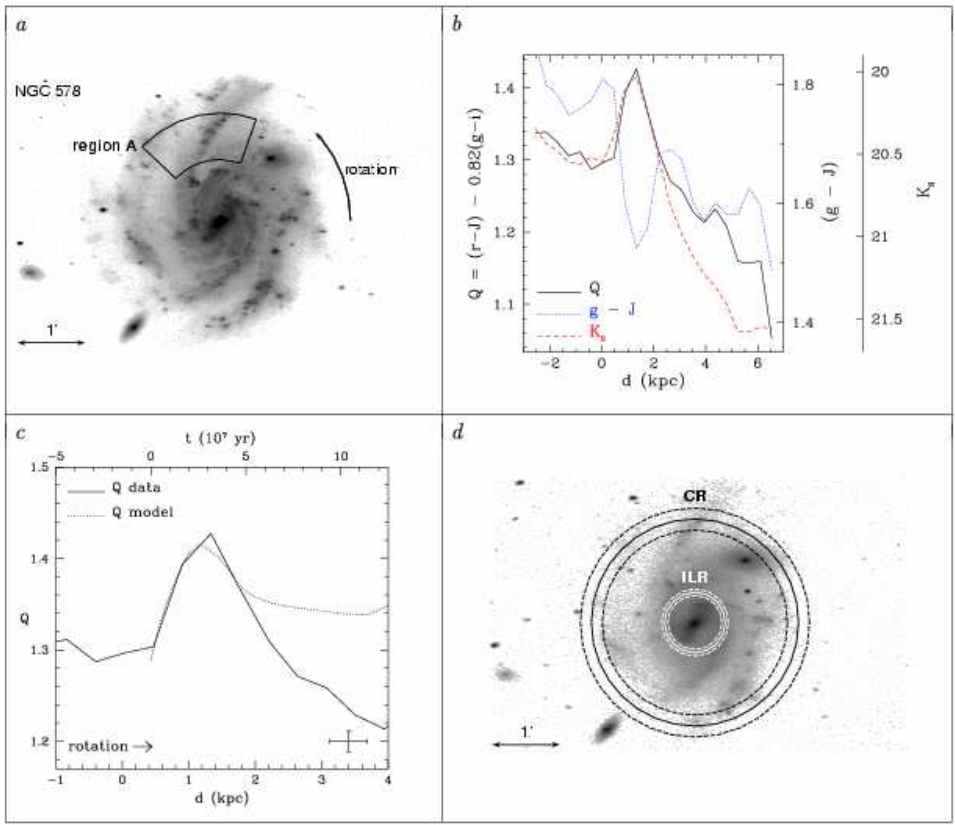}
\caption[f40.eps]{Region NGC 578 A.
Panel $d$,
{\it solid line circles:}
location of the ILR and corotation radius,
as obtained from the comparison between data and
SPS model shown in panel $c$.
\label{REG_578_A}}
\end{figure}

\begin{figure}
\centering
\epsscale{1.0}
\plotone{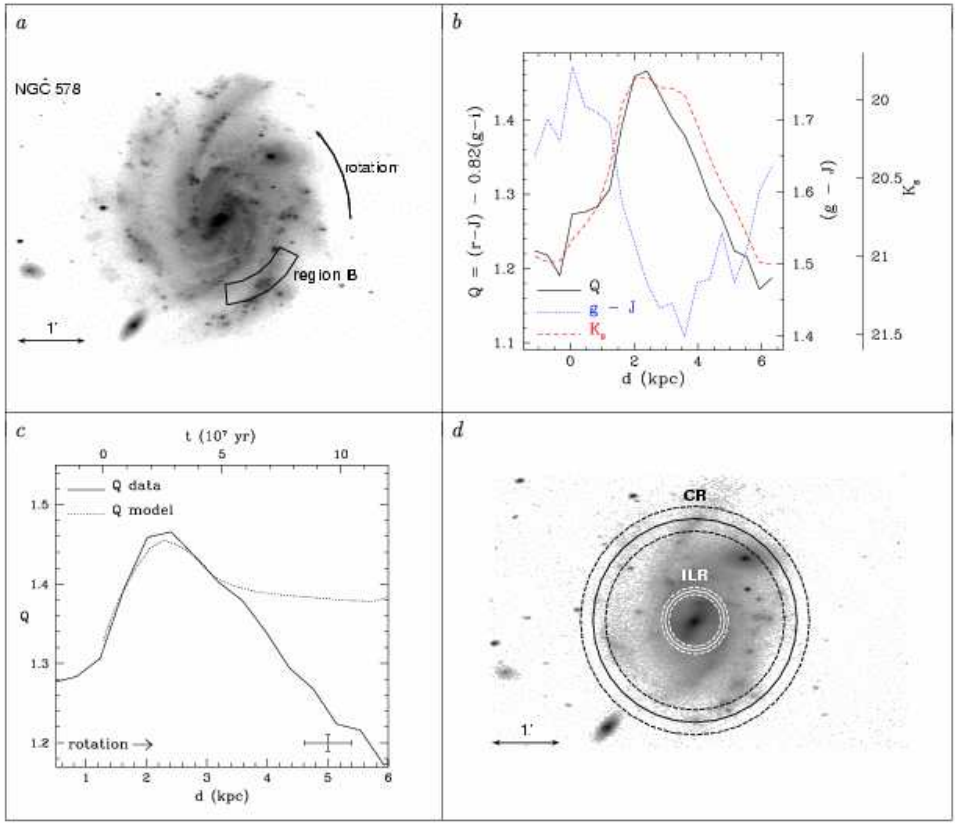}
\caption[f41.eps]{Region NGC 578 B.
Panel $d$,
$K_s$-band deprojected mosaic of spiral galaxy NGC~578.
{\it solid line circles:}
location of the ILR and corotation radius,
as obtained from the comparison between data and
SPS model shown in panel $c$.
\label{REG_578_B}}
\end{figure}

\end{document}